\documentclass[twocolumn,showpacs,superscriptaddress,floatfix,prl]{revtex4-1}
\usepackage{graphicx,amsfonts,amssymb,amsmath,hyperref,hypcap,enumerate}

\usepackage{amsthm}
\usepackage{multirow}
\usepackage{graphicx}
\usepackage{dcolumn}   
\usepackage{bm}        
\usepackage{amssymb}   
\usepackage{amsmath}
\usepackage{epstopdf}
\usepackage{color}
\usepackage{subfigure}
\usepackage{diagbox}

\usepackage{multirow}

\DeclareMathAlphabet{\mathitb}{OT1}{cmr}{bx}{sl}

\begin{document}
	\title{Universality classes of the Anderson Transitions Driven by non-Hermitian Disorder}

\author{Xunlong Luo}
\email{luoxunlong@pku.edu.cn}
\affiliation{Science and Technology on Surface Physics and Chemistry Laboratory, Mianyang 621907, China}

\author{Tomi Ohtsuki}
\email{ohtsuki@sophia.ac.jp}
\affiliation{Physics Division, Sophia University, Chiyoda-ku, Tokyo 102-8554, Japan}

\author{Ryuichi Shindou}
\email{rshindou@pku.edu.cn}
\affiliation{International Center for Quantum Materials, Peking University, Beijing 100871, China}
\affiliation{Collaborative Innovation Center of Quantum Matter, Beijing 100871, China}

\date{\today}
	\begin{abstract}
An interplay between non-Hermiticity and disorder plays an important role in condensed matter physics.
Here, we report the universal critical behaviors of the Anderson transitions driven by non-Hermitian disorders 
for three dimensional (3D) Anderson model and 3D U(1) model, which belong to 
3D class ${\rm AI}^{\dagger}$ and 3D class A in the classification of non-Hermitian systems, respectively. 
Based on level statistics and finite-size scaling analysis, the critical exponent for length scale is estimated as 
$\nu=0.99\pm 0.05$ for class ${\rm AI}^{\dagger}$, and $\nu=1.09\pm 0.05 $ for class A, both of which 
are clearly distinct from the critical exponents for 3D orthogonal and 3D unitary classes, respectively. 
In addition, spectral rigidity, level spacing distribution, and level spacing ratio distribution are studied. 
These critical behaviors strongly support that the non-Hermiticity changes the universality classes 
of the Anderson transitions.
	\end{abstract}
\maketitle
\textit{Introduction}---
Continuous quantum phase transitions are universally characterized by critical exponent (CE) and scaling functions for 
physical observables around the critical point~\cite{sondhi97}. The CE and scaling functions 
represent scaling properties of an underlying effective theory that describes the phase transition, and 
classify the phase transitions in different models in terms of the universality class. The universality class of the 
Anderson transition (AT)~\cite{Anderson58} is determined only by the spatial 
dimension and symmetry of a system~\cite{Abrahams79,Wegner76,Efetov80,Hikami80,Hikami81,
Altland97,Evers08,Kramer93,Slevin14,Slevin97,Slevin16,Asada05,Slevin09,Asada04,Luo19QMCT,Luo20}. 
Recently, the AT in non-Hermitian (NH) system attracts a lot of 
attentions~\cite{Xu16,Tzortzakakis20,Wang20,Huang20,Huang20SR}.
NH systems and localization phenomena therein 
are remarkably ubiquitous in nature, such as random lasers~\cite{Cao99,Wiersma08,Wiersma13}, 
non-equilibrium open systems with gain and/or loss 
\cite{Konotop16,Feng17,El-Ganainy18,Ozdemir19,Miri19}, and correlated quantum many-particle systems of  
quasiparticles with finite life-time \cite{Shen18,Papaj19}. Hatano and Nelson's pioneering work introduced 
a one-dimensional (1D) NH Anderson model with asymmetric hopping potentials \cite{Hatano96}.  
The 1D NH model shows a delocalization-localization transition, contrary to the absence of the AT in 1D Hermitian 
system, indicating that the transition belongs to a new universality class~\cite{Kawabata20}. 
According to recent studies, the non-Hermiticity 
enriches the ten-fold classification scheme of the Hermitian system by Altland and Zirnbauer~\cite{Altland97}
into 38-fold symmetry classes~\cite{Kawabata19,Zhou19}.  

A natural question arises whether the AT in each of these 38-fold symmetry classes in the NH 
system belongs to a new universality class or not, compared with the known universality classes in the Hermitian 
system. A recent work~\cite{Xu16} shows that a NH spin ice model belongs to the same 
universality class as two-dimensional (2D) quantum Hall universality class of the Hermitian system. 
Another recent work~\cite{Huang20} indicates the CE $\nu$ of three-dimensional (3D) NH 
Anderson model to be the same as the CE of the Hermitian Anderson model~\cite{Huang20}. 
These works, at first sight, suggest that the non-Hermiticity does not change the universality class 
of the AT, and the AT in the NH system with the enriched symmetry classes share the same universal critical 
properties as the AT in the corresponding symmetry classes in the Hermitian system. 

In this paper, we show that the non-Hermiticity does change the universality class of the AT. 
By precise estimates of the CE $\nu$ as well as critical level statistics such as 
spectral compressibility, level spacing distribution and level spacing ratio (LSR) distribution, the universal 
critical properties of the AT in the NH systems are shown to be significantly different 
from any of the Hermitian symmetry classes. Here, two symmetry classes are studied 
as an example; 3D class AI$^{\dagger}$ and 3D class A in the NH classification scheme. 
By an accurate calculation of the LSR \cite{Oganesyan07,Sa20,Huang20} and 
polynomial fitting of the data\cite{Slevin14}, $\nu$ is estimated to be $0.99\pm0.05$ for the 
class AI$^\dagger$ and $1.09\pm0.05$ for the class A, which are clearly distinct from the CE of the 3D AT in the orthogonal \cite{Slevin18} and unitary classes\cite{Slevin16} of the Hermitian system, respectively.
We further study the spectral rigidity, level spacing distribution, and LSR distribution. 
These critical level statistics strongly support that non-Hermiticity changes the universality 
class of the AT in 3D class AI$^{\dagger}$ and 3D class A. This paper paves a solid 
path toward a new research paradigm of quantum phase transitions in NH systems, which will 
give a bridge between non-Hermitian random matrix theory and different branches in physics. 

\begin{table*}[t]
	\centering
	\caption{Polynomial fitting results for the level spacing ratio (LSR) around the Anderson transition in 3D class 
        AI$^{\dagger}$ and 3D class A models. The goodness of fit (GOF), critical disorder $W_{c}$, critical 
        exponent $\nu$, the scaling dimension of the least irrelevant scaling variable $-y$, and the critical LSR $\langle r\rangle_c$ are shown for various system sizes and disorder ranges and for different 
		orders of the Taylor expansion of the scaling function for the LSR: $(m_1,n_1,m_2,n_2)$. 
		The square bracket is the 95\% confidence interval.}
	\begin{tabular}{cccccccccccc}
		\hline
		 Symmetry&	$L$& $W$&	$m_1$& $n_1$ & $m_2$ & $n_2$ & GOF & $W_c$ & $\nu$ & $y$&$\langle r\rangle_c$   \\
		\hline
		\multirow{2}*{Class AI$^{\dagger}$}	&8-24& [6, 7.12]&   3&3&0&1&0.11&6.28[6.26, 6.30]&1.046[1.012, 1.086]&1.75[1.65, 1.84]&0.7169[0.7163, 0.7177]	\\
		&10-24& [6, 7.19]&  3&3&0&1&0.15&6.32[6.30, 6.34]&0.990[0.945, 1.040]&2.10[1.87, 2.35]&0.7155[0.7146, 0.7164]	\\
		\hline
		\hline
		\multirow{4}*{Class A}	&8-24& [7, 7.56]& 1&3&0&1&0.32&7.14[7.13, 7.15]&1.065[1.036, 1.100]&2.60[2.31, 2.89]&0.7178[0.7171, 0.7188]\\
		 &	8-24& [7, 7.56]& 2&3&0&1&0.43&7.15[7.14, 7.16]&1.068[1.034, 1.105]&2.63[2.35, 2.92]&0.7177[0.7169, 0.7186]\\
		 &	8-24& [7, 7.56]& 3&3&0&1&0.49&7.15[7.14, 7.16]&1.065[1.031, 1.103]&2.64[2.35, 2.92]&0.7177[0.7169, 0.7186]\\
		 &	10-24& [6.8, 7.6]&3&3&0&1&0.12&7.14[7.12, 7.16]&1.091[1.050, 1.151]&2.50[1.88, 3.16]&0.7187[0.7170, 0.7201]\\
		\hline
	\end{tabular}
	\label{table}
\end{table*}

\textit{Model and numerical method}---
We study the following tight-binding model on a 3D cubic lattice,
\begin{align}
{\cal H}=\sum_i \varepsilon_i c_i^{\dagger} c_i+\sum_{\langle i,j \rangle} e^{2\pi i\cdot \theta_{i,j}}c_i^{\dagger}c_j,
\end{align}
where $c_i^\dagger$ ($c_i$) is the creation (annihilation) operator, and 
$\langle i,j\rangle$ means the nearest neighbor sites with $\theta_{i,j}=-\theta_{j,i}$. 
The AT driven by real-valued random potentials $\varepsilon_i$ belongs to the 3D orthogonal universality 
class with $\theta_{i,j}=0$ and 3D unitary universality class with 
$\theta_{i,j}$ random number in $[0,1)$. In this paper, 
we consider NH disorder, set $\varepsilon_{j}=w_{j}^{r}\,+\,i\, w_{j}^i$ with 
the imaginary unit $i$, where $w_{j}^{r}$ and $w_{j}^{i}$ are independent random numbers 
with identical uniform distribution in $[-W/2,W/2]$ at site $j$. Hence 
${\cal H}\neq {\cal H}^{\dagger}$. The NH random potentials can be physically 
realized in random lasers in random dissipation and amplification region \cite{Cao99,Wiersma08,Wiersma13}.
According to the symmetry classification for NH system \cite{Kawabata19,Zhou19}, the model belongs to 
3D class AI$^{\dagger}$ with $\theta_{i,j}=0$ and 3D class A with $\theta_{i,j}$ random number in $[0,1)$. 
The time reversal symmetry (TRS) is broken (${\cal H}^*\neq {\cal H}$) in the both classes, whereas the 
transposition symmetry (${\cal H}^{\rm T}={\cal H}$), namely TRS$^\dagger$, holds true 
in the class AI$^{\dagger}$.

The AT can be characterized by the energy level statistics 
\cite{Wigner51,Dyson62,Dyson62TFW}. The level statistics in NH disordered 
systems are known in the two limiting cases; it belongs to the Poisson ensemble in the 
localized phase \cite{Grobe88}, while it belongs to the Ginibre ensemble in the 
delocalized phase \cite{Ginibre65}. In this paper, we analyze scaling 
behaviors \cite{Shklovskii93} of the energy level statistics \cite{Wigner51,Dyson62,Dyson62TFW} 
around the AT in the NH systems, where a narrow energy window 
$\{E_i\}$ is set with an assumption that all eigenstates 
within the energy window have a similar critical disorder strength. Eigenvalues of the NH 
system are complex numbers, except for a system with a special symmetry, such as 
${\cal PT}$ symmetry \cite{El-Ganainy18}. Thus, an energy level spacing is defined by 
$s_i \equiv |E_i-E_{\rm NN}|$, where $E_{\rm NN}$ is a complex-valued eigenvalue nearest 
to $E_i$ in the complex Euler plane. In order to exclude an effect of 
the density of states, a procedure called unfolding is often used 
in the literature~\cite{Shklovskii93}. However, the unfolding process causes additional errors, that are crucial for 
our precise estimation of the CE. We thus introduce another dimensionless variable that characterizes 
the AT, the LSR \cite{Oganesyan07,Huang20,Sa20},
$r_i \equiv |z_i|$ with $z_{i}\equiv  \frac{E_i-E_{\rm NN}}{E_i-E_{\rm NNN}}$. 
Here $E_{\rm NNN}$ is a complex-valued eigenvalue that is the next nearest neighbor to $E_i$ in the 
Euler plane. $r_i$ is averaged over the energy window and over $M$ realizations 
of disordered systems, giving a precise mean value $\langle r\rangle$ with a standard deviation 
$\sigma^2_{\langle r\rangle} \equiv \frac{1}{M-1}(\langle r^2\rangle-\langle r\rangle^2)$. 
\begin{figure}
	\centering
	\includegraphics[width=1\linewidth]{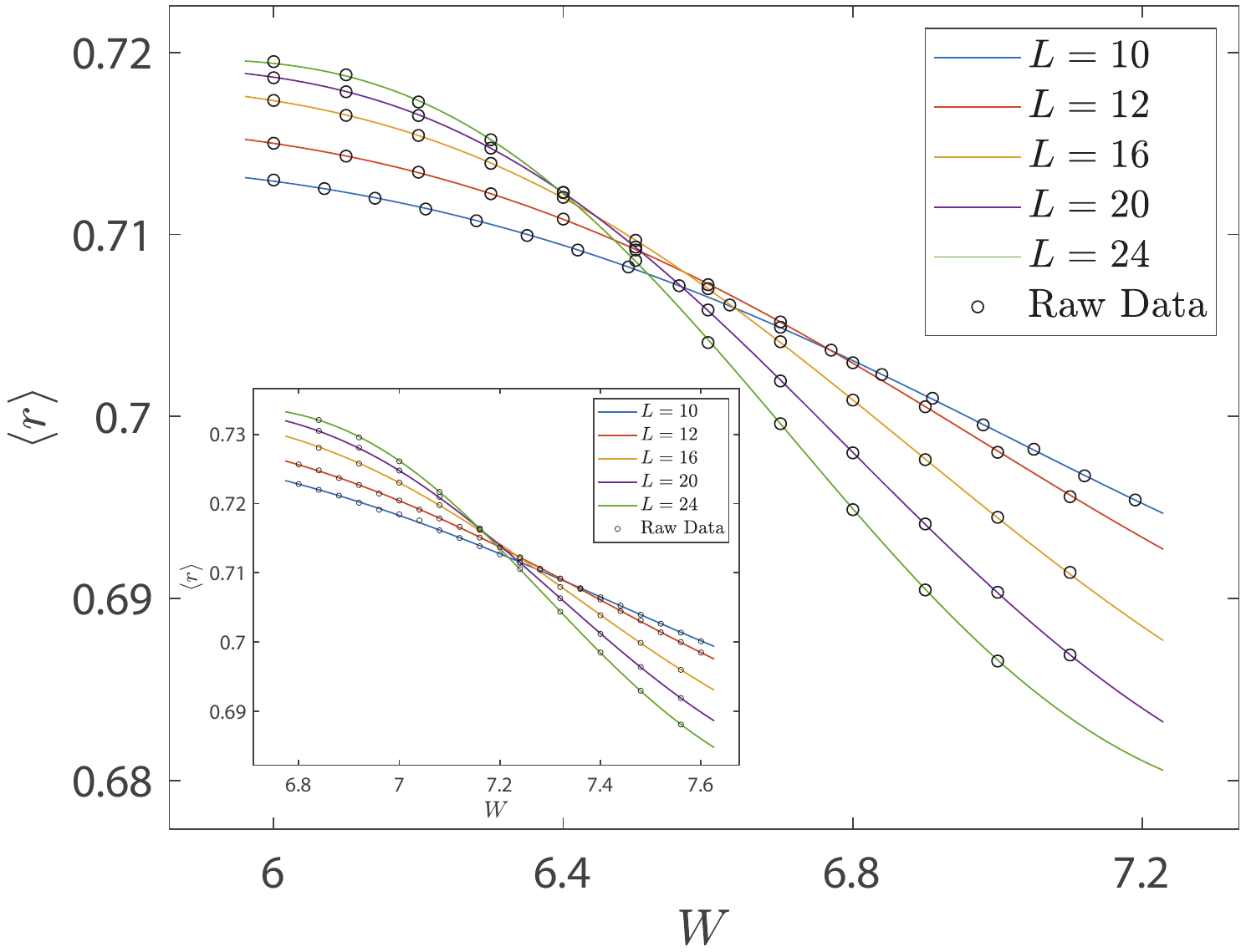}
	\caption{Level spacing ratio $\langle r \rangle$ as a function of the disorder strength $W$ for 
        the class AI$^\dagger$ model. The circles are for raw data of $\langle r \rangle$, where an error is smaller 
        than the circle size. The curves are from the polynomial fitting results with $m_1, n_1 , m_2, n_2$=(3, 3, 0, 1).
	   Inset: the same plot for class A.}
	\label{O1_U1_r}
\end{figure}

\textit{Numerical result and polynomial fitting}---
In order to obtain large number of eigenvalues for the level 
statistics and also guarantee that their eigenstates share almost similar critical disorder, 
we choose the energy window to be $10\%$ eigenvalues around $E=0$ in the complex Euler plane. 
$M$ is chosen in such a way that the total number 
of the eigenvalues reaches $5\times 10^7$ ($L < 24$) and 
$10^7$ ($L=24$) for the class AI$^\dagger$, and $10^7$ for the 
class A \cite{supplemental}. Fig.~\ref{O1_U1_r} shows a plot of $\langle r \rangle$ as a function of disorder strength with the various system sizes. 
The plots for both class AI$^{\dagger}$ and class A models
show critical points $W_c$, where the scale-invariant quantity 
$\langle r\rangle$ does not change with the system size $L$. 
We note that $W_c$ in the class A model is larger than that in the class AI$^{\dagger}$ model
even though the former contains more randomness in the transfer.  This is similar
to the AT in Hermitian systems, and indicates that the AT in NH systems is
also caused by quantum interference.

For localized phase ($W>W_c$), different energy levels have less correlations 
because of  exponentially small overlap between eigenfunctions. In the thermodynamic 
limit, the nearest neighbor and next nearest neighbor levels become independent, and 
$z_i$ is equally distributed within a circle with radius one in the complex Euler plane. 
$\langle r\rangle_{\rm insulator} =\int_{0}^{1}r\rho(r)dr=2/3$ with 
$\rho(r)=2r$ the density of $r$ in the complex plane. We confirmed 
$\langle r\rangle\approx 0.66$ for strong disorder for  both symmetry 
classes~\cite{supplemental}. On the other hand, the energy levels are correlated in 
delocalized phase ($W<W_c$) because of a spatial overlap between eigenfunctions. The 
overlap causes an level repulsion between the energy levels, which generally 
makes $\rho(r)$ near $r=0$ to be smaller in the delocalized phase than in the localized phase. 
Thus, $\langle r\rangle_{\rm metal}$ tends to be larger than $\langle r\rangle_{\rm insulator}$.  
We observed that $\langle r\rangle$ reaches a constant value in the metal phase 
in  both models, where the constant value increases with the system size~\cite{supplemental}. 
In the thermodynamic limit,  $\langle r\rangle$ in the metal phase reaches a certain universal value.  
This is analogous to metal phases of Hermitian systems in the three Wigner-Dyson (WD) 
classes~\cite{Atas13}. 
Our calculation with the largest system size shows 
$\langle r\rangle_{\rm metal}\approx 0.720$ for the class AI$^{\dagger}$ model 
and $\langle r\rangle_{\rm metal}\approx0.736$ for the class A model~\cite{supplemental}. 
The different values of $\langle r \rangle_{\rm metal}$ in the thermodynamic limit 
indicates that the two models belong to the different classes.  

The LSR $\langle r\rangle$ takes a size-independent universal value at the 
critical point $W=W_c$ (TABLE \ref{table}). The critical LSR as well as the 
CE are evaluated in terms of the polynomial fitting method~\cite{Slevin14}.  
The criticality in each model is controlled by a saddle-point 
fixed point of a renormalization group equation for a certain effective theory, which describes 
the AT of the model.  A standard scaling argument around the saddle-point fixed point 
gives  $\langle r\rangle$ near the critical point by a universal function 
$\langle r\rangle  = F(\phi_1,\phi_2)$. Thereby, $\phi_1 \equiv u_1(w) L^{1/\nu}$ 
and $\phi_2 \equiv u_2(w) L^{-y}$ stand for a relevant and the least irrelevant scaling variable 
around the postulated saddle-point fixed point; $1/\nu \!\ (>0)$ and $-y \!\ (<0)$ are the scaling 
dimensions of the relevant and the irrelevant scaling variables around the fixed point. $w$ is a 
normalized distance from the critical point; $w \equiv (W-W_c)/W_c$. 
When $W$ is close enough to the critical disorder strength $W_c$, $u_{1}(w)$ and $u_{2}(w)$ can 
be Taylor expanded in small $w$. By definition, the expansions take forms of 
$u_{i}(w) \equiv \sum^{m_i}_{j=0} b_{i,j} w^{j}$ with $i=1,2$, $b_{1,0}=0$ and $b_{2,0} \ne 0$. 
For smaller $w$ and larger $L$, the universal function can be further expanded in small $\phi_1$ and 
$\phi_2$ as $F = \sum^{n_1}_{j_1=0} \sum^{n_2}_{j_2=0} a_{j_1,j_2} \phi^{j_1}_1 \phi^{j_2}_2$.
For a given set of $(n_1,n_2,m_1,m_2)$, 
$\chi^2 \equiv \sum^{N_D}_{k=1} (F_{k}-\langle r\rangle _{k})^2/\sigma^2_{\langle r\rangle_{k}}$
is minimized in terms of $W_c$, $\nu$, $-y$, $a_{i,j}$ and $b_{i,j}$ ($a_{1,0}=a_{0,1}=1$). 
Here each data point $k$ ($k=1,\cdots, N_D$) is specified by $L$ and $W$. 
$\langle r\rangle _{k}$ and $\sigma_{\langle r\rangle_{k}}$ are the mean value and the standard deviation at $k=(L,W)$, 
respectively, while $F_{k}$ is a fitting value from the polynomial expansion of $F$ at $k=(L,W)$. 
Fittings are carried out for several different $(n_1,n_2,m_1,m_2)$. Table~\ref{table} shows the fitting results 
with goodness of fit greater than 0.1. The 95$\%$ confidence intervals are determined by 1000 sets of $N_D$ 
number of synthetic data that are generated from the mean value and the standard deviation.   
$W_c$, $\nu$, $y$ and $\langle r\rangle_c$ are shown to be robust against the change of the 
expansion order and various system size and disorder range. We also confirm that our estimation 
is stable against changing the size of the energy windows \cite{supplemental}.

The CE $\nu$ of the AT is evaluated as $\nu=0.99 \pm 0.05$ for the 
3D class AI$^{\dagger}$ and $\nu=1.09 \pm 0.05$ for the 3D class A model, 
which are clearly distinct from $\nu=1.57\pm0.01$ for the 3D orthogonal class 
\cite{Slevin18}, and $\nu=1.44\pm0.01$ for the 3D unitary class \cite{Slevin97, Slevin16} respectively.
This unambiguously concludes that the AT in 3D class AI$^{\dagger}$ as well as 3D class A belongs to 
a new universality class that is different from any of the WD universality classes and in this 
respect, our result has confirmed that the non-Hermiticity changes the universality classes of the AT.
It is also intriguing to see whether the AT in the 3D class AI$^{\dagger}$ and that in the 3D class A 
belong to the same universality class or not. However, our estimation of $\nu$ and 
$\langle r\rangle_c$ are quite close to each other within the 95$\%$ confidence intervals and it 
is hard to give a definite answer to this question. To answer this important question, we study in the 
following the spectral rigidity, level spacing distribution, and LSR distribution at the 
critical points of the two models.
 
\textit{Spectral rigidity}--- The spectral rigidity is defined by number variance
$\Sigma_{2}\equiv \langle \delta N^2 \rangle=\langle (N-\langle N\rangle )^2 \rangle $,
where $N$ is the number of eigenvalues in a fixed energy window and 
$\langle N \rangle$ stands for $N$ averaged over different disorder realizations.
The spectral compressibility $\chi$ can be extracted by 
$\chi\equiv \lim_{L\rightarrow\infty} \lim_{N\rightarrow\infty}
\frac{d \Sigma_2(N)}{d \langle N\rangle}$.  
Energy levels in insulator phase have less correlations and they 
show $\Sigma_2=\langle N\rangle$ in the thermodynamic limit. In metal phase, 
energy levels show the repulsive correlation, where 
$\Sigma_2 \sim \ln(\langle N\rangle)$ and $\chi$ goes to the zero in the large 
$N$ limit.
\begin{figure}[t]
	\centering
	\includegraphics[width=1\linewidth]{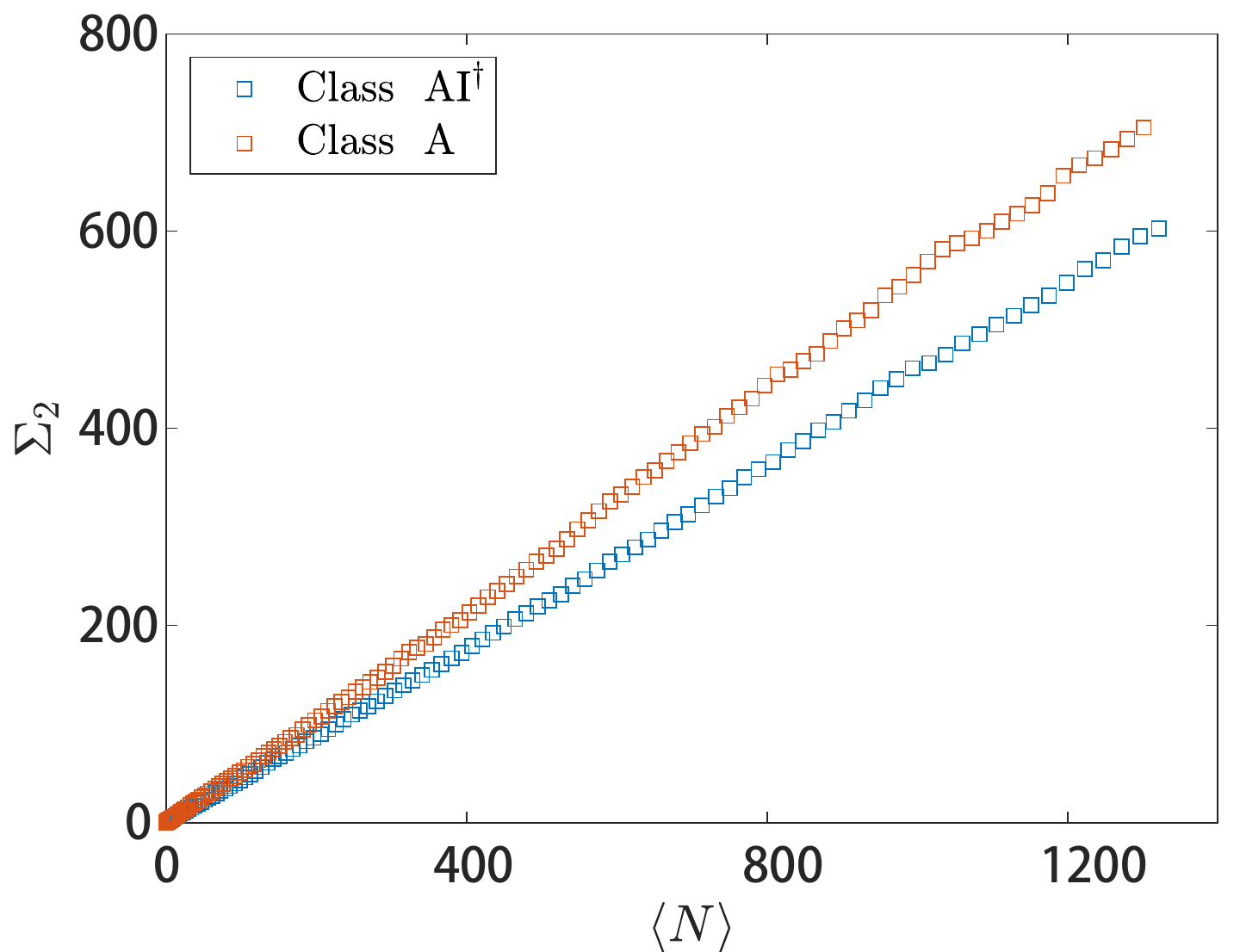}
	\caption{Number variance $\Sigma_{2}$ as a function of averaged level number $\langle N\rangle$ 
     at the critical point ($W=6.3$ for class AI$^{\dagger}$ and $W=7.16$ for class A), and 
     $\Sigma_{2} = \chi \langle N\rangle$ with $\chi\approx 0.46$ for class AI$^{\dagger}$ 
      and $\chi\approx 0.55$ for class A. The plot comes from $10^4$ samples for class 
      AI$^{\dagger}$ and 6400 samples for class A with $L=24$. Variable $\langle N\rangle$ is 
      obtained by changing the energy window within the $10\%$ eigenvalues around $E=0$. 
     The linear relationship holds true for the system size $L\ge 8$ with a consistent $\chi$~\cite{supplemental}.}
	\label{varN}
\end{figure}
At the critical point, $\chi$ takes a universal value and it has been conjectured 
that $\chi$ is related with multifractal dimensions $D_q$ \cite{Rodriguez11} as $2\chi+D_2/d=1$ \cite{Chalker96,Chalker96PRL,Chalker96_SR} 
and $\chi+D_1/d=1$ \cite{Bogomolny11}.
Fig.~\ref{varN} shows that $\Sigma_{2}$ at the critical point for the both NH systems 
is indeed linear in $\langle N\rangle$ in the the large $N$ limit. $\chi$ is extracted by a linear fitting, as 
$\chi\approx 0.46$ for the class AI$^{\dagger}$ case~\cite{Huang20SR}, and $\chi\approx 0.55$ 
for the class A case. These two values are clearly different from each other, and they are also distinct 
from the Hermitian cases; $\chi\approx0.28$ 
for 3D orthogonal class~\cite{Braun98,Ndawana02,Bogomolny11,Ghosh17,supplemental}, 
and $\chi\approx 0.31$ for 3D unitary class~\cite{supplemental}.

\textit{Level spacing distribution}---
A level spacing distribution $P(s)$ plays an essential role in characterizing the AT in the Hermitian systems.  
$P(s)$ in metal phase can be described by the WD surmise \cite{Wigner51,Dyson62,Dyson62TFW} in random matrix theory, 
$P(s)=a_{\beta} s^{\beta} e^{ -b_{\beta} s^2}$, 
where the Dyson index $\beta=1,2,4$ for orthogonal, unitary, and symplectic class, respectively.
At the critical point, $P(s) \propto s^{\beta_c}$ for small $s$ region, where 
$\beta_c$ for each of the three classes are almost the same as the respective Dyson index $\beta$ in 
the metal phase \cite{Zharekeshev95,Kawarabayashi96}. For larger $s$ region, 
$P(s) \propto e^{-\alpha s}$ with almost an identical value of $\alpha$ for these three WD 
symmetry classes; $\alpha= 1.8\pm 0.1$~\cite{supplemental,Batsch96,Hofstetter96,Zharekeshev95}.

\begin{figure}
	\centering
	\includegraphics[width=1\linewidth]{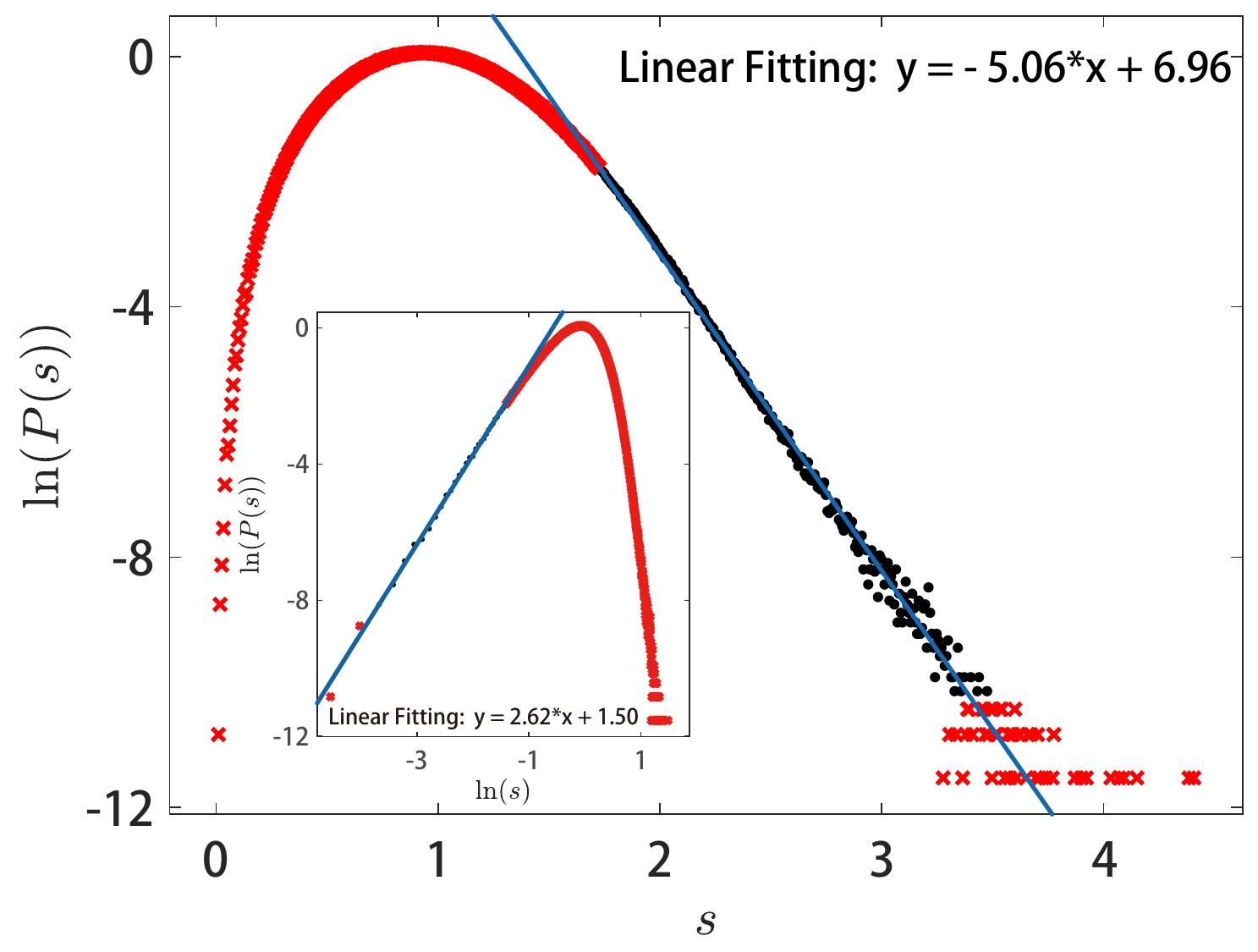}
	\caption{Critical level spacing distribution $P(s)$ at large $s$ and small $s$ (inset) for class AI$^{\dagger}$ at $W_c=6.3$. 
		The small-$s$ behavior of $P(s)$ is fitted by $P(s) \propto s^{\beta_c}$ with $\beta_c\approx 2.62$ (blue solid line in the 
        inset), and the large-$s$ behavior of $P(s)$ is fitted by $P(s) \propto e^{-\alpha s}$ with $\alpha\approx 5.06$ (blue solid 
        line). The distribution is obtained from $10\%$ eigenvalues around $E=0$ of $10^4$ disorder realizations with $L=24$. 
        Red crosses are data excluded from the linear fitting. Similar critical behaviors but with different 
        $\beta_c$ and $\alpha$ are also observed for class A \cite{supplemental}.}
	\label{Ps}
\end{figure}

For the NH systems, things become more interesting. Our numerical results of $P(s)$ in insulator 
phase shows a 2D Poisson distribution 
\cite{Grobe88}, $P_{P}^{2D}(s)=\frac{\pi}{2}se^{-\pi s^2/4}$ 
for both classes \cite{supplemental}. In metal phase, $P(s)$ for 
class A case \cite{supplemental} follows the statistics of Ginibre ensemble\cite{Ginibre65} with cubic repulsion ($\beta=3$) for 
small $s$ \cite{Grobe88,Grobe89,Akemann19}, but not for the class AI$^{\dagger}$ case 
\cite{supplemental}. This implies that the two classes belong to different symmetry classes 
according to level spacing distribution \cite{Hamazaki20}. At the critical point, the same asymptotic behaviors 
of $P(s)$ at small and large $s$ regions as in the Hermitian case hold true for the NH case with 
different values of $\alpha$ and $\beta_c$ (Fig. \ref{Ps}).  Our numerical result shows that 
$\alpha=5.0\pm0.1$ for class AI$^{\dagger}$, $\alpha=4.5\pm0.1$ for class A \cite{supplemental}, 
which are larger than those for the three WD classes \cite{supplemental,Batsch96,Hofstetter96,Zharekeshev95}.
We also find $\beta_c=2.6\pm 0.05$ for class AI$^{\dagger}$ and $\beta_c=2.9\pm 0.05$ for class A, 
which are also different from $\beta\approx 1$ for 3D orthogonal class and $\beta\approx 2$ for 
3D unitary class, respectively~\cite{supplemental}.

\textit{level spacing ratio distribution}---
The complex ratio $z_i$ contains information of its modulus $r_i \equiv |z_i|$ and angle $\theta_i \equiv {\rm arg}(z_i)$.
In insulator phase, $z_i$ is equally distributed in the complex plane due to the absence of the energy level correlation; 
$P(r)=2r$, and $P(\theta)=\frac{1}{2\pi}$ for  both of the classes \cite{supplemental}. 
In metal phase, $P(r)$ and $P(\theta)$ for the class A case is
consistent with that for the Ginibre ensemble, but not for the  
class AI$^{\dagger}$ \cite{supplemental}. The behaviors of $P(r)$ and $P(\theta)$ here 
are similar to $P(s)$, and all the three distributions 
exhibit the unique universal features in the metal phases of the 
 AI$^{\dagger}$ and A classes. At the critical point, both $P(r)$ and $P(\theta)$ are independent of the system sizes
for both classes \cite{supplemental}, except for $P(\theta)$ with small deviation at two edges caused 
by the boundary effect \cite{Sa20}. We found it hard to distinguish the universality class of the AT in the 
class AI$^{\dagger}$ and that in the class A by the critical distributions of $P(r)$ and $P(\theta)$.

\textit{Summary}---
The Anderson transition driven by non-Hermitian disorder is studied by level statistics for 3D 
class AI$^{\dagger}$ and class A models. Critical exponents $\nu$ are estimated from the 
LSR by the polynomial fitting method and the estimated values conclude that 
the AT in these NH systems belong to new universality classes. Our estimation of $\nu$ for 
class AI$^{\dagger}$ is at variance with the preceding study~\cite{Huang20}, and we believe 
the discrepancy comes from the insufficient accuracy. 
Critical spectral compressibility is evaluated as $\chi\approx 0.46$ for class AI$^{\dagger}$ and 
$\chi\approx 0.55$ for class A, which are larger than those for 3D orthogonal and unitary classes.
How the multifractal dimensions for the NH systems are related to the spectral compressibility at the critical point is
an interesting open problem left for the future.
The critical behavior of $P(s)$ at small and large $s$ regions in the NH systems are characterized by exponents 
$\beta_c$ and $\alpha$ as in the Hermitian case. Our numerical result for $\beta_c$ and $\alpha$ in the class AI$^{\dagger}$ and A  are clearly distinct from each other and they are larger than those 
for 3D orthogonal and unitary classes. $P(s)$, $P(r)$ and $P(\theta)$ of class A in the metal phase 
are consistent with the statistics of the Ginibre ensemble, but those of the class AI$^{\dagger}$ are not. 
All the estimated critical values of $\nu$, $\chi$, $\beta_c$ and $\alpha$ conclude that the non-Hermiticity 
changes the universality class of the AT for 3D class AI$^{\dagger}$ and 3D class A.	

\textit{Acknowledgment}--- 
X. L. thanks fruitful discussions with Dr. Yi Huang. X. L. was supported by National Natural Science Foundation of China of Grant No.51701190.
T. O. was supported by JSPS KAKENHI Grants No. 16H06345 and 19H00658. R. S. was supported by 
the National Basic Research Programs of China (No. 2019YFA0308401) and by National Natural Science 
Foundation of China (No.11674011 and No. 12074008). 

\bibliography{paper}

\begin{thebibliography}{63}%
\makeatletter
\providecommand \@ifxundefined [1]{%
 \@ifx{#1\undefined}
}%
\providecommand \@ifnum [1]{%
 \ifnum #1\expandafter \@firstoftwo
 \else \expandafter \@secondoftwo
 \fi
}%
\providecommand \@ifx [1]{%
 \ifx #1\expandafter \@firstoftwo
 \else \expandafter \@secondoftwo
 \fi
}%
\providecommand \natexlab [1]{#1}%
\providecommand \enquote  [1]{``#1''}%
\providecommand \bibnamefont  [1]{#1}%
\providecommand \bibfnamefont [1]{#1}%
\providecommand \citenamefont [1]{#1}%
\providecommand \href@noop [0]{\@secondoftwo}%
\providecommand \href [0]{\begingroup \@sanitize@url \@href}%
\providecommand \@href[1]{\@@startlink{#1}\@@href}%
\providecommand \@@href[1]{\endgroup#1\@@endlink}%
\providecommand \@sanitize@url [0]{\catcode `\\12\catcode `\$12\catcode
  `\&12\catcode `\#12\catcode `\^12\catcode `\_12\catcode `\%12\relax}%
\providecommand \@@startlink[1]{}%
\providecommand \@@endlink[0]{}%
\providecommand \url  [0]{\begingroup\@sanitize@url \@url }%
\providecommand \@url [1]{\endgroup\@href {#1}{\urlprefix }}%
\providecommand \urlprefix  [0]{URL }%
\providecommand \Eprint [0]{\href }%
\providecommand \doibase [0]{http://dx.doi.org/}%
\providecommand \selectlanguage [0]{\@gobble}%
\providecommand \bibinfo  [0]{\@secondoftwo}%
\providecommand \bibfield  [0]{\@secondoftwo}%
\providecommand \translation [1]{[#1]}%
\providecommand \BibitemOpen [0]{}%
\providecommand \bibitemStop [0]{}%
\providecommand \bibitemNoStop [0]{.\EOS\space}%
\providecommand \EOS [0]{\spacefactor3000\relax}%
\providecommand \BibitemShut  [1]{\csname bibitem#1\endcsname}%
\let\auto@bib@innerbib\@empty
\bibitem [{\citenamefont {Sondhi}\ \emph {et~al.}(1997)\citenamefont {Sondhi},
  \citenamefont {Girvin}, \citenamefont {Carini},\ and\ \citenamefont
  {Shahar}}]{sondhi97}%
  \BibitemOpen
  \bibfield  {author} {\bibinfo {author} {\bibfnamefont {S.~L.}\ \bibnamefont
  {Sondhi}}, \bibinfo {author} {\bibfnamefont {S.~M.}\ \bibnamefont {Girvin}},
  \bibinfo {author} {\bibfnamefont {J.~P.}\ \bibnamefont {Carini}}, \ and\
  \bibinfo {author} {\bibfnamefont {D.}~\bibnamefont {Shahar}},\ }\href
  {\doibase 10.1103/RevModPhys.69.315} {\bibfield  {journal} {\bibinfo
  {journal} {Rev. Mod. Phys.}\ }\textbf {\bibinfo {volume} {69}},\ \bibinfo
  {pages} {315} (\bibinfo {year} {1997})}\BibitemShut {NoStop}%
\bibitem [{\citenamefont {Anderson}(1958)}]{Anderson58}%
  \BibitemOpen
  \bibfield  {author} {\bibinfo {author} {\bibfnamefont {P.~W.}\ \bibnamefont
  {Anderson}},\ }\href {\doibase 10.1103/PhysRev.109.1492} {\bibfield
  {journal} {\bibinfo  {journal} {Phys. Rev.}\ }\textbf {\bibinfo {volume}
  {109}},\ \bibinfo {pages} {1492} (\bibinfo {year} {1958})}\BibitemShut
  {NoStop}%
\bibitem [{\citenamefont {Abrahams}\ \emph {et~al.}(1979)\citenamefont
  {Abrahams}, \citenamefont {Anderson}, \citenamefont {Licciardello},\ and\
  \citenamefont {Ramakrishnan}}]{Abrahams79}%
  \BibitemOpen
  \bibfield  {author} {\bibinfo {author} {\bibfnamefont {E.}~\bibnamefont
  {Abrahams}}, \bibinfo {author} {\bibfnamefont {P.~W.}\ \bibnamefont
  {Anderson}}, \bibinfo {author} {\bibfnamefont {D.~C.}\ \bibnamefont
  {Licciardello}}, \ and\ \bibinfo {author} {\bibfnamefont {T.~V.}\
  \bibnamefont {Ramakrishnan}},\ }\href {\doibase 10.1103/PhysRevLett.42.673}
  {\bibfield  {journal} {\bibinfo  {journal} {Phys. Rev. Lett.}\ }\textbf
  {\bibinfo {volume} {42}},\ \bibinfo {pages} {673} (\bibinfo {year}
  {1979})}\BibitemShut {NoStop}%
\bibitem [{\citenamefont {Wegner}(1976)}]{Wegner76}%
  \BibitemOpen
  \bibfield  {author} {\bibinfo {author} {\bibfnamefont {F.~J.}\ \bibnamefont
  {Wegner}},\ }\href {\doibase 10.1007/BF01315248} {\bibfield  {journal}
  {\bibinfo  {journal} {Zeitschrift fur Physik B}\ }\textbf {\bibinfo {volume}
  {25}},\ \bibinfo {pages} {327} (\bibinfo {year} {1976})}\BibitemShut
  {NoStop}%
\bibitem [{\citenamefont {Effetov}\ \emph {et~al.}(1980)\citenamefont
  {Effetov}, \citenamefont {Larkin},\ and\ \citenamefont
  {Khmel'nitsukii}}]{Efetov80}%
  \BibitemOpen
  \bibfield  {author} {\bibinfo {author} {\bibfnamefont {K.~B.}\ \bibnamefont
  {Effetov}}, \bibinfo {author} {\bibfnamefont {A.~I.}\ \bibnamefont {Larkin}},
  \ and\ \bibinfo {author} {\bibfnamefont {D.~E.}\ \bibnamefont
  {Khmel'nitsukii}},\ }\href@noop {} {\bibfield  {journal} {\bibinfo  {journal}
  {Soviet Phys. JETP}\ }\textbf {\bibinfo {volume} {52}},\ \bibinfo {pages}
  {568} (\bibinfo {year} {1980})}\BibitemShut {NoStop}%
\bibitem [{\citenamefont {Hikami}\ \emph {et~al.}(1980)\citenamefont {Hikami},
  \citenamefont {Larkin},\ and\ \citenamefont {Nagaoka}}]{Hikami80}%
  \BibitemOpen
  \bibfield  {author} {\bibinfo {author} {\bibfnamefont {S.}~\bibnamefont
  {Hikami}}, \bibinfo {author} {\bibfnamefont {A.~I.}\ \bibnamefont {Larkin}},
  \ and\ \bibinfo {author} {\bibfnamefont {Y.}~\bibnamefont {Nagaoka}},\ }\href
  {\doibase 10.1143/PTP.63.707} {\bibfield  {journal} {\bibinfo  {journal}
  {Progress of Theoretical Physics}\ }\textbf {\bibinfo {volume} {63}},\
  \bibinfo {pages} {707} (\bibinfo {year} {1980})}\BibitemShut {NoStop}%
\bibitem [{\citenamefont {Hikami}(1981)}]{Hikami81}%
  \BibitemOpen
  \bibfield  {author} {\bibinfo {author} {\bibfnamefont {S.}~\bibnamefont
  {Hikami}},\ }\href@noop {} {\bibfield  {journal} {\bibinfo  {journal} {Phys.
  Rev. B}\ }\textbf {\bibinfo {volume} {24}},\ \bibinfo {pages} {2671}
  (\bibinfo {year} {1981})}\BibitemShut {NoStop}%
\bibitem [{\citenamefont {Altland}\ and\ \citenamefont
  {Zirnbauer}(1997)}]{Altland97}%
  \BibitemOpen
  \bibfield  {author} {\bibinfo {author} {\bibfnamefont {A.}~\bibnamefont
  {Altland}}\ and\ \bibinfo {author} {\bibfnamefont {M.~R.}\ \bibnamefont
  {Zirnbauer}},\ }\href {\doibase 10.1103/PhysRevB.55.1142} {\bibfield
  {journal} {\bibinfo  {journal} {Phys. Rev. B}\ }\textbf {\bibinfo {volume}
  {55}},\ \bibinfo {pages} {1142} (\bibinfo {year} {1997})}\BibitemShut
  {NoStop}%
\bibitem [{\citenamefont {Evers}\ and\ \citenamefont {Mirlin}(2008)}]{Evers08}%
  \BibitemOpen
  \bibfield  {author} {\bibinfo {author} {\bibfnamefont {F.}~\bibnamefont
  {Evers}}\ and\ \bibinfo {author} {\bibfnamefont {A.~D.}\ \bibnamefont
  {Mirlin}},\ }\href@noop {} {\bibfield  {journal} {\bibinfo  {journal} {Rev.
  Mod. Phys.}\ }\textbf {\bibinfo {volume} {80}},\ \bibinfo {pages} {1355}
  (\bibinfo {year} {2008})}\BibitemShut {NoStop}%
\bibitem [{\citenamefont {Kramer}\ and\ \citenamefont
  {MacKinnon}(1993)}]{Kramer93}%
  \BibitemOpen
  \bibfield  {author} {\bibinfo {author} {\bibfnamefont {B.}~\bibnamefont
  {Kramer}}\ and\ \bibinfo {author} {\bibfnamefont {A.}~\bibnamefont
  {MacKinnon}},\ }\href {http://stacks.iop.org/0034-4885/56/i=12/a=001}
  {\bibfield  {journal} {\bibinfo  {journal} {Reports on Progress in Physics}\
  }\textbf {\bibinfo {volume} {56}},\ \bibinfo {pages} {1469} (\bibinfo {year}
  {1993})}\BibitemShut {NoStop}%
\bibitem [{\citenamefont {Slevin}\ and\ \citenamefont
  {Ohtsuki}(2014)}]{Slevin14}%
  \BibitemOpen
  \bibfield  {author} {\bibinfo {author} {\bibfnamefont {K.}~\bibnamefont
  {Slevin}}\ and\ \bibinfo {author} {\bibfnamefont {T.}~\bibnamefont
  {Ohtsuki}},\ }\href {http://stacks.iop.org/1367-2630/16/i=1/a=015012}
  {\bibfield  {journal} {\bibinfo  {journal} {New Journal of Physics}\ }\textbf
  {\bibinfo {volume} {16}},\ \bibinfo {pages} {015012} (\bibinfo {year}
  {2014})}\BibitemShut {NoStop}%
\bibitem [{\citenamefont {Slevin}\ and\ \citenamefont
  {Ohtsuki}(1997)}]{Slevin97}%
  \BibitemOpen
  \bibfield  {author} {\bibinfo {author} {\bibfnamefont {K.}~\bibnamefont
  {Slevin}}\ and\ \bibinfo {author} {\bibfnamefont {T.}~\bibnamefont
  {Ohtsuki}},\ }\href {\doibase 10.1103/PhysRevLett.78.4083} {\bibfield
  {journal} {\bibinfo  {journal} {Phys. Rev. Lett.}\ }\textbf {\bibinfo
  {volume} {78}},\ \bibinfo {pages} {4083} (\bibinfo {year}
  {1997})}\BibitemShut {NoStop}%
\bibitem [{\citenamefont {Slevin}\ and\ \citenamefont
  {Ohtsuki}(2016)}]{Slevin16}%
  \BibitemOpen
  \bibfield  {author} {\bibinfo {author} {\bibfnamefont {K.}~\bibnamefont
  {Slevin}}\ and\ \bibinfo {author} {\bibfnamefont {T.}~\bibnamefont
  {Ohtsuki}},\ }\href {\doibase 10.7566/JPSJ.85.104712} {\bibfield  {journal}
  {\bibinfo  {journal} {Journal of the Physical Society of Japan}\ }\textbf
  {\bibinfo {volume} {85}},\ \bibinfo {pages} {104712} (\bibinfo {year}
  {2016})}\BibitemShut {NoStop}%
\bibitem [{\citenamefont {Asada}\ \emph {et~al.}(2005)\citenamefont {Asada},
  \citenamefont {Slevin},\ and\ \citenamefont {Ohtsuki}}]{Asada05}%
  \BibitemOpen
  \bibfield  {author} {\bibinfo {author} {\bibfnamefont {Y.}~\bibnamefont
  {Asada}}, \bibinfo {author} {\bibfnamefont {K.}~\bibnamefont {Slevin}}, \
  and\ \bibinfo {author} {\bibfnamefont {T.}~\bibnamefont {Ohtsuki}},\ }\href
  {\doibase 10.1143/JPSJS.74S.238} {\bibfield  {journal} {\bibinfo  {journal}
  {Journal of the Physical Society of Japan}\ }\textbf {\bibinfo {volume}
  {74}},\ \bibinfo {pages} {238} (\bibinfo {year} {2005})}\BibitemShut
  {NoStop}%
\bibitem [{\citenamefont {Slevin}\ and\ \citenamefont
  {Ohtsuki}(2009)}]{Slevin09}%
  \BibitemOpen
  \bibfield  {author} {\bibinfo {author} {\bibfnamefont {K.}~\bibnamefont
  {Slevin}}\ and\ \bibinfo {author} {\bibfnamefont {T.}~\bibnamefont
  {Ohtsuki}},\ }\href {\doibase 10.1103/PhysRevB.80.041304} {\bibfield
  {journal} {\bibinfo  {journal} {Phys. Rev. B}\ }\textbf {\bibinfo {volume}
  {80}},\ \bibinfo {pages} {041304} (\bibinfo {year} {2009})}\BibitemShut
  {NoStop}%
\bibitem [{\citenamefont {Asada}\ \emph {et~al.}(2004)\citenamefont {Asada},
  \citenamefont {Slevin},\ and\ \citenamefont {Ohtsuki}}]{Asada04}%
  \BibitemOpen
  \bibfield  {author} {\bibinfo {author} {\bibfnamefont {Y.}~\bibnamefont
  {Asada}}, \bibinfo {author} {\bibfnamefont {K.}~\bibnamefont {Slevin}}, \
  and\ \bibinfo {author} {\bibfnamefont {T.}~\bibnamefont {Ohtsuki}},\ }\href
  {\doibase 10.1103/PhysRevB.70.035115} {\bibfield  {journal} {\bibinfo
  {journal} {Phys. Rev. B}\ }\textbf {\bibinfo {volume} {70}},\ \bibinfo
  {pages} {035115} (\bibinfo {year} {2004})}\BibitemShut {NoStop}%
\bibitem [{\citenamefont {Luo}\ \emph {et~al.}(2018)\citenamefont {Luo},
  \citenamefont {Xu}, \citenamefont {Ohtsuki},\ and\ \citenamefont
  {Shindou}}]{Luo19QMCT}%
  \BibitemOpen
  \bibfield  {author} {\bibinfo {author} {\bibfnamefont {X.}~\bibnamefont
  {Luo}}, \bibinfo {author} {\bibfnamefont {B.}~\bibnamefont {Xu}}, \bibinfo
  {author} {\bibfnamefont {T.}~\bibnamefont {Ohtsuki}}, \ and\ \bibinfo
  {author} {\bibfnamefont {R.}~\bibnamefont {Shindou}},\ }\href {\doibase
  10.1103/PhysRevB.97.045129} {\bibfield  {journal} {\bibinfo  {journal} {Phys.
  Rev. B}\ }\textbf {\bibinfo {volume} {97}},\ \bibinfo {pages} {045129}
  (\bibinfo {year} {2018})}\BibitemShut {NoStop}%
\bibitem [{\citenamefont {Luo}\ \emph {et~al.}(2020)\citenamefont {Luo},
  \citenamefont {Xu}, \citenamefont {Ohtsuki},\ and\ \citenamefont
  {Shindou}}]{Luo20}%
  \BibitemOpen
  \bibfield  {author} {\bibinfo {author} {\bibfnamefont {X.}~\bibnamefont
  {Luo}}, \bibinfo {author} {\bibfnamefont {B.}~\bibnamefont {Xu}}, \bibinfo
  {author} {\bibfnamefont {T.}~\bibnamefont {Ohtsuki}}, \ and\ \bibinfo
  {author} {\bibfnamefont {R.}~\bibnamefont {Shindou}},\ }\href {\doibase
  10.1103/PhysRevB.101.020202} {\bibfield  {journal} {\bibinfo  {journal}
  {Phys. Rev. B}\ }\textbf {\bibinfo {volume} {101}},\ \bibinfo {pages}
  {020202} (\bibinfo {year} {2020})}\BibitemShut {NoStop}%
\bibitem [{\citenamefont {Xu}\ \emph {et~al.}(2016)\citenamefont {Xu},
  \citenamefont {Ohtsuki},\ and\ \citenamefont {Shindou}}]{Xu16}%
  \BibitemOpen
  \bibfield  {author} {\bibinfo {author} {\bibfnamefont {B.}~\bibnamefont
  {Xu}}, \bibinfo {author} {\bibfnamefont {T.}~\bibnamefont {Ohtsuki}}, \ and\
  \bibinfo {author} {\bibfnamefont {R.}~\bibnamefont {Shindou}},\ }\href
  {\doibase 10.1103/PhysRevB.94.220403} {\bibfield  {journal} {\bibinfo
  {journal} {Phys. Rev. B}\ }\textbf {\bibinfo {volume} {94}},\ \bibinfo
  {pages} {220403} (\bibinfo {year} {2016})}\BibitemShut {NoStop}%
\bibitem [{\citenamefont {Tzortzakakis}\ \emph {et~al.}(2020)\citenamefont
  {Tzortzakakis}, \citenamefont {Makris},\ and\ \citenamefont
  {Economou}}]{Tzortzakakis20}%
  \BibitemOpen
  \bibfield  {author} {\bibinfo {author} {\bibfnamefont {A.~F.}\ \bibnamefont
  {Tzortzakakis}}, \bibinfo {author} {\bibfnamefont {K.~G.}\ \bibnamefont
  {Makris}}, \ and\ \bibinfo {author} {\bibfnamefont {E.~N.}\ \bibnamefont
  {Economou}},\ }\href {\doibase 10.1103/PhysRevB.101.014202} {\bibfield
  {journal} {\bibinfo  {journal} {Phys. Rev. B}\ }\textbf {\bibinfo {volume}
  {101}},\ \bibinfo {pages} {014202} (\bibinfo {year} {2020})}\BibitemShut
  {NoStop}%
\bibitem [{\citenamefont {Wang}\ and\ \citenamefont {Wang}(2020)}]{Wang20}%
  \BibitemOpen
  \bibfield  {author} {\bibinfo {author} {\bibfnamefont {C.}~\bibnamefont
  {Wang}}\ and\ \bibinfo {author} {\bibfnamefont {X.~R.}\ \bibnamefont
  {Wang}},\ }\href {\doibase 10.1103/PhysRevB.101.165114} {\bibfield  {journal}
  {\bibinfo  {journal} {Phys. Rev. B}\ }\textbf {\bibinfo {volume} {101}},\
  \bibinfo {pages} {165114} (\bibinfo {year} {2020})}\BibitemShut {NoStop}%
\bibitem [{\citenamefont {Huang}\ and\ \citenamefont
  {Shklovskii}(2020{\natexlab{a}})}]{Huang20}%
  \BibitemOpen
  \bibfield  {author} {\bibinfo {author} {\bibfnamefont {Y.}~\bibnamefont
  {Huang}}\ and\ \bibinfo {author} {\bibfnamefont {B.~I.}\ \bibnamefont
  {Shklovskii}},\ }\href {\doibase 10.1103/PhysRevB.101.014204} {\bibfield
  {journal} {\bibinfo  {journal} {Phys. Rev. B}\ }\textbf {\bibinfo {volume}
  {101}},\ \bibinfo {pages} {014204} (\bibinfo {year}
  {2020}{\natexlab{a}})}\BibitemShut {NoStop}%
\bibitem [{\citenamefont {Huang}\ and\ \citenamefont
  {Shklovskii}(2020{\natexlab{b}})}]{Huang20SR}%
  \BibitemOpen
  \bibfield  {author} {\bibinfo {author} {\bibfnamefont {Y.}~\bibnamefont
  {Huang}}\ and\ \bibinfo {author} {\bibfnamefont {B.~I.}\ \bibnamefont
  {Shklovskii}},\ }\href {\doibase 10.1103/PhysRevB.102.064212} {\bibfield
  {journal} {\bibinfo  {journal} {Phys. Rev. B}\ }\textbf {\bibinfo {volume}
  {102}},\ \bibinfo {pages} {064212} (\bibinfo {year}
  {2020}{\natexlab{b}})}\BibitemShut {NoStop}%
\bibitem [{\citenamefont {Cao}\ \emph {et~al.}(1999)\citenamefont {Cao},
  \citenamefont {Zhao}, \citenamefont {Ho}, \citenamefont {Seelig},
  \citenamefont {Wang},\ and\ \citenamefont {Chang}}]{Cao99}%
  \BibitemOpen
  \bibfield  {author} {\bibinfo {author} {\bibfnamefont {H.}~\bibnamefont
  {Cao}}, \bibinfo {author} {\bibfnamefont {Y.~G.}\ \bibnamefont {Zhao}},
  \bibinfo {author} {\bibfnamefont {S.~T.}\ \bibnamefont {Ho}}, \bibinfo
  {author} {\bibfnamefont {E.~W.}\ \bibnamefont {Seelig}}, \bibinfo {author}
  {\bibfnamefont {Q.~H.}\ \bibnamefont {Wang}}, \ and\ \bibinfo {author}
  {\bibfnamefont {R.~P.~H.}\ \bibnamefont {Chang}},\ }\href {\doibase
  10.1103/PhysRevLett.82.2278} {\bibfield  {journal} {\bibinfo  {journal}
  {Phys. Rev. Lett.}\ }\textbf {\bibinfo {volume} {82}},\ \bibinfo {pages}
  {2278} (\bibinfo {year} {1999})}\BibitemShut {NoStop}%
\bibitem [{\citenamefont {Wiersma}(2008)}]{Wiersma08}%
  \BibitemOpen
  \bibfield  {author} {\bibinfo {author} {\bibfnamefont {D.~S.}\ \bibnamefont
  {Wiersma}},\ }\href {\doibase 10.1038/nphys971} {\bibfield  {journal}
  {\bibinfo  {journal} {Nature Physics}\ }\textbf {\bibinfo {volume} {4}},\
  \bibinfo {pages} {359} (\bibinfo {year} {2008})}\BibitemShut {NoStop}%
\bibitem [{\citenamefont {Wiersma}(2013)}]{Wiersma13}%
  \BibitemOpen
  \bibfield  {author} {\bibinfo {author} {\bibfnamefont {D.~S.}\ \bibnamefont
  {Wiersma}},\ }\href {\doibase 10.1038/nphoton.2013.29} {\bibfield  {journal}
  {\bibinfo  {journal} {Nature Photonics}\ }\textbf {\bibinfo {volume} {7}},\
  \bibinfo {pages} {188} (\bibinfo {year} {2013})}\BibitemShut {NoStop}%
\bibitem [{\citenamefont {Konotop}\ \emph {et~al.}(2016)\citenamefont
  {Konotop}, \citenamefont {Yang},\ and\ \citenamefont {Zezyulin}}]{Konotop16}%
  \BibitemOpen
  \bibfield  {author} {\bibinfo {author} {\bibfnamefont {V.~V.}\ \bibnamefont
  {Konotop}}, \bibinfo {author} {\bibfnamefont {J.}~\bibnamefont {Yang}}, \
  and\ \bibinfo {author} {\bibfnamefont {D.~A.}\ \bibnamefont {Zezyulin}},\
  }\href {\doibase 10.1103/RevModPhys.88.035002} {\bibfield  {journal}
  {\bibinfo  {journal} {Rev. Mod. Phys.}\ }\textbf {\bibinfo {volume} {88}},\
  \bibinfo {pages} {035002} (\bibinfo {year} {2016})}\BibitemShut {NoStop}%
\bibitem [{\citenamefont {Feng}\ \emph {et~al.}(2017)\citenamefont {Feng},
  \citenamefont {El-Ganainy},\ and\ \citenamefont {Ge}}]{Feng17}%
  \BibitemOpen
  \bibfield  {author} {\bibinfo {author} {\bibfnamefont {L.}~\bibnamefont
  {Feng}}, \bibinfo {author} {\bibfnamefont {R.}~\bibnamefont {El-Ganainy}}, \
  and\ \bibinfo {author} {\bibfnamefont {L.}~\bibnamefont {Ge}},\ }\href
  {\doibase 10.1038/s41566-017-0031-1} {\bibfield  {journal} {\bibinfo
  {journal} {Nature Photonics}\ }\textbf {\bibinfo {volume} {11}},\ \bibinfo
  {pages} {752} (\bibinfo {year} {2017})}\BibitemShut {NoStop}%
\bibitem [{\citenamefont {El-Ganainy}\ \emph {et~al.}(2018)\citenamefont
  {El-Ganainy}, \citenamefont {Makris}, \citenamefont {Khajavikhan},
  \citenamefont {Musslimani}, \citenamefont {Rotter},\ and\ \citenamefont
  {Christodoulides}}]{El-Ganainy18}%
  \BibitemOpen
  \bibfield  {author} {\bibinfo {author} {\bibfnamefont {R.}~\bibnamefont
  {El-Ganainy}}, \bibinfo {author} {\bibfnamefont {K.~G.}\ \bibnamefont
  {Makris}}, \bibinfo {author} {\bibfnamefont {M.}~\bibnamefont {Khajavikhan}},
  \bibinfo {author} {\bibfnamefont {Z.~H.}\ \bibnamefont {Musslimani}},
  \bibinfo {author} {\bibfnamefont {S.}~\bibnamefont {Rotter}}, \ and\ \bibinfo
  {author} {\bibfnamefont {D.~N.}\ \bibnamefont {Christodoulides}},\ }\href
  {\doibase 10.1038/nphys4323} {\bibfield  {journal} {\bibinfo  {journal}
  {Nature Physics}\ }\textbf {\bibinfo {volume} {14}},\ \bibinfo {pages} {11}
  (\bibinfo {year} {2018})}\BibitemShut {NoStop}%
\bibitem [{\citenamefont {Ozdemir}\ \emph {et~al.}(2019)\citenamefont
  {Ozdemir}, \citenamefont {Rotter}, \citenamefont {Nori},\ and\ \citenamefont
  {Yang}}]{Ozdemir19}%
  \BibitemOpen
  \bibfield  {author} {\bibinfo {author} {\bibfnamefont {S.~K.}\ \bibnamefont
  {Ozdemir}}, \bibinfo {author} {\bibfnamefont {S.}~\bibnamefont {Rotter}},
  \bibinfo {author} {\bibfnamefont {F.}~\bibnamefont {Nori}}, \ and\ \bibinfo
  {author} {\bibfnamefont {L.}~\bibnamefont {Yang}},\ }\href {\doibase
  10.1038/s41563-019-0304-9} {\bibfield  {journal} {\bibinfo  {journal} {Nat
  Mater}\ }\textbf {\bibinfo {volume} {18}},\ \bibinfo {pages} {783} (\bibinfo
  {year} {2019})}\BibitemShut {NoStop}%
\bibitem [{\citenamefont {Miri}\ and\ \citenamefont {Al{\`u}}(2019)}]{Miri19}%
  \BibitemOpen
  \bibfield  {author} {\bibinfo {author} {\bibfnamefont {M.-A.}\ \bibnamefont
  {Miri}}\ and\ \bibinfo {author} {\bibfnamefont {A.}~\bibnamefont {Al{\`u}}},\
  }\href {\doibase 10.1126/science.aar7709} {\bibfield  {journal} {\bibinfo
  {journal} {Science}\ }\textbf {\bibinfo {volume} {363}} (\bibinfo {year}
  {2019}),\ 10.1126/science.aar7709}\BibitemShut {NoStop}%
\bibitem [{\citenamefont {Shen}\ and\ \citenamefont {Fu}(2018)}]{Shen18}%
  \BibitemOpen
  \bibfield  {author} {\bibinfo {author} {\bibfnamefont {H.}~\bibnamefont
  {Shen}}\ and\ \bibinfo {author} {\bibfnamefont {L.}~\bibnamefont {Fu}},\
  }\href {\doibase 10.1103/PhysRevLett.121.026403} {\bibfield  {journal}
  {\bibinfo  {journal} {Phys. Rev. Lett.}\ }\textbf {\bibinfo {volume} {121}},\
  \bibinfo {pages} {026403} (\bibinfo {year} {2018})}\BibitemShut {NoStop}%
\bibitem [{\citenamefont {Papaj}\ \emph {et~al.}(2019)\citenamefont {Papaj},
  \citenamefont {Isobe},\ and\ \citenamefont {Fu}}]{Papaj19}%
  \BibitemOpen
  \bibfield  {author} {\bibinfo {author} {\bibfnamefont {M.}~\bibnamefont
  {Papaj}}, \bibinfo {author} {\bibfnamefont {H.}~\bibnamefont {Isobe}}, \ and\
  \bibinfo {author} {\bibfnamefont {L.}~\bibnamefont {Fu}},\ }\href {\doibase
  10.1103/PhysRevB.99.201107} {\bibfield  {journal} {\bibinfo  {journal} {Phys.
  Rev. B}\ }\textbf {\bibinfo {volume} {99}},\ \bibinfo {pages} {201107}
  (\bibinfo {year} {2019})}\BibitemShut {NoStop}%
\bibitem [{\citenamefont {Hatano}\ and\ \citenamefont
  {Nelson}(1996)}]{Hatano96}%
  \BibitemOpen
  \bibfield  {author} {\bibinfo {author} {\bibfnamefont {N.}~\bibnamefont
  {Hatano}}\ and\ \bibinfo {author} {\bibfnamefont {D.~R.}\ \bibnamefont
  {Nelson}},\ }\href {\doibase 10.1103/PhysRevLett.77.570} {\bibfield
  {journal} {\bibinfo  {journal} {Phys. Rev. Lett.}\ }\textbf {\bibinfo
  {volume} {77}},\ \bibinfo {pages} {570} (\bibinfo {year} {1996})}\BibitemShut
  {NoStop}%
\bibitem [{\citenamefont {Kawabata}\ and\ \citenamefont
  {Ryu}(2020)}]{Kawabata20}%
  \BibitemOpen
  \bibfield  {author} {\bibinfo {author} {\bibfnamefont {K.}~\bibnamefont
  {Kawabata}}\ and\ \bibinfo {author} {\bibfnamefont {S.}~\bibnamefont {Ryu}},\
  }\href@noop {} {\enquote {\bibinfo {title} {Nonunitary scaling theory of
  non-hermitian localization},}\ } (\bibinfo {year} {2020}),\ \Eprint
  {http://arxiv.org/abs/arXiv:2005.00604} {arXiv:2005.00604} \BibitemShut
  {NoStop}%
\bibitem [{\citenamefont {Kawabata}\ \emph {et~al.}(2019)\citenamefont
  {Kawabata}, \citenamefont {Shiozaki}, \citenamefont {Ueda},\ and\
  \citenamefont {Sato}}]{Kawabata19}%
  \BibitemOpen
  \bibfield  {author} {\bibinfo {author} {\bibfnamefont {K.}~\bibnamefont
  {Kawabata}}, \bibinfo {author} {\bibfnamefont {K.}~\bibnamefont {Shiozaki}},
  \bibinfo {author} {\bibfnamefont {M.}~\bibnamefont {Ueda}}, \ and\ \bibinfo
  {author} {\bibfnamefont {M.}~\bibnamefont {Sato}},\ }\href {\doibase
  10.1103/PhysRevX.9.041015} {\bibfield  {journal} {\bibinfo  {journal} {Phys.
  Rev. X}\ }\textbf {\bibinfo {volume} {9}},\ \bibinfo {pages} {041015}
  (\bibinfo {year} {2019})}\BibitemShut {NoStop}%
\bibitem [{\citenamefont {Zhou}\ and\ \citenamefont {Lee}(2019)}]{Zhou19}%
  \BibitemOpen
  \bibfield  {author} {\bibinfo {author} {\bibfnamefont {H.}~\bibnamefont
  {Zhou}}\ and\ \bibinfo {author} {\bibfnamefont {J.~Y.}\ \bibnamefont {Lee}},\
  }\href {\doibase 10.1103/PhysRevB.99.235112} {\bibfield  {journal} {\bibinfo
  {journal} {Phys. Rev. B}\ }\textbf {\bibinfo {volume} {99}},\ \bibinfo
  {pages} {235112} (\bibinfo {year} {2019})}\BibitemShut {NoStop}%
\bibitem [{\citenamefont {Oganesyan}\ and\ \citenamefont
  {Huse}(2007)}]{Oganesyan07}%
  \BibitemOpen
  \bibfield  {author} {\bibinfo {author} {\bibfnamefont {V.}~\bibnamefont
  {Oganesyan}}\ and\ \bibinfo {author} {\bibfnamefont {D.~A.}\ \bibnamefont
  {Huse}},\ }\href {\doibase 10.1103/PhysRevB.75.155111} {\bibfield  {journal}
  {\bibinfo  {journal} {Phys. Rev. B}\ }\textbf {\bibinfo {volume} {75}},\
  \bibinfo {pages} {155111} (\bibinfo {year} {2007})}\BibitemShut {NoStop}%
\bibitem [{\citenamefont {S\'a}\ \emph {et~al.}(2020)\citenamefont {S\'a},
  \citenamefont {Ribeiro},\ and\ \citenamefont {Prosen}}]{Sa20}%
  \BibitemOpen
  \bibfield  {author} {\bibinfo {author} {\bibfnamefont {L.}~\bibnamefont
  {S\'a}}, \bibinfo {author} {\bibfnamefont {P.}~\bibnamefont {Ribeiro}}, \
  and\ \bibinfo {author} {\bibfnamefont {T.~c.~v.}\ \bibnamefont {Prosen}},\
  }\href {\doibase 10.1103/PhysRevX.10.021019} {\bibfield  {journal} {\bibinfo
  {journal} {Phys. Rev. X}\ }\textbf {\bibinfo {volume} {10}},\ \bibinfo
  {pages} {021019} (\bibinfo {year} {2020})}\BibitemShut {NoStop}%
\bibitem [{\citenamefont {Slevin}\ and\ \citenamefont
  {Ohtsuki}(2018)}]{Slevin18}%
  \BibitemOpen
  \bibfield  {author} {\bibinfo {author} {\bibfnamefont {K.}~\bibnamefont
  {Slevin}}\ and\ \bibinfo {author} {\bibfnamefont {T.}~\bibnamefont
  {Ohtsuki}},\ }\href {\doibase 10.7566/JPSJ.87.094703} {\bibfield  {journal}
  {\bibinfo  {journal} {Journal of the Physical Society of Japan}\ }\textbf
  {\bibinfo {volume} {87}},\ \bibinfo {pages} {094703} (\bibinfo {year}
  {2018})},\ \Eprint
  {http://arxiv.org/abs/https://doi.org/10.7566/JPSJ.87.094703}
  {https://doi.org/10.7566/JPSJ.87.094703} \BibitemShut {NoStop}%
\bibitem [{\citenamefont {Wigner}(1951)}]{Wigner51}%
  \BibitemOpen
  \bibfield  {author} {\bibinfo {author} {\bibfnamefont {E.~P.}\ \bibnamefont
  {Wigner}},\ }\href {\doibase 10.1017/S0305004100027237} {\bibfield  {journal}
  {\bibinfo  {journal} {Mathematical Proceedings of the Cambridge Philosophical
  Society}\ }\textbf {\bibinfo {volume} {47}},\ \bibinfo {pages}
  {790^^e2^^80^^93798} (\bibinfo {year} {1951})}\BibitemShut {NoStop}%
\bibitem [{\citenamefont {Dyson}(1962{\natexlab{a}})}]{Dyson62}%
  \BibitemOpen
  \bibfield  {author} {\bibinfo {author} {\bibfnamefont {F.~J.}\ \bibnamefont
  {Dyson}},\ }\href@noop {} {\bibfield  {journal} {\bibinfo  {journal} {Journal
  of Mathematical Physics}\ }\textbf {\bibinfo {volume} {3}},\ \bibinfo {pages}
  {140} (\bibinfo {year} {1962}{\natexlab{a}})}\BibitemShut {NoStop}%
\bibitem [{\citenamefont {Dyson}(1962{\natexlab{b}})}]{Dyson62TFW}%
  \BibitemOpen
  \bibfield  {author} {\bibinfo {author} {\bibfnamefont {F.~J.}\ \bibnamefont
  {Dyson}},\ }\href@noop {} {\bibfield  {journal} {\bibinfo  {journal} {Journal
  of Mathematical Physics}\ }\textbf {\bibinfo {volume} {3}},\ \bibinfo {pages}
  {1199} (\bibinfo {year} {1962}{\natexlab{b}})}\BibitemShut {NoStop}%
\bibitem [{\citenamefont {Grobe}\ \emph {et~al.}(1988)\citenamefont {Grobe},
  \citenamefont {Haake},\ and\ \citenamefont {Sommers}}]{Grobe88}%
  \BibitemOpen
  \bibfield  {author} {\bibinfo {author} {\bibfnamefont {R.}~\bibnamefont
  {Grobe}}, \bibinfo {author} {\bibfnamefont {F.}~\bibnamefont {Haake}}, \ and\
  \bibinfo {author} {\bibfnamefont {H.-J.}\ \bibnamefont {Sommers}},\ }\href
  {\doibase 10.1103/PhysRevLett.61.1899} {\bibfield  {journal} {\bibinfo
  {journal} {Phys. Rev. Lett.}\ }\textbf {\bibinfo {volume} {61}},\ \bibinfo
  {pages} {1899} (\bibinfo {year} {1988})}\BibitemShut {NoStop}%
\bibitem [{\citenamefont {Ginibre}(1965)}]{Ginibre65}%
  \BibitemOpen
  \bibfield  {author} {\bibinfo {author} {\bibfnamefont {J.}~\bibnamefont
  {Ginibre}},\ }\href {\doibase 10.1063/1.1704292} {\bibfield  {journal}
  {\bibinfo  {journal} {Journal of Mathematical Physics}\ }\textbf {\bibinfo
  {volume} {6}},\ \bibinfo {pages} {440} (\bibinfo {year} {1965})},\ \Eprint
  {http://arxiv.org/abs/https://doi.org/10.1063/1.1704292}
  {https://doi.org/10.1063/1.1704292} \BibitemShut {NoStop}%
\bibitem [{\citenamefont {Shklovskii}\ \emph {et~al.}(1993)\citenamefont
  {Shklovskii}, \citenamefont {Shapiro}, \citenamefont {Sears}, \citenamefont
  {Lambrianides},\ and\ \citenamefont {Shore}}]{Shklovskii93}%
  \BibitemOpen
  \bibfield  {author} {\bibinfo {author} {\bibfnamefont {B.~I.}\ \bibnamefont
  {Shklovskii}}, \bibinfo {author} {\bibfnamefont {B.}~\bibnamefont {Shapiro}},
  \bibinfo {author} {\bibfnamefont {B.~R.}\ \bibnamefont {Sears}}, \bibinfo
  {author} {\bibfnamefont {P.}~\bibnamefont {Lambrianides}}, \ and\ \bibinfo
  {author} {\bibfnamefont {H.~B.}\ \bibnamefont {Shore}},\ }\href {\doibase
  10.1103/PhysRevB.47.11487} {\bibfield  {journal} {\bibinfo  {journal} {Phys.
  Rev. B}\ }\textbf {\bibinfo {volume} {47}},\ \bibinfo {pages} {11487}
  (\bibinfo {year} {1993})}\BibitemShut {NoStop}%
\bibitem [{sup()}]{supplemental}%
  \BibitemOpen
  \href@noop {} {}\bibinfo {note} {See Supplemental Material at [URL will be
  inserted by publisher].}\BibitemShut {Stop}%
\bibitem [{\citenamefont {Atas}\ \emph {et~al.}(2013)\citenamefont {Atas},
  \citenamefont {Bogomolny}, \citenamefont {Giraud},\ and\ \citenamefont
  {Roux}}]{Atas13}%
  \BibitemOpen
  \bibfield  {author} {\bibinfo {author} {\bibfnamefont {Y.~Y.}\ \bibnamefont
  {Atas}}, \bibinfo {author} {\bibfnamefont {E.}~\bibnamefont {Bogomolny}},
  \bibinfo {author} {\bibfnamefont {O.}~\bibnamefont {Giraud}}, \ and\ \bibinfo
  {author} {\bibfnamefont {G.}~\bibnamefont {Roux}},\ }\href {\doibase
  10.1103/PhysRevLett.110.084101} {\bibfield  {journal} {\bibinfo  {journal}
  {Phys. Rev. Lett.}\ }\textbf {\bibinfo {volume} {110}},\ \bibinfo {pages}
  {084101} (\bibinfo {year} {2013})}\BibitemShut {NoStop}%
\bibitem [{\citenamefont {Rodriguez}\ \emph {et~al.}(2011)\citenamefont
  {Rodriguez}, \citenamefont {Vasquez}, \citenamefont {Slevin},\ and\
  \citenamefont {R\"omer}}]{Rodriguez11}%
  \BibitemOpen
  \bibfield  {author} {\bibinfo {author} {\bibfnamefont {A.}~\bibnamefont
  {Rodriguez}}, \bibinfo {author} {\bibfnamefont {L.~J.}\ \bibnamefont
  {Vasquez}}, \bibinfo {author} {\bibfnamefont {K.}~\bibnamefont {Slevin}}, \
  and\ \bibinfo {author} {\bibfnamefont {R.~A.}\ \bibnamefont {R\"omer}},\
  }\href {\doibase 10.1103/PhysRevB.84.134209} {\bibfield  {journal} {\bibinfo
  {journal} {Phys. Rev. B}\ }\textbf {\bibinfo {volume} {84}},\ \bibinfo
  {pages} {134209} (\bibinfo {year} {2011})}\BibitemShut {NoStop}%
\bibitem [{\citenamefont {Chalker}\ \emph
  {et~al.}(1996{\natexlab{a}})\citenamefont {Chalker}, \citenamefont {Lerner},\
  and\ \citenamefont {Smith}}]{Chalker96}%
  \BibitemOpen
  \bibfield  {author} {\bibinfo {author} {\bibfnamefont {J.~T.}\ \bibnamefont
  {Chalker}}, \bibinfo {author} {\bibfnamefont {I.~V.}\ \bibnamefont {Lerner}},
  \ and\ \bibinfo {author} {\bibfnamefont {R.~A.}\ \bibnamefont {Smith}},\
  }\href {\doibase 10.1063/1.531676} {\bibfield  {journal} {\bibinfo  {journal}
  {Journal of Mathematical Physics}\ }\textbf {\bibinfo {volume} {37}},\
  \bibinfo {pages} {5061} (\bibinfo {year} {1996}{\natexlab{a}})}\BibitemShut
  {NoStop}%
\bibitem [{\citenamefont {Chalker}\ \emph
  {et~al.}(1996{\natexlab{b}})\citenamefont {Chalker}, \citenamefont {Lerner},\
  and\ \citenamefont {Smith}}]{Chalker96PRL}%
  \BibitemOpen
  \bibfield  {author} {\bibinfo {author} {\bibfnamefont {J.~T.}\ \bibnamefont
  {Chalker}}, \bibinfo {author} {\bibfnamefont {I.~V.}\ \bibnamefont {Lerner}},
  \ and\ \bibinfo {author} {\bibfnamefont {R.~A.}\ \bibnamefont {Smith}},\
  }\href {\doibase 10.1103/PhysRevLett.77.554} {\bibfield  {journal} {\bibinfo
  {journal} {Phys. Rev. Lett.}\ }\textbf {\bibinfo {volume} {77}},\ \bibinfo
  {pages} {554} (\bibinfo {year} {1996}{\natexlab{b}})}\BibitemShut {NoStop}%
\bibitem [{\citenamefont {Chalker}\ \emph
  {et~al.}(1996{\natexlab{c}})\citenamefont {Chalker}, \citenamefont
  {Kravtsov},\ and\ \citenamefont {Lerner}}]{Chalker96_SR}%
  \BibitemOpen
  \bibfield  {author} {\bibinfo {author} {\bibfnamefont {J.~T.}\ \bibnamefont
  {Chalker}}, \bibinfo {author} {\bibfnamefont {V.~E.}\ \bibnamefont
  {Kravtsov}}, \ and\ \bibinfo {author} {\bibfnamefont {I.~V.}\ \bibnamefont
  {Lerner}},\ }\href {\doibase 10.1134/1.567208} {\bibfield  {journal}
  {\bibinfo  {journal} {Journal of Experimental and Theoretical Physics
  Letters}\ }\textbf {\bibinfo {volume} {64}},\ \bibinfo {pages} {386}
  (\bibinfo {year} {1996}{\natexlab{c}})}\BibitemShut {NoStop}%
\bibitem [{\citenamefont {Bogomolny}\ and\ \citenamefont
  {Giraud}(2011)}]{Bogomolny11}%
  \BibitemOpen
  \bibfield  {author} {\bibinfo {author} {\bibfnamefont {E.}~\bibnamefont
  {Bogomolny}}\ and\ \bibinfo {author} {\bibfnamefont {O.}~\bibnamefont
  {Giraud}},\ }\href {\doibase 10.1103/PhysRevLett.106.044101} {\bibfield
  {journal} {\bibinfo  {journal} {Phys. Rev. Lett.}\ }\textbf {\bibinfo
  {volume} {106}},\ \bibinfo {pages} {044101} (\bibinfo {year}
  {2011})}\BibitemShut {NoStop}%
\bibitem [{\citenamefont {Braun}\ \emph {et~al.}(1998)\citenamefont {Braun},
  \citenamefont {Montambaux},\ and\ \citenamefont {Pascaud}}]{Braun98}%
  \BibitemOpen
  \bibfield  {author} {\bibinfo {author} {\bibfnamefont {D.}~\bibnamefont
  {Braun}}, \bibinfo {author} {\bibfnamefont {G.}~\bibnamefont {Montambaux}}, \
  and\ \bibinfo {author} {\bibfnamefont {M.}~\bibnamefont {Pascaud}},\ }\href
  {\doibase 10.1103/PhysRevLett.81.1062} {\bibfield  {journal} {\bibinfo
  {journal} {Phys. Rev. Lett.}\ }\textbf {\bibinfo {volume} {81}},\ \bibinfo
  {pages} {1062} (\bibinfo {year} {1998})}\BibitemShut {NoStop}%
\bibitem [{\citenamefont {Ndawana}\ \emph {et~al.}(2002)\citenamefont
  {Ndawana}, \citenamefont {Romer},\ and\ \citenamefont
  {Schreiber}}]{Ndawana02}%
  \BibitemOpen
  \bibfield  {author} {\bibinfo {author} {\bibfnamefont {M.~L.}\ \bibnamefont
  {Ndawana}}, \bibinfo {author} {\bibfnamefont {R.~A.}\ \bibnamefont {Romer}},
  \ and\ \bibinfo {author} {\bibfnamefont {M.}~\bibnamefont {Schreiber}},\
  }\href {\doibase 10.1140/epjb/e2002-00171-4} {\bibfield  {journal} {\bibinfo
  {journal} {The European Physical Journal B - Condensed Matter and Complex
  Systems}\ }\textbf {\bibinfo {volume} {27}},\ \bibinfo {pages} {399}
  (\bibinfo {year} {2002})}\BibitemShut {NoStop}%
\bibitem [{\citenamefont {Ghosh}\ \emph {et~al.}(2017)\citenamefont {Ghosh},
  \citenamefont {Miniatura}, \citenamefont {Cherroret},\ and\ \citenamefont
  {Delande}}]{Ghosh17}%
  \BibitemOpen
  \bibfield  {author} {\bibinfo {author} {\bibfnamefont {S.}~\bibnamefont
  {Ghosh}}, \bibinfo {author} {\bibfnamefont {C.}~\bibnamefont {Miniatura}},
  \bibinfo {author} {\bibfnamefont {N.}~\bibnamefont {Cherroret}}, \ and\
  \bibinfo {author} {\bibfnamefont {D.}~\bibnamefont {Delande}},\ }\href
  {\doibase 10.1103/PhysRevA.95.041602} {\bibfield  {journal} {\bibinfo
  {journal} {Phys. Rev. A}\ }\textbf {\bibinfo {volume} {95}},\ \bibinfo
  {pages} {041602} (\bibinfo {year} {2017})}\BibitemShut {NoStop}%
\bibitem [{\citenamefont {Zharekeshev}\ and\ \citenamefont
  {Kramer}(1995)}]{Zharekeshev95}%
  \BibitemOpen
  \bibfield  {author} {\bibinfo {author} {\bibfnamefont {I.~K.}\ \bibnamefont
  {Zharekeshev}}\ and\ \bibinfo {author} {\bibfnamefont {B.}~\bibnamefont
  {Kramer}},\ }\href {\doibase 10.1143/jjap.34.4361} {\bibfield  {journal}
  {\bibinfo  {journal} {Japanese Journal of Applied Physics}\ }\textbf
  {\bibinfo {volume} {34}},\ \bibinfo {pages} {4361} (\bibinfo {year}
  {1995})}\BibitemShut {NoStop}%
\bibitem [{\citenamefont {Kawarabayashi}\ \emph {et~al.}(1996)\citenamefont
  {Kawarabayashi}, \citenamefont {Ohtsuki}, \citenamefont {Slevin},\ and\
  \citenamefont {Ono}}]{Kawarabayashi96}%
  \BibitemOpen
  \bibfield  {author} {\bibinfo {author} {\bibfnamefont {T.}~\bibnamefont
  {Kawarabayashi}}, \bibinfo {author} {\bibfnamefont {T.}~\bibnamefont
  {Ohtsuki}}, \bibinfo {author} {\bibfnamefont {K.}~\bibnamefont {Slevin}}, \
  and\ \bibinfo {author} {\bibfnamefont {Y.}~\bibnamefont {Ono}},\ }\href
  {\doibase 10.1103/PhysRevLett.77.3593} {\bibfield  {journal} {\bibinfo
  {journal} {Phys. Rev. Lett.}\ }\textbf {\bibinfo {volume} {77}},\ \bibinfo
  {pages} {3593} (\bibinfo {year} {1996})}\BibitemShut {NoStop}%
\bibitem [{\citenamefont {Batsch}\ \emph {et~al.}(1996)\citenamefont {Batsch},
  \citenamefont {Schweitzer}, \citenamefont {Zharekeshev},\ and\ \citenamefont
  {Kramer}}]{Batsch96}%
  \BibitemOpen
  \bibfield  {author} {\bibinfo {author} {\bibfnamefont {M.}~\bibnamefont
  {Batsch}}, \bibinfo {author} {\bibfnamefont {L.}~\bibnamefont {Schweitzer}},
  \bibinfo {author} {\bibfnamefont {I.~K.}\ \bibnamefont {Zharekeshev}}, \ and\
  \bibinfo {author} {\bibfnamefont {B.}~\bibnamefont {Kramer}},\ }\href
  {\doibase 10.1103/PhysRevLett.77.1552} {\bibfield  {journal} {\bibinfo
  {journal} {Phys. Rev. Lett.}\ }\textbf {\bibinfo {volume} {77}},\ \bibinfo
  {pages} {1552} (\bibinfo {year} {1996})}\BibitemShut {NoStop}%
\bibitem [{\citenamefont {Hofstetter}(1996)}]{Hofstetter96}%
  \BibitemOpen
  \bibfield  {author} {\bibinfo {author} {\bibfnamefont {E.}~\bibnamefont
  {Hofstetter}},\ }\href {\doibase 10.1103/PhysRevB.54.4552} {\bibfield
  {journal} {\bibinfo  {journal} {Phys. Rev. B}\ }\textbf {\bibinfo {volume}
  {54}},\ \bibinfo {pages} {4552} (\bibinfo {year} {1996})}\BibitemShut
  {NoStop}%
\bibitem [{\citenamefont {Grobe}\ and\ \citenamefont {Haake}(1989)}]{Grobe89}%
  \BibitemOpen
  \bibfield  {author} {\bibinfo {author} {\bibfnamefont {R.}~\bibnamefont
  {Grobe}}\ and\ \bibinfo {author} {\bibfnamefont {F.}~\bibnamefont {Haake}},\
  }\href {\doibase 10.1103/PhysRevLett.62.2893} {\bibfield  {journal} {\bibinfo
   {journal} {Phys. Rev. Lett.}\ }\textbf {\bibinfo {volume} {62}},\ \bibinfo
  {pages} {2893} (\bibinfo {year} {1989})}\BibitemShut {NoStop}%
\bibitem [{\citenamefont {Akemann}\ \emph {et~al.}(2019)\citenamefont
  {Akemann}, \citenamefont {Kieburg}, \citenamefont {Mielke},\ and\
  \citenamefont {Prosen}}]{Akemann19}%
  \BibitemOpen
  \bibfield  {author} {\bibinfo {author} {\bibfnamefont {G.}~\bibnamefont
  {Akemann}}, \bibinfo {author} {\bibfnamefont {M.}~\bibnamefont {Kieburg}},
  \bibinfo {author} {\bibfnamefont {A.}~\bibnamefont {Mielke}}, \ and\ \bibinfo
  {author} {\bibfnamefont {T.~c.~v.}\ \bibnamefont {Prosen}},\ }\href {\doibase
  10.1103/PhysRevLett.123.254101} {\bibfield  {journal} {\bibinfo  {journal}
  {Phys. Rev. Lett.}\ }\textbf {\bibinfo {volume} {123}},\ \bibinfo {pages}
  {254101} (\bibinfo {year} {2019})}\BibitemShut {NoStop}%
\bibitem [{\citenamefont {Hamazaki}\ \emph {et~al.}(2020)\citenamefont
  {Hamazaki}, \citenamefont {Kawabata}, \citenamefont {Kura},\ and\
  \citenamefont {Ueda}}]{Hamazaki20}%
  \BibitemOpen
  \bibfield  {author} {\bibinfo {author} {\bibfnamefont {R.}~\bibnamefont
  {Hamazaki}}, \bibinfo {author} {\bibfnamefont {K.}~\bibnamefont {Kawabata}},
  \bibinfo {author} {\bibfnamefont {N.}~\bibnamefont {Kura}}, \ and\ \bibinfo
  {author} {\bibfnamefont {M.}~\bibnamefont {Ueda}},\ }\href {\doibase
  10.1103/PhysRevResearch.2.023286} {\bibfield  {journal} {\bibinfo  {journal}
  {Phys. Rev. Research}\ }\textbf {\bibinfo {volume} {2}},\ \bibinfo {pages}
  {023286} (\bibinfo {year} {2020})}\BibitemShut {NoStop}%
\end{thebibliography}%

\begin{widetext}
\section{supplemental materials}
\subsection{Polynomial fitting for non-Hermitian Anderson model and U(1) model}
We study the tight-binding model on a three-dimensional cubic lattice (Anderson model, AM),
\begin{align}
{\cal H}=\sum_{i}\varepsilon_i c^\dagger_{i}c_{i}+\sum_{\langle i,j \rangle}V_{i,j}c^\dagger_{i}c_{j}\label{AM}
\end{align}
and U(1) model,
\begin{align}
{\cal H}=\sum_i \varepsilon_i c_i^{\dagger} c_i+\sum_{\langle i,j \rangle} e^{2\pi i\cdot \theta_{i,j}}V_{i,j}c_i^{\dagger}c_j\label{U1}
\end{align}
where $c_i^\dagger$ ($c_i$) is the creation (annihilation) operator for electrons at site $i$ and $\varepsilon_i$ is random onsite potential. 
$\langle i,j\rangle$ means that $i$ and $j$ are the nearest neighbor lattice sites to each other, 
$V_{i,j}=V_{j,i}=1$ are the nearest neighbor hopping term, and $\theta_{i,j}=-\theta_{j,i}$ is the random phase 
that distributes uniformly within $[0,1)$. To study the Anderson transition (AT) in the 
non-Hermitian (NH) system, we set $\varepsilon_{j}=w_{j}^{r}+i w_{j}^i$ with the imaginary unit $i$. 
$w_{j}^{r}$ and $w_{j}^{i}$ are independent random numbers, both of which distribute uniformly within 
$[-W/2,W/2]$ for a given disorder strength $W$.

\begin{table}[b]
	\caption{Polynomial fitting results for (a) Anderson model and (b) U(1) model with non-Hermitian (NH) disorders. 
        The goodness of fit (GOF), critical disorder $W_{c}$, critical exponent $\nu$, 
		the scaling dimension of the least irrelevant scaling variable $-y$, and critical level spacing ratio $\langle r\rangle_c$ are shown for different 
		orders of the Taylor expansion of the scaling function $(m_1,n_1,m_2,n_2)$, for different system size range, and for 
        different disorder range. The square bracket stands for the 95\% confidence interval for each fitting result. Polynomial fittings with 
        various energy windows (`percent'), expansion orders, system size range, and disorder range have been carried out, for the purpose of proving 
        the stability of the fitting results against these changes.}
	\begin{tabular}{cccccccccccc}
		&&&&&&&&&&&\\
		\multicolumn{12}{l}{(a)NH Anderson model} \\
		\hline
percent &	$L$& $W$&	$m_1$& $n_1$ & $m_2$ & $n_2$ & GOF & $W_c$ & $\nu$ & $y$&$\langle r\rangle_c$   \\
\hline
10\%&	8-24& [6, 7.12]&   3&3&0&1&0.11&6.28[6.26, 6.30]&1.046[1.012, 1.086]&1.75[1.65, 1.84]&0.7169[0.7163, 0.7177]\\
10\%&	10-24& [6, 7.19]&  3&3&0&1&0.15&6.32[6.30, 6.34]&0.990[0.945, 1.040]&2.10[1.87, 2.35]&0.7155[0.7146, 0.7164]\\
10\%&	12-24& [5.9, 7.2]& 3&3&0&1&0.13&6.34[6.32, 6.36]&0.942[0.897, 0.989]&2.53[2.14, 2.90]&0.7145[0.7138, 0.7154]\\
\hline
5\%&	8-24& [6.14, 7.3]	&3&3&0&1&0.21&6.38[6.35, 6.40]&0.977[0.938, 1.022]&1.82[1.73, 1.90]&0.7157[0.7149, 0.7165]\\
5\%&	10-24& [6.14, 7.26] &3&3&0&1&0.17&6.37[6.33, 6.41]&0.948[0.878, 1.039]&1.77[1.54, 2.02]&0.7159[0.7143, 0.7179]\\
5\%&	12-24& [5.9, 7.3]   &3&3&0&1&0.12&6.42[6.39, 6.45]&0.825[0.765, 0.893]&2.27[1.89, 2.60]&0.7134[0.7124, 0.7151]\\
\hline
&&&&&&&&&&&\\
\multicolumn{12}{l}{(b)NH U(1) model} \\
\hline
percent  &	$L$& $W$&	$m_1$& $n_1$ & $m_2$ & $n_2$ & GOF & $W_c$ & $\nu$ & $y$ & $\langle r\rangle_c$  \\
\hline
10\% &	8-24& [7, 7.56]& 1&3&0&1&0.32&7.14[7.13, 7.15]&1.065[1.036, 1.100]&2.60[2.31, 2.89]&0.7178[0.7171, 0.7188]\\
10\% &	8-24& [7, 7.56]& 2&3&0&1&0.43&7.15[7.14, 7.16]&1.068[1.034, 1.105]&2.63[2.35, 2.92]&0.7177[0.7169, 0.7186]\\
10\% &	8-24& [7, 7.56]& 3&3&0&1&0.49&7.15[7.14, 7.16]&1.065[1.031, 1.103]&2.64[2.35, 2.92]&0.7177[0.7169, 0.7186]\\
10\% &	10-24& [6.8, 7.6]&3&3&0&1&0.12&7.14[7.12, 7.16]&1.091[1.050, 1.151]&2.50[1.88, 3.16]&0.7187[0.7170, 0.7201]\\
10\% &	12-24& [6.68, 7.64]&3&3&0&1&0.46&7.13[7.06, 7.16]&1.133[1.065, 1.411]&2.29[0.83, 4.62]&0.7187[0.7166, 0.7281]\\
\hline
5\% & 8-24 & [7, 7.8]	&2&3&0&1&0.09&7.23[7.22, 7.24]&1.028[1.000, 1.059]&2.37[2.18, 2.57]&0.7165[0.7155, 0.7176]\\
5\% & 8-24 & [7, 7.8]	&3&3&0&1&0.18&7.23[7.22, 7.24]&1.012[0.979, 1.048]&2.35[2.15, 2.54]&0.7167[0.7155, 0.7178]\\
5\% & 10-24 & [6.8, 8] &3&3&0&1&0.23&7.23[7.22, 7.25]&1.012[0.979, 1.047]&2.40[2.09, 2.73]&0.7164[0.7152, 0.7179]\\
5\% & 12-24 & [6.8, 8]	&3&3&0&1&0.23&7.24[7.22, 7.25]&1.027[0.974, 1.079]&2.68[2.09, 3.38]&0.7158[0.7143, 0.7177]\\
\hline
	\end{tabular}
	\label{table}
\end{table}

Level spacing ratio $r_i$ for each complex-valued eigenvalue $E_i$ as 
\begin{align}
	r_i\equiv  |z_i|,
\end{align}
with
\begin{align}
z_{i}\equiv  \frac{E_i-E_{\rm NN}}{E_i-E_{\rm NNN}}, \label{z_i}
\end{align} 
where $E_{\rm NN}$ and $E_{\rm NNN}$ are the nearest neighbor and next nearest neighbor to $E_i$ in the complex Euler plane. 
$z_i$ is a complex number with modulus and angle. Here we focus on the modulus of $z_i$ first. $r_i$ will be averaged within 
an energy window (see below) and then averaged over $M$ realizations of disordered systems. This gives a precise value of 
$\langle r\rangle$ with a standard deviation $\sigma^2_{\langle r\rangle}=\frac{1}{M-1}(\langle r^2\rangle-\langle r\rangle^2)$.

In order to have enough  energy levels whose critical $W$ for the AT are sufficiently close to that for $E=0$, 
we take only $10\%$ eigenvalues near $E=0$ in the Euler plane as the energy window, and calculate them for each disorder realization. Here, the periodic boundary condition is imposed in the 
three directions for both of the two NH models. We prepare $M$ disorder 
realizations so that the total number of eigenvalues for the statistics ($M\ \times10\% \times L^3$) 
reaches $5\times 10^7$ ($L<24$) or $10^7$ ($L=24$) for the NH AM and $10^7$ ($L\le 24$) for the NH U(1) model; 
$M=10^6$, $5\times 10^5$, $3\times 10^5$, $1.2\times 10^5,$ $6\times 10^4$, $10^4$ for $L=8,\!\ 10, \!\ 12, \!\ 16, \!\ 20, \!\ 24$  
in the AM, and $M=2\times 10^5$, $10^5$, $6\times 10^4$, $2.5\times 10^4$, $1.2\times 10^4$, 
$6.4\times 10^3$ for $L=8,\!\ 10,\!\ 12,\!\ 16,\!\ 20,\!\ 24$ in the U(1) model.

\begin{figure}[t]
	\centering
	\includegraphics[width=0.8\linewidth]{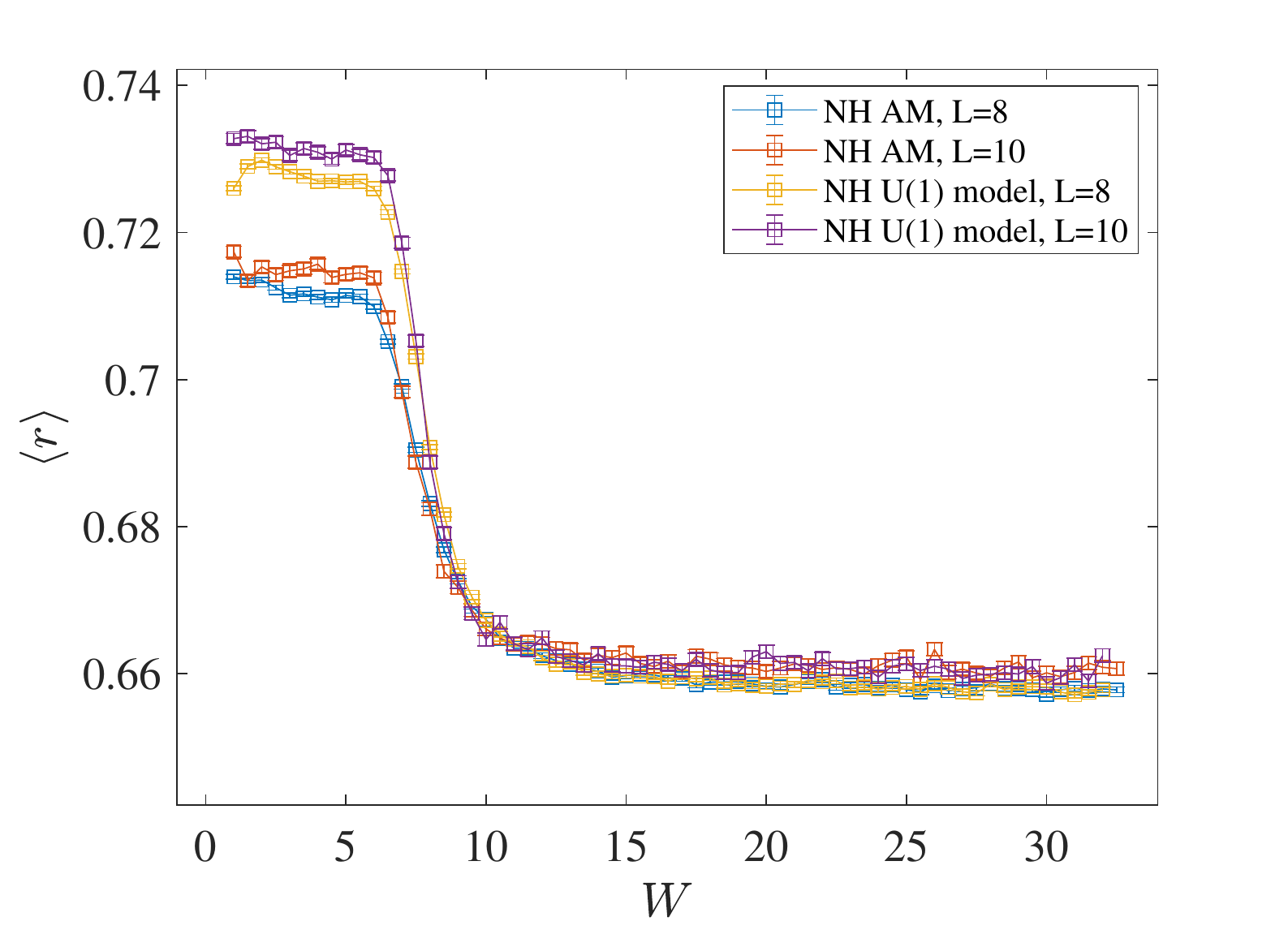}
	\caption{Averaged level spacing ratio $\langle r\rangle$ as a function of disorder. The level spacing ratio $r$ calculated from 
		$10\%$ eigenvalues around $E=0$ is averaged over $M$ number of different disorder realizations. 
		We take $M=10^4$, $10^3$ for the system size $L=8$, $10$, respectively, in non-Hermitian (NH) Anderson 
		model (AM) and NH U(1) model. When $W$ is either very small or very large, 
		$\langle r\rangle $ approaches  constant values.}
	\label{r_constant}
\end{figure}

$\langle r\rangle$ goes to almost constant values when $W$ is  far away from the critical point (FIG. \ref{r_constant}).
Consider first the very strong disorder region; $W\gg W_c$. In the thermodynamic limit, energy levels are distributed randomly
without any correlations with others. Thus, $r$ could be any value within $0<r<1$ with  equal probability. Hence 
\begin{align}
	\langle r\rangle_{\rm insulator}=\int_{0}^{1}r\rho(r) dr=\frac{2}{3}, \label{r_insulator}
\end{align}
where the density for $r$ is determined from $\langle \rangle_{\rm insulator} =1$; $\rho(r)=2r$.
Both models with the very strong disorder indeed show $\langle r\rangle\approx 0.66$ (FIG. \ref{r_constant}).

Consider next the metal phase; $W\ll W_c$. The energy levels in the metal phase have 
repulsive interactions with others, because of spatial overlaps between the extended eigenfunctions. Accordingly, the density 
of $r$ near zero will be smaller than in the insulator phase, hence
$\langle r\rangle_{\rm metal}>\langle r\rangle_{\rm insulator}$. 

For the Hermitian case (Gaussian ensemble), a mean value of $r_i\equiv s_i/s_{i-1}$, where $s_i \equiv E_i-E_{i-1}$ on  
the real axis with $\{E_i\}$ ordered, takes a certain constant value in the metal phase. The value depends on the symmetry class in the 
three Wigner-Dyson (WD) classes \cite{Atas13}. For the non-Hermitian case, we found that 
$\langle r\rangle$ reaches a constant value in the weaker disorder region 
($W\le W_c$), and the value increases with the system size for both models (FIG. \ref{r_constant}). For example, 
$\langle r\rangle=0.7182$, $0.7201$ for $L=12$, $20$ at $W=1$ for the NH AM and 
$\langle r\rangle=0.7307$, $0.7329$, $0.7353$, $0.7363$ for $L=10$, $12$, $16$, $20$ at $W=1$ for the NH U(1) model.  
By an extrapolation, we speculate $\langle r\rangle_{\rm metal}$ in the thermodynamic limit as 
$0.720$, and $0.736$ for the NH AM and U(1) models, respectively. Thus, the distributions of $r$
in the metal phases are different in the two NH models (FIG. \ref{Pr_NH_O1_U1_Metal}).

In order to characterize the Anderson transition, the scale-invariant quantity $\langle r\rangle$ is adopted. 
We estimate the critical exponent (CE) $\nu$ by polynomial fitting method \cite{Slevin14}, with 
various system size range, disorder range and expansion orders in the polynomial fitting.
Moreover, we narrow the energy window from the $10\%$ eigenvalues around $E=0$ to $5\%$ to see the stability of the polynomial fitting results. 
The results are shown in TABLE \ref{table} (a) for NH AM 
and (b) for NH U(1) model. 

CEs are consistent with each other for data sets with various system size range, disorder range, and 
expansion orders. This implies that our results are stable and precise. 
Moreover, CEs estimated from the 
$5\%$ eigenvalues energy window are consistent with that from the $10\%$ eigenvalues energy window, 
although they have a tendency to become smaller.
CEs estimated from the $10\%$ eigenvalues energy window is preferable because of the abundant energy levels. 
We conclude $\nu=0.99\pm 0.05$ for NH AM and $\nu=1.09\pm 0.05$ for NH U(1) model.

When the energy window is narrowed from 10\% to 5\% eigenvalues, the critical disorders in the both models 
tend to get larger; delocalized states at the band center are more robust against the disorder.
Besides, the CE for the NH AM changes to smaller values, when the data points for the smaller system 
sizes are excluded. The change of the CE in the NH AM becomes more prominent with 
the 5\% energy window than with the 10\% energy window. This means that the CE for the NH AM 
suffers from a systematic error by the choice of the energy windows. 
For the NH U(1) model, on the one hand, CEs extracted from the 5\% energy window stay nearly constant 
against the exclusion of the data from the smaller system sizes; 
the fitting of the NH U(1) model is much more stable than that of the NH AM.  

The difference of the stability in the fittings between the two models could be explained as follows.  
In the limit of the strongly localized phase, the level spacing ratio reaches the same value for the both models; 
$\langle r\rangle= 2/3$, while $\langle r\rangle$ in the limit of the delocalized phase goes to two 
different constant values in the two models respectively (FIG. \ref{r_constant}); there are two plateau regimes of 
$\langle r\rangle$ as a function of the disorder strength in these models. Note that data points near the plateau regimes 
should {\it not} be included for the scaling analysis, for they are likely outside the critical regime.
FIG. 1 in the main text shows that for the NH AM, the intersection of curves of 
$\langle r\rangle$ is quite close to the plateau value of $\langle r\rangle$ in the limit of the delocalized phase.    
Therefore, it is expected that a valid data range of the polynomial fitting in the NH AM becomes 
quite small in the side of the metal phase. On the other hand, the intersection for the NH U(1) model 
is relatively far away from the plateau value in the delocalized phase; 
the valid data range of the fitting becomes much wider in the NH U(1) model.

\subsection{Spectral rigidity}
The spectral rigidity is defined by a number variance within an energy window 
\begin{align}
\Sigma_{2}\equiv \langle \delta N^2 \rangle=\langle (N-\langle N\rangle )^2 \rangle,
\end{align}
where $N$ is the number of eigenvalues within the same energy window, and 
$\langle \cdots \rangle$ stands for the average over $M$ disorder realizations. $M$ is 
typically on the order of or larger than $10^4$. The spectral compressibility $\chi$ can be extracted by 
\begin{align}
\chi\equiv \lim_{L \rightarrow \infty} 
\lim_{N\rightarrow\infty}\frac{d \Sigma_2(N)}{d\langle N\rangle}. 
\end{align}
Here, we set a circular energy window around $E=0$ as $\{|E|<E_{\rm bound}\}$ with $E_{\rm bound}>0$. To calculate the spectral compressibility, 
we decrease the energy window (reduce $E_{\rm bound}$) for a fixed (but sufficiently large) system size.

Let us focus on the spectral compressibility at a critical point for the AT; $W=W_c$. 
From TABLE \ref{table}, We choose $W_c=6.3$ for the NH AM and $W_c=7.16$ for the NH U(1) model. We prepare
$M$ disorder realizations with $M=2\times 10^5$, $10^5$, $3\times 10^5$, $1.2\times 10^5$, $6\times10^4$, 
$10^4$ for $L=8$, $10$, $12$, $16$, $20$, $24$ for the NH AM  
and $N=2\times 10^5$, $10^5$, $6\times10^4$, $2.5\times 10^4$, $1.2\times10^4$, $6.4\times10^3$ for 
$L=8$, $10$, $12$, $16$, $20$, $24$ for the NH U(1) model.
In order to extract $\chi$, we change the energy window, namely $E_{\rm bound}$, within $10\%$ eigenvalues for a fixed system size $L$, and 
obtain various $\langle N\rangle$ and $\Sigma_2$ for each system size $L$. Then we carry out the linear fitting for 
$\Sigma_{2}$ vs. $\langle N\rangle$ for each $L$ (TABLE \ref{table_chi}). 
From TABLE \ref{table_chi}, we found that $\chi$ thus obtained is stable against changing the 
system size $L$. We thus choose $\chi$ as of the largest system size in TABLE \ref{table_chi}; 
$\chi\approx0.46$ for the NH AM and $\chi\approx0.55$ for the NH U(1) model.

For the 3D Hermitian AM, the critical spectral compressibility $\chi$ has been already studied by others 
\cite{Ndawana02,Bogomolny11,Ghosh17}. For the comparison with the NH cases, we recalculate the same quantity for 
the Hermitian case for much smaller size system ($L=10$). The 10$\%$ eigenvalues around $E=0$ are calculated at the 
critical disorder strength of the AT ($W=W_c$), and the average number and number variance are taken over $10^5$ disorder 
realizations for the AM ($W_c=16.5$, see \cite{Slevin18}), and over $6400$ disorder realizations for the U(1) model ($W_c=18.8$, see \cite{Slevin16}).
We narrow the energy window, obtain the various $\langle N\rangle$ and $\Sigma_2$, and carry out the linear fitting 
for $\Sigma_{2}$ vs. $\langle N\rangle$. This gives $\chi\approx 0.28$ \cite{Bogomolny11} for the Hermitian AM 
and $\chi\approx 0.31$ for the Hermitian U(1) model at the critical point of the AT.

\begin{table}[b]
	\centering
	\setlength{\tabcolsep}{6.5mm}
	\caption{Linear fitting results of critical spectral compressibility 
    $\chi$ of various system size $L$ for non-Hermitian (NH) Anderson model and U(1) model. }
	\begin{tabular}{c|cccccc}
		\hline
		\diagbox{model}{$L$}& 8 & 10 & 12 & 16 & 20 & 24 \\
		\hline
		NH Anderson model & 0.4658 & 0.461 & 0.4593 & 0.4585 & 0.457 & 0.4592 \\
		NH U(1) model & 0.5439 & 0.5425 & 0.5456 & 0.5469 & 0.5397 & 0.5533 \\
		\hline
	\end{tabular}\label{table_chi}
\end{table}

\subsection{Level spacing distribution for Gaussian random matrix}
For the Hermitian case, 
the level spacing distribution $P(s)$ in a metal phase 
is well described by the Wigner-Dyson (WD) surmise \cite{Wigner51} in random matrix theory; 
\begin{align}
P_{\rm GOE}(s)&=\frac{\pi}{2}s e^{-\frac{\pi}{4}s^2},  \  \ \beta=1 \nonumber \\
P_{\rm GUE}(s)&=\frac{32}{\pi^2}s^2e^{-\frac{4}{\pi}s^2},  \  \ \beta=2 \nonumber \\
P_{\rm GSE}(s)&=\frac{2^{18}}{3^6\pi^3}s^4e^{-\frac{64}{9\pi}s^2},  \  \ \beta=4 
\end{align}
where $\beta$ is the power exponent for smaller $s$ region (the Dyson index), and $\beta=1,2,4$ for Gaussian 
Orthogonal ensemble (GOE),Gaussian Unitary ensemble (GUE), and Gaussian Symplectic ensemble (GSE), respectively.
$P(s)\propto s^{\beta}$ for the smaller $s$ region is caused by repulsive interactions between the energy levels in the metal phase. 

\begin{figure}[b]
	\centering
	\includegraphics[width=0.65\linewidth]{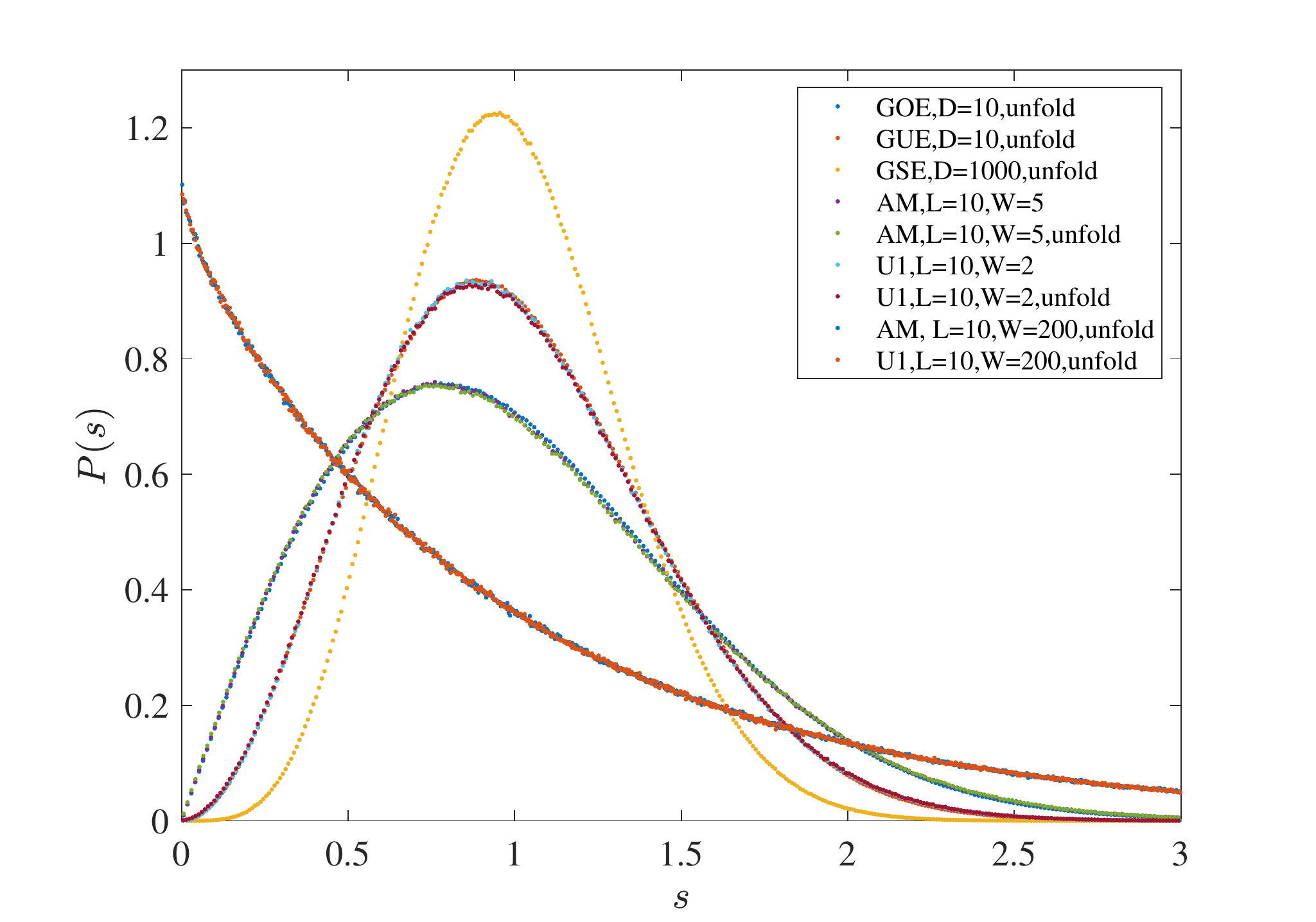}
	\caption{Level spacing distribution $P(s)$ for the Hermitian case. 
		$W=5, 200$ for Hermitian Anderson model (AM), and $W=2, 200$ 
         for U(1) models, corresponding to metal and insulator phase 
        ($W_c=16.5$ for the AM, and $W_c=18.8$ for the U(1) model). 
         The $10\%$ eigenvalues around $E=0$ 
         are calculated over $10^5$ samples for Hermitian AM and U(1) models ($L=10$). All the eigenvalues of 
         the random matrix are calculated over $6\times 10^6$ samples with the matrix dimension
        $D=10$ for the GOE, GUE, and over $64000$ samples with the matrix dimension
        $D=1000$ for GSE. $P(s)$ with $s$ that is not unfolded are also tried for comparison and unfolding makes no difference here, 
    because density of states is almost constant in the region calculated.}
	\label{Ps_HermitianSystem}
\end{figure}

In insulator phase, $P(s)$ obeys the Poisson distribution,
\begin{align}
 P(s)=e^{-s},
\end{align}
since energy levels are uncorrelated with others. At the critical point of the AT, 
\begin{align}
P(s) \propto s^{\beta_c},
\end{align}
for smaller $s$ region and 
\begin{align}
P(s) \propto e^{-\alpha s}, 
\end{align}
for larger $s$ region.

For the purpose of the comparison, we recalculated the level spacing distribution $P(s)$ of the Hermitian WD random matrix.
We prepare the Hermitian random matrix $H$ as, 
\begin{align}
	H=(A+A^{\dagger})/2, \label{dagger}
\end{align}
where $A$ is a random matrix with a restriction according to the symmetry.
For GOE, $H^*=H$, so $A$ is a real random matrix. 
For GUE, there is no restriction for $H$, so that $A$ is a complex random matrix.  
For GSE, $\Sigma_y H^* \Sigma_y=H$, with
\begin{align}
	\Sigma_y=
	\begin{pmatrix}
	 0& -i\\
	 i& 0\\
	\end{pmatrix},
\end{align}
 so that $A$ has the following structure, 
\begin{align}
    A=\begin{pmatrix}
    X & Y \\
    -Y^* & X^* \\
    \end{pmatrix}, \label{symplectic}
\end{align} 
where $X$, $Y$ are complex random matrices. Each element of the real random matrix 
is produced by the same Gaussian distribution, and  is independent. Real and imaginary parts of each element 
of the complex random matrix are produced by the same Gaussian distribution, and are independent. 
Eigenvalues in the GSE are doubly degenerate by the symmetry. We excluded this trivial degeneracy in the energy level 
statistics of $P(s)$.
We calculate all the eigenvalues of the random matrix with the matrix dimension $D=10$
and take energy level statistics over $6\times 10^6$ 
different realizations of the random matrix for GOE and GUE, while energy level statistics of the random matrix
with the matrix dimension $D=1000$ is taken over $64000$ different realizations of the random matrix for GSE. For the unfolded energy levels in FIG.~\ref{Ps_HermitianSystem}, 
we first calculate an averaged density of state out of many samples. In terms of the 
averaged density of states $\rho(E)$, we map an energy level in each sample into an 
integrated density of states (IDOS); 
\begin{align}
{\rm IDOS}(E_i)=\int_{}^{E_i}\rho(E)dE. 
\end{align} 
The level spacing corrected by the averaged density of states, $s_i$, 
is given by a difference between the neighboring IDOSs,
\begin{align}
	s_i \equiv {\rm IDOS}(E_{i+1})-{\rm IDOS}(E_i).
\end{align}

Our recalculation in FIG. \ref{Ps_HermitianSystem} reproduces the Dyson index;   
$\beta\approx 1$ for GOE, $\beta\approx 2$ for GUE, $\beta\approx 4$ for GSE.

\subsection{Level spacing distribution for Anderson model and U(1) model with Hermitian disorder}
For comparison, we also (re)calculate the level spacing distribution for Hermitian 
AM [Eq. (\ref{AM})] and U(1) models [Eq. (\ref{U1})] with real random onsite potentials. 
We calculate $10\%$ of the whole eigenvalues near $E=0$ with the system size $L=10$ for 
every disorder realization, and take the statistics over $10^5$ different disorder realizations.
FIG. \ref{Ps_HermitianSystem} shows that $P(s)$ in metal phase thus obtained  
for AM and U(1) model are consistent with $P(s)$ for GOE and GUE, respectively. 
Moreover, $P(s)$ with and without unfolding are almost identical to each other in metal phase. 
$P(s)$ with and without unfolding are so close to each other,  because the density of states 
within the $10\%$ energy windows is nearly constant in energy in metal phase. 
The unfolding process produces an error in $P(s)$ for small $s$ region, breaking a linear relationship; $\ln P(s) \propto \ln s$. 
We do not use the unfolding process when focusing on the behaviors of $P(s)$ at small-$s$ region. 

At the critical point, $P(s) \propto s^{\beta_c}$ for small $s$ region, where critical power-law exponent $\beta_c$ takes almost 
the same exponent as the corresponding Dyson index $\beta$ for every WD classes; 
$\beta_c \approx \beta$ \cite{Kawarabayashi96}. $P(s) \propto e^{-\alpha s}$ for large $s$ region, where $\alpha$ takes  
almost the same value for the three WD classes; $\alpha = 1.8\pm 0.1$~\cite{Batsch96,Hofstetter96,Zharekeshev95}.
In our calculation (FIG. \ref{Ps_O1_U1_H_large_s}), $\beta_c\approx0.98$, and $\alpha\approx 1.79$ at the critical point 
of the Hermitian AM ($W_c=16.5$), and $\beta_c\approx2.01$ and $\alpha\approx 1.71$ at critical point of the Hermitian U(1) model 
($W_c=18.8$).

\begin{figure}[t]
	\centering
	\subfigure[Hermitian Anderson model, $W_c$=16.5;  $\beta_c \approx 0.98$]{
		\begin{minipage}[t]{0.5\linewidth}
			\centering
			\includegraphics[width=1\linewidth]{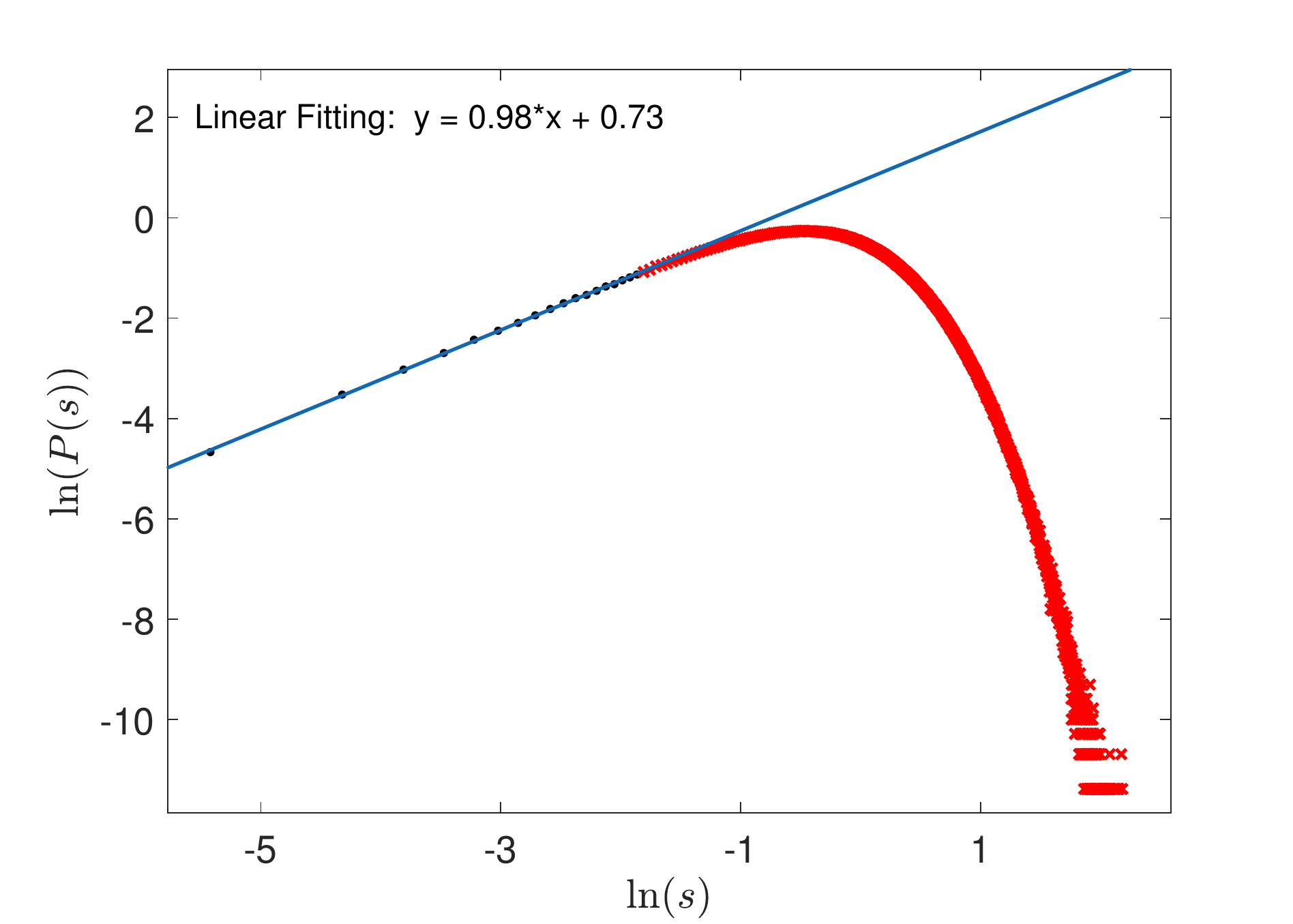}
		\end{minipage}%
	}%
	\subfigure[Hermitian U(1) model, at $W_c$=18.8;  $\beta_c \approx 2.01$]{
		\begin{minipage}[t]{0.5\linewidth}
			\centering
			\includegraphics[width=1\linewidth]{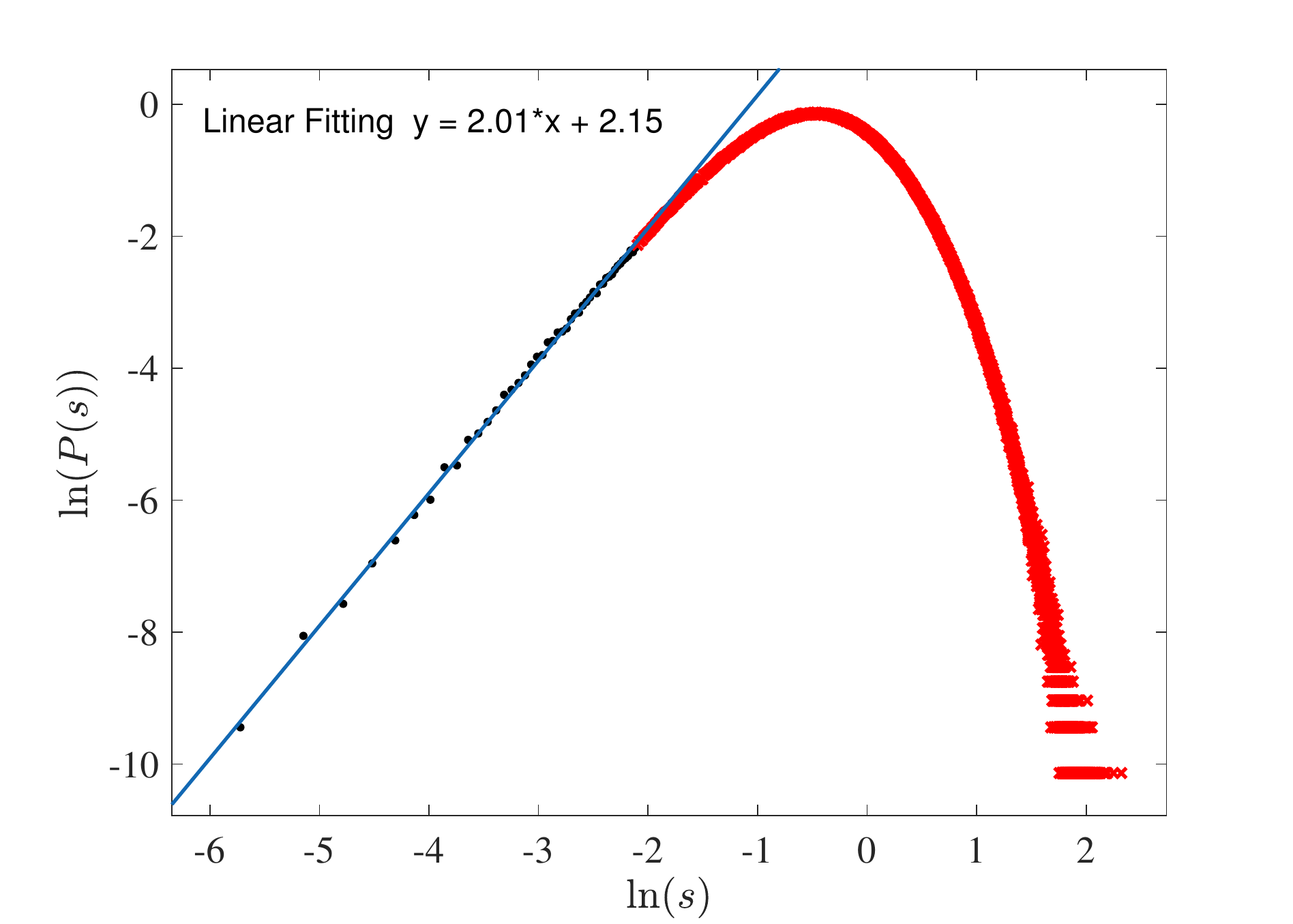}
		\end{minipage}%
	}%
	
	\subfigure[Hermitian Anderson model, at $W_c$=16.5; $\alpha\approx 1.79$]{
		\begin{minipage}[t]{0.5\linewidth}
			\centering
			\includegraphics[width=1\linewidth]{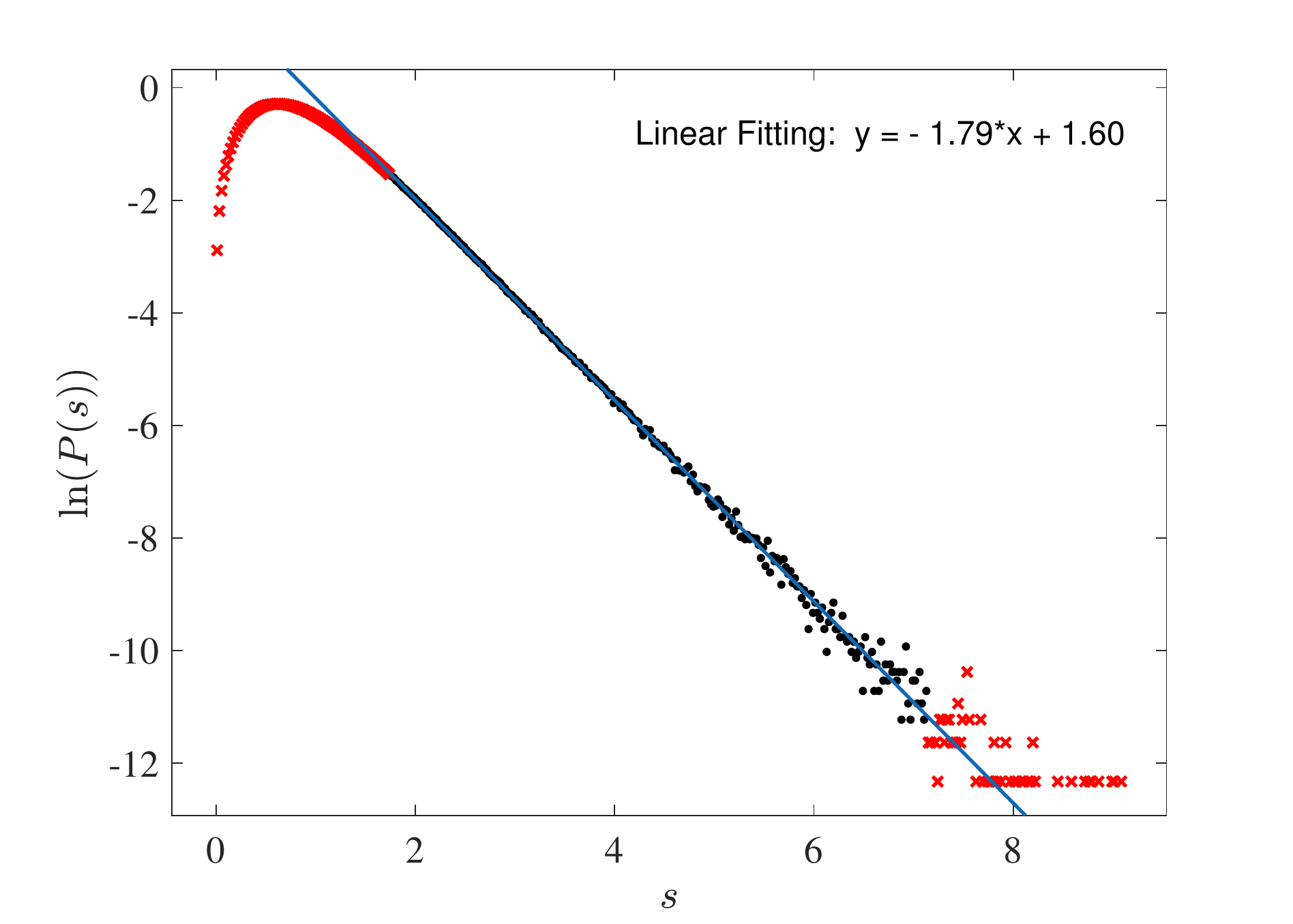}
		\end{minipage}%
	}%
	\subfigure[Hermitian U(1) model, $W_c$=18.8;  $\alpha\approx 1.71$]{
		\begin{minipage}[t]{0.5\linewidth}
			\centering
			\includegraphics[width=1\linewidth]{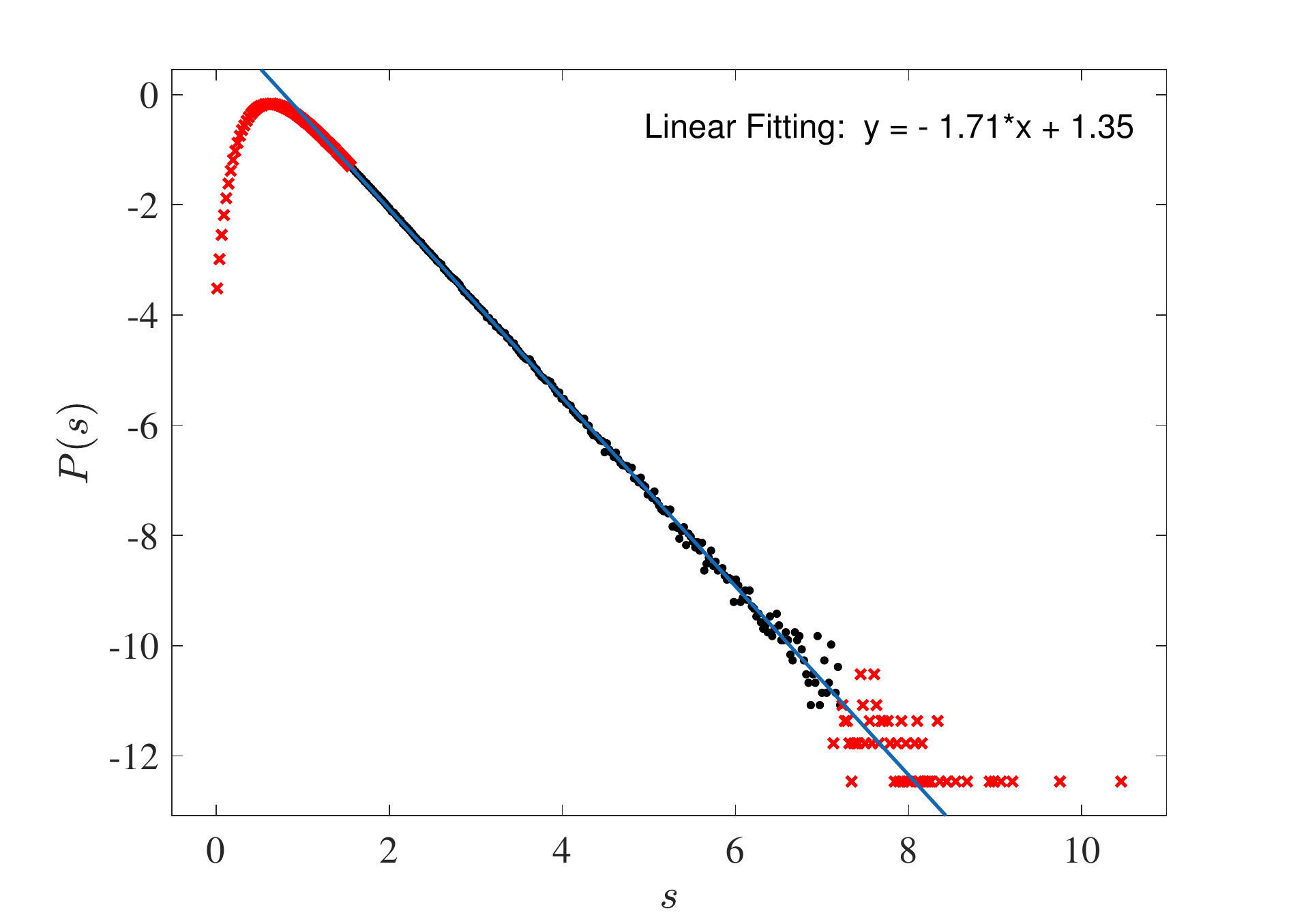}
		\end{minipage}%
	}%
	\caption{Level spacing distribution $P(s)$ for small $s$ ((a), (b); without unfolding), and for large $s$ ((c), (d); with unfolding). 
		The data comes from $10^5$ disorder realizations with $L=10$ for both models. The data are fitted by  
		$P(s)\propto s^{\beta_c}$ at small s and $P(s) \propto e^{-\alpha s}$ at large s. Red points are data excluded 
		for the linear fitting.}
	\label{Ps_O1_U1_H_large_s}
\end{figure}

\subsection{Level spacing distribution for Ginibre ensemble}
Ginibre ensembles are classes of ensembles for non-Hermitian random matrices~\cite{Ginibre65}, and therefore they might be useful for understanding the 
energy level statistics and the AT in a non-Hermitian disorder system.
According to the classification, there exist three kinds of the Ginibre ensembles;  
Ginibre Orthogonal ensemble (GinOE) ($H^*=H$), Ginibre Unitary ensemble (GinUE) (no restriction on $H$), and Ginibre Symplectic ensemble (GinSE) ($\Sigma_y H^* \Sigma_y=H$).  
The matrix $A$ for GOE, GUE, and GSE without Eq.~(\ref{dagger})  
corresponds to random matrix in GinOE, GinUE, and GinSE, respectively. 
GinOE, GinUE, GinSE correspond to the symmetry class AI, A, AII in the 
classification for the NH system. Because of the symmetry, 
eigenvalues in GinOE and GinSE come in pairs; 
$\{E_i, E_i^*\}$, and eigenvalues in the upper-half Euler 
plane are sufficient for the energy level statistics. The double degeneracy 
on the real axis needs to be excluded. We thus use only those eigenvalues 
whose imaginary parts are greater than 1, to determine the energy level 
statistics. Now the eigenvalues are complex number and the level 
spacing $s$ is defined by
\begin{align}
	s_i=|E_i-E_{\rm NN}|
\end{align}
where $E_{\rm NN}$ is the nearest neighbor for $E_i$.
Here the density of states in complex Euler plane is almost constant in the 
region calculated, so that we omit the unfolding process.
For small $s$, $P(s)$ of all these three kinds of the random matrix, GinOE, GinUE, and GinSE,  
obeys the same distribution (FIG. \ref{Ps_NH_insulator}(b)) with a cubic repulsion; $P(s) \propto s^3$~\cite{Hamazaki20}.

\subsection{Level spacing distribution for Anderson model and U(1) model with non-Hermitian disorder}
Level spacing distribution are calculated for the NH AM (Eq. \ref{AM}) and U(1) models (Eq. \ref{U1}). 
In insulator, $P(s)$ takes the 2D Poisson distribution \cite{Grobe88},
\begin{align}
	P_{P}^{2D}(s)=\frac{\pi}{2}se^{-\pi s^2/4}.
\end{align}
To test this formula, we calculates $10\%$ eigenvalues around $E=0$ in the complex Euler plane 
for the NH AM and U(1) model with $L=12$ at $W=100$, where the eigenstates in both models 
are in insulator phase. We determine $P(s)$ out of the $10\%$ eigenvalues calculated over $6\times 10^4$ 
different disorder realizations. $P(s)$ thus determined takes the same 2D Poisson distribution in the 
both models (FIG. \ref{Ps_NH_insulator}(a)).

To calculate $P(s)$ in metal phase, we calculate the $10\%$ eigenvalues around $E=0$  
for NH AM and U(1) model with $L=16$ at $W=3$, where it is guaranteed that 
all the eigenstates within the $10\%$ circular energy window are in the metal phase. We find that 
$P(s)$ for NH U(1) model is consistent with that for GinOE, GinUE, GinSE (FIG. \ref{Ps_NH_insulator}(b)), 
but $P(s)$ for NH AM deviates from that for GinOE, GinUE, GinSE.
In small $s$ region, $P(s)$ behaves as $s^{\beta}$, where $\beta\approx 2.74$ 
for NH AM and $\beta\approx 2.93$ for NH U(1) model 
(FIG. \ref{Ps_O1_U1_NH_c}). This is consistent with ref. \cite{Hamazaki20} where the 
class AI$^{\dagger}$ shows unique $P(s)$, different from $P(s)$ for the 
class A (FIG. \ref{Ps_NH_insulator}(b)).

To calculate $P(s)$ at the critical point,  we calculate the $10\%$ eigenvalues around $E=0$  
for the NH AM and U(1) model at $W=W_c$ ($W_c \approx 6.3$ for NH AM and $W_c\approx 7.16$ for NH U(1) model). 
The system size $L$ and the number of disorder realizations $M$ are set as 
$M=2\times 10^5$, $10^5$, $3\times 10^5$, $1.2\times 10^5$, $6\times10^4$, $10^4$ for $L=8$, $10$, $12$, 
$16$, $20$, $24$ for the NH AM, and $M=2\times 10^5$, $10^5$, $6\times10^4$, $2.5\times 10^4$, $1.2\times10^4$, 
$6.4\times10^3$ for $L=8$, $10$, $12$, $16$, $20$, $24$ for the NH U(1) model. For each $L$, $P(s)$ is determined  
from the $10\%$ eigenvalues calculated over the $M$ different disorder realizations. We fit $P(s)$ thus obtained 
as $s^{\beta_c}$ for small $s$ and as $e^{-\alpha s}$ for large $s$. $\beta_c$ and $\alpha$ are estimated for 
each system size $L$ (FIG. \ref{Ps_O1_U1_NH_c} and TABLE \ref{table_ps_c}). We find that the fitted values of 
$\beta_c$ and $\alpha$ are robust against the change of system size; $\beta_c= 2.6\pm 0.05$ , 
$\alpha=5.0\pm0.1$ for NH AM and $\beta_c= 2.9\pm 0.05$ , $\alpha= 4.5\pm 0.1$ for NH U(1) model.

\begin{table}[h]
	\centering
	\setlength{\tabcolsep}{5.5mm}
	\caption{Linear fitting result of critical level spacing distribution $P(s)$ for small and 
large $s$ regions. The fitting by $e^{-\alpha s}$ for the large $s$ region gives $\alpha$ and 
the fitting by $s^{\beta_c}$ for the small $s$ region gives $\beta_c$.}
	\begin{tabular}{c|c|cccccc}
		\hline
		Model & \diagbox{quantity}{$L$} & 8 & 10 & 12 & 16 & 20 & 24 \\
		\hline
		NH AM & $\alpha$ & 4.75 & 4.8 & 4.9 & 4.9 & 5.0 & 5.06 \\
		NH AM & $\beta_c$ & 2.61 & 2.6 & 2.63 & 2.6 & 2.56 & 2.62 \\
	NH	U(1) model & $\alpha$ & 4.1 & 4.1 & 4.3 & 4.4 & 4.46 & 4.61 \\
	NH 	U(1) model & $\beta_c$ & 2.96 & 2.98 & 2.96 & 2.95 & 2.88 & 2.84 \\
		\hline
	\end{tabular}\label{table_ps_c}
\end{table}

\begin{figure}[h]
	\centering
				\subfigure[ $P(s)$ for insulator phase]{
		\begin{minipage}[t]{0.5\linewidth}
			\centering
			\includegraphics[width=1\linewidth]{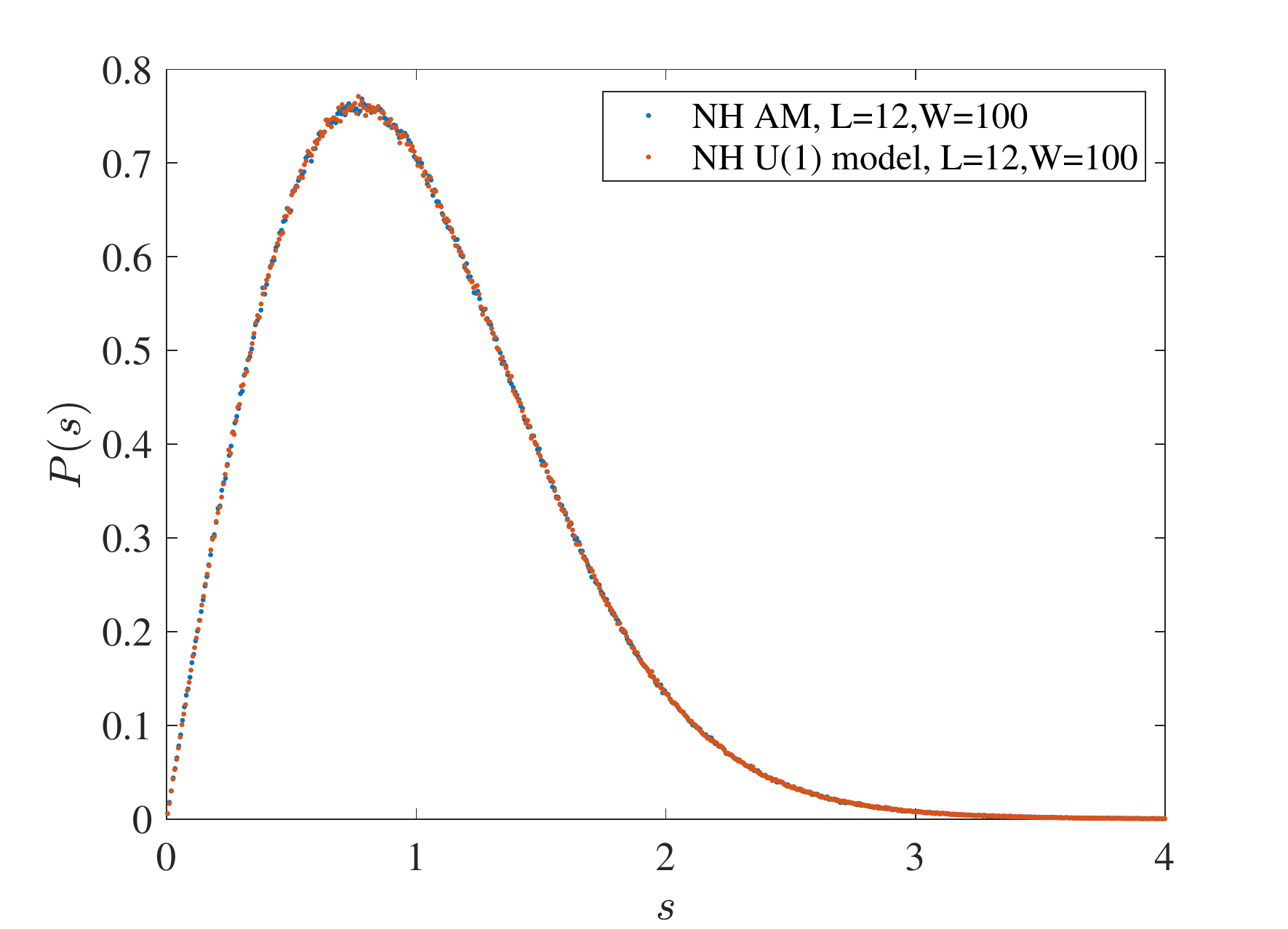}
		\end{minipage}%
	}%
	\subfigure[ $P(s)$ for metal phase and Ginibre ensembles]{
		\begin{minipage}[t]{0.5\linewidth}
			\centering
			\includegraphics[width=1\linewidth]{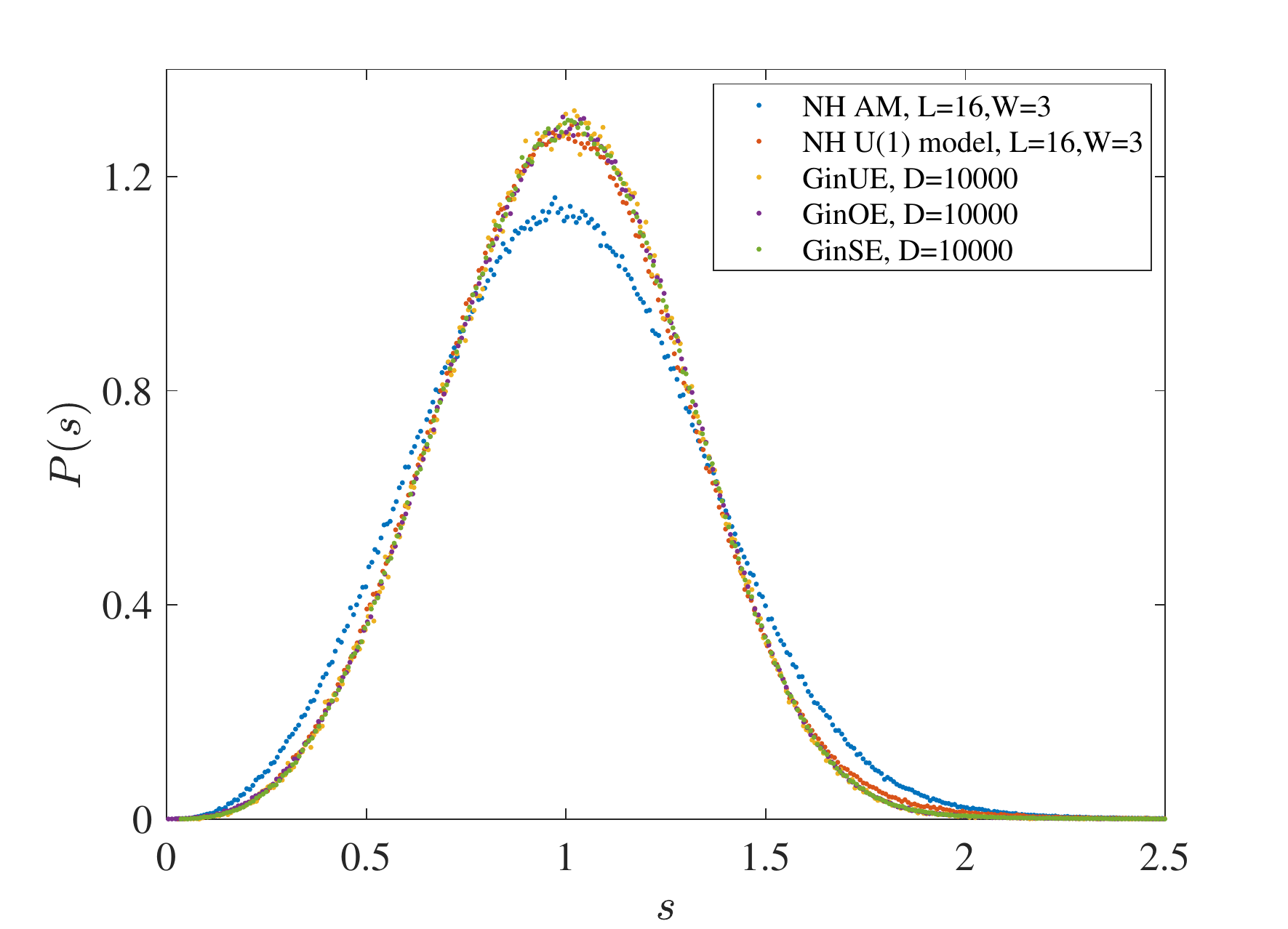}
		\end{minipage}%
	}%
	\caption{Level spacing distribution $P(s)$ in (a) insulator phase ($W=100$) and (b) metal phase ($W=3$) of non-Hermitian (NH) Anderson model (AM) and U(1) model.
		In (b)$P(s)$'s for GinUE, GinOE, GinSE with matrix dimension $D=10^4$ are also shown; 64 samples for GinUE and 640 samples for GinOE and GinSE.
		The distributions for insulator phase are constructed out of the $10\%$ eigenvalues around $E=0$ over $6\times 10^4$ samples for  both models ($L=12$).
	The distributions for metal phase are calculated from $10\%$ eigenvalues around $E=0$ over $6400$ samples for both models ($L=16$). }
	\label{Ps_NH_insulator}
\end{figure}

 \begin{figure}
	\centering
			\subfigure[NH Anderson model at $W=3$ (metal phase); $\beta \approx  2.74$]{
		\begin{minipage}[t]{0.5\linewidth}
			\centering
			\includegraphics[width=1\linewidth]{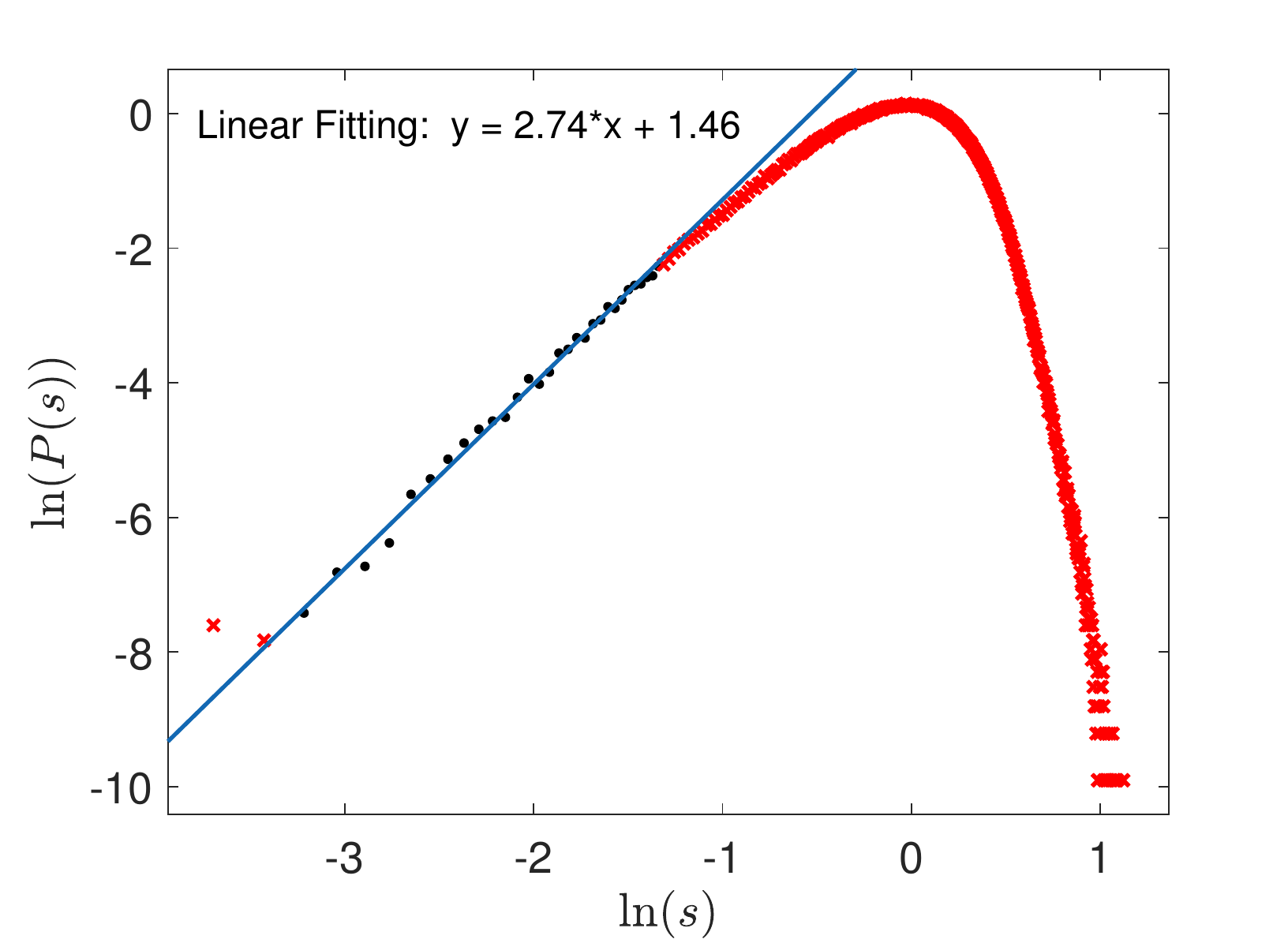}
		\end{minipage}%
	}%
	\subfigure[NH U(1) model at $W=3$ (metal phase); $\beta \approx 2.93 $]{
		\begin{minipage}[t]{0.5\linewidth}
			\centering
			\includegraphics[width=1\linewidth]{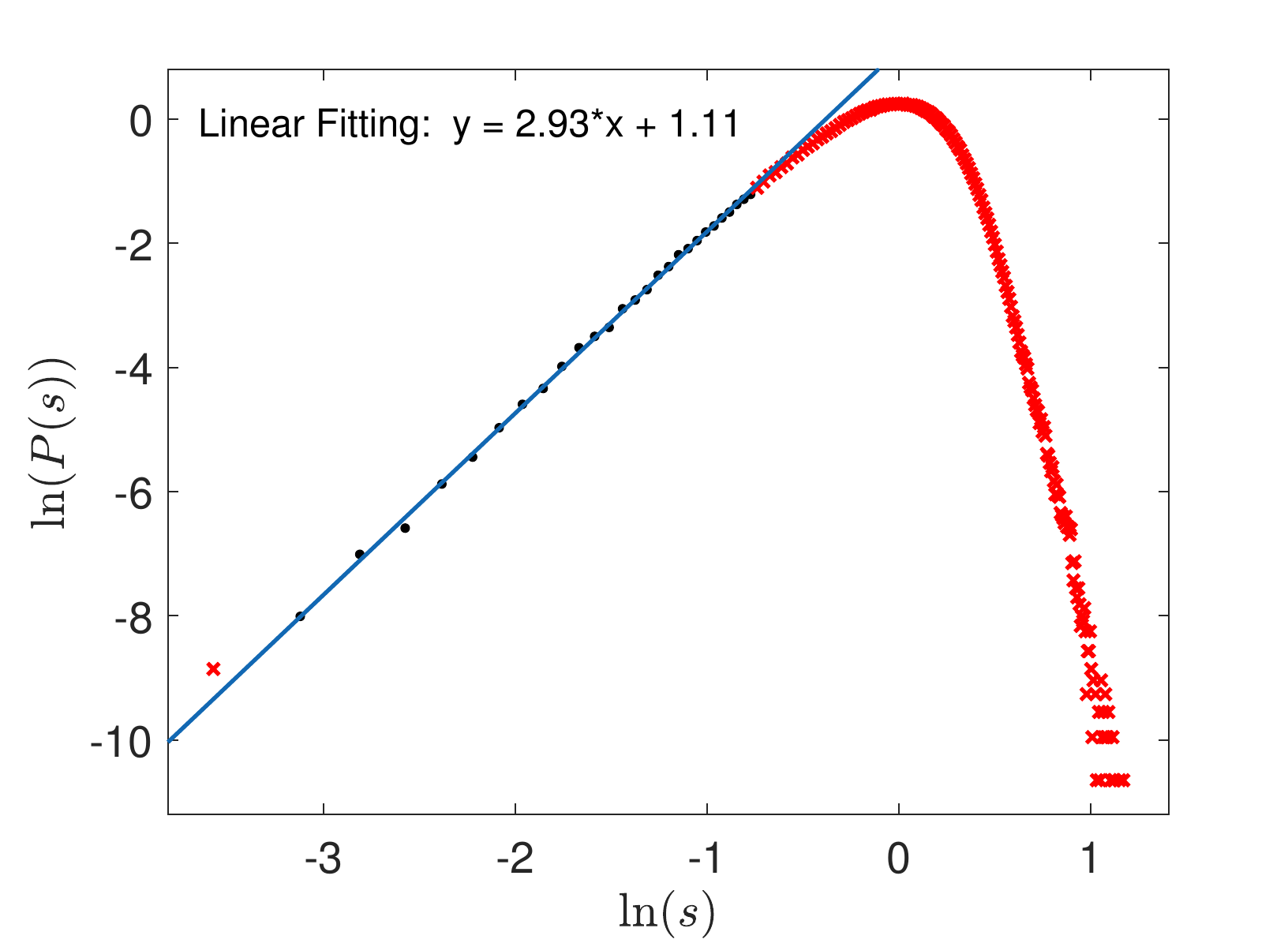}
		\end{minipage}%
	}%

		\subfigure[NH Anderson model at $W_c=6.3$; $\beta_c \approx  2.62$]{
		\begin{minipage}[t]{0.5\linewidth}
			\centering
			\includegraphics[width=1\linewidth]{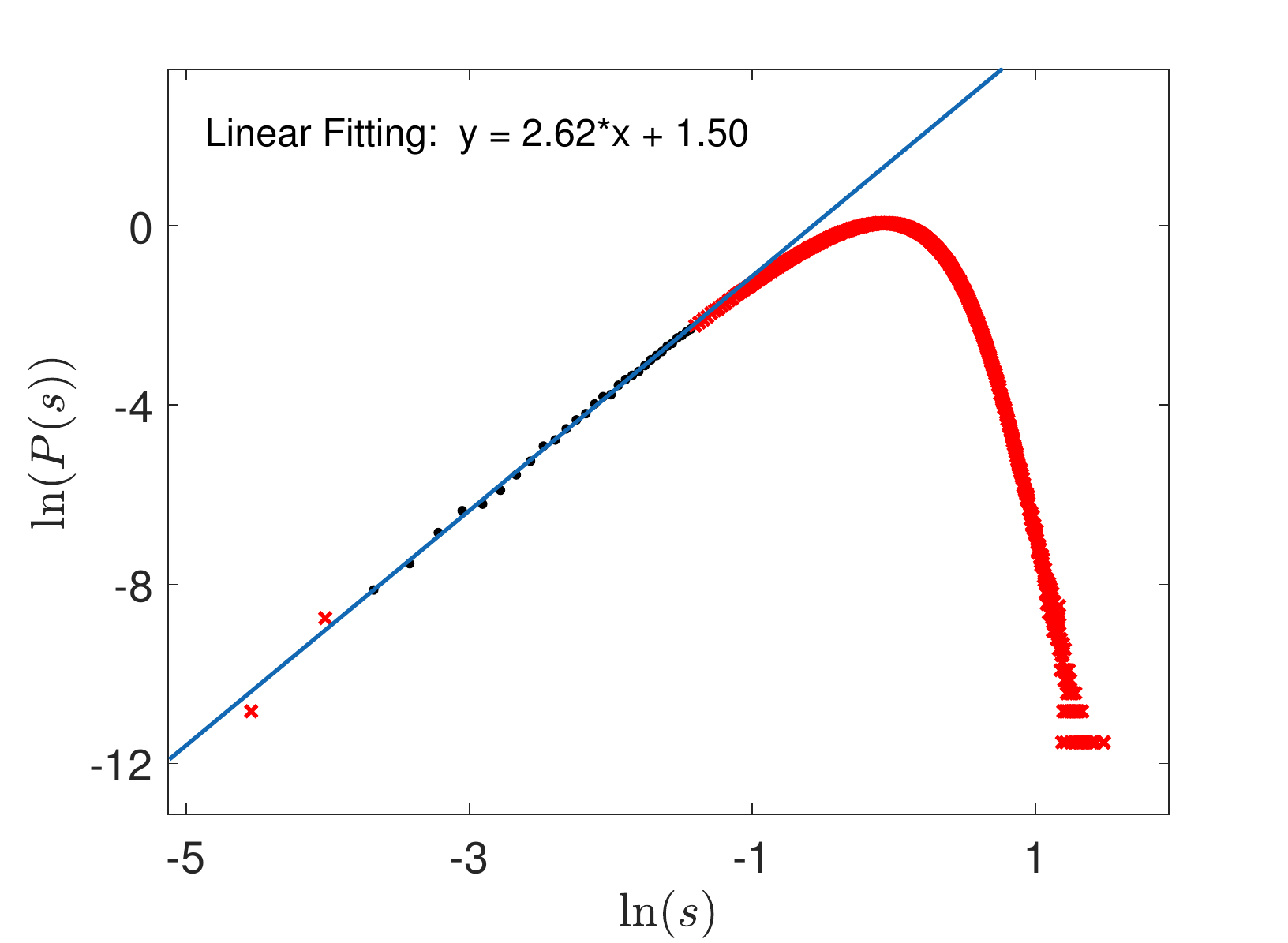}
		\end{minipage}%
	}%
	\subfigure[NH U(1) model at $W_c=7.16$; $\beta_c \approx 2.84 $]{
		\begin{minipage}[t]{0.5\linewidth}
			\centering
			\includegraphics[width=1\linewidth]{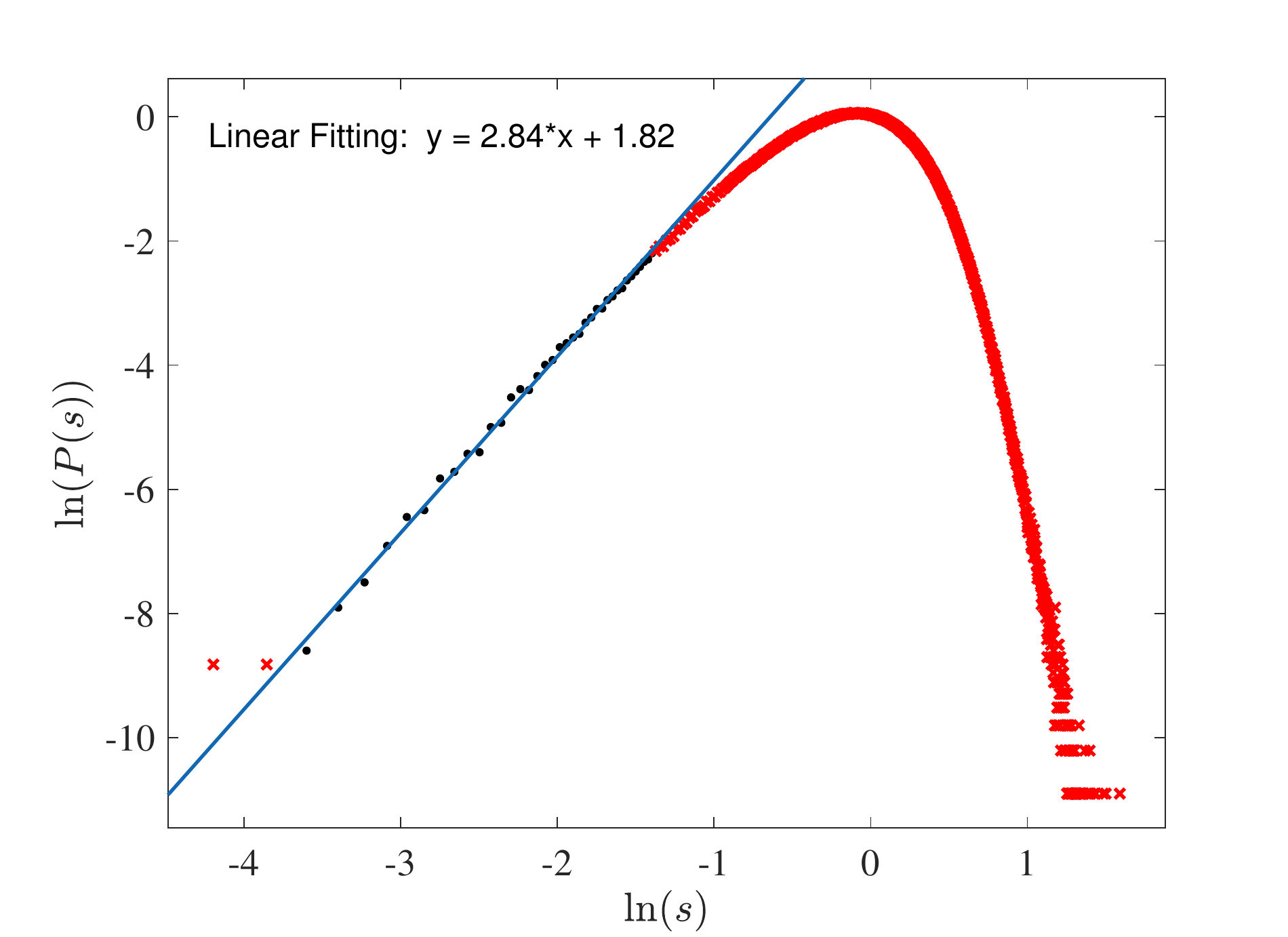}
		\end{minipage}%
	}%

	\subfigure[NH Anderson model at $W_c=6.3$; $\alpha\approx 5.06 $]{
	\begin{minipage}[t]{0.5\linewidth}
		\centering
		\includegraphics[width=1\linewidth]{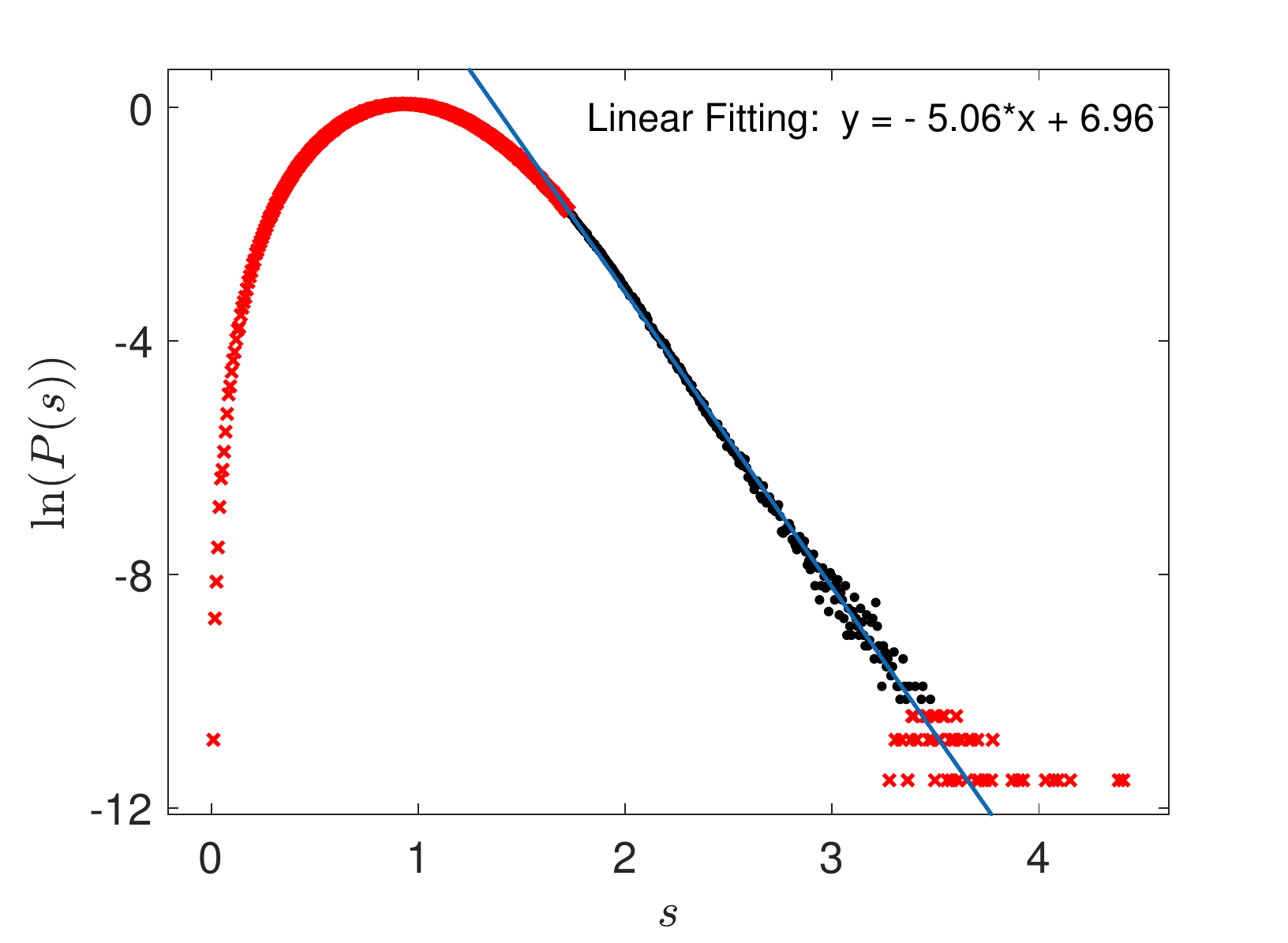}
	\end{minipage}%
}%
\subfigure[NH U(1) model at $W_c=7.16$; $\alpha\approx 4.61$]{
	\begin{minipage}[t]{0.5\linewidth}
		\centering
		\includegraphics[width=1\linewidth]{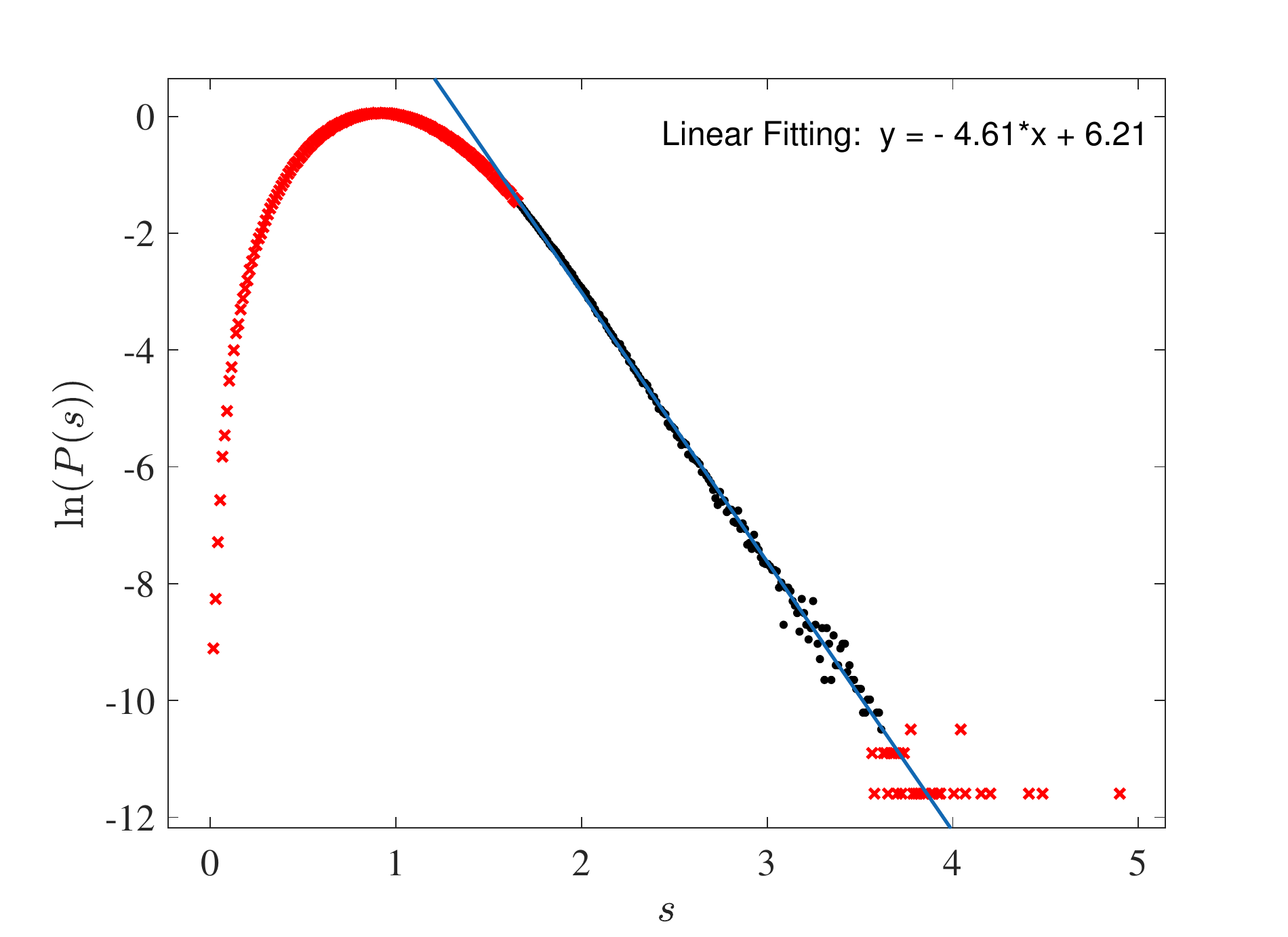}
	\end{minipage}%
}%
	\caption{Level spacing distribution $P(s)$ at small $s$ [(a)-(d)] and large $s$ [(e), (f)]. The blue lines are fitting curves by
    $P(s)\propto s^{\beta_{c}}$ at small $s$ and $P(s)\propto e^{-\alpha s}$ at large $s$. 
    In metal phase [(a), (b); $W=3$], the data comes from $6400$ disorder realizations with $L=16$. 
    At the critical point [(c)-(f); $W=W_c$], we take $W_c=6.3$, $L=24$ and $10^4$ samples for the 
   non-Hermitian (NH) Anderson model, and $W_c=7.16$, $L=24$ and $6400$ samples for the NH U(1) model. 
   Red points are data excluded for the linear fitting. }
	\label{Ps_O1_U1_NH_c}
\end{figure}

\clearpage
\subsection{Level spacing ratio distribution for the Hermitian system}
For the Hermitian case, we consider a distribution of level spacing ratio $r$, that is defined by\cite{Oganesyan07}
\begin{align}
r_i\equiv {\rm min}\Big(\frac{E_{i+1}-E_i}{E_i-E_{i-1}}, \frac{E_{i}-E_{i-1}}{E_{i+1}-E_{i}}\Big). 
\end{align}
Here $\{E_i\}$ are ordered in the ascending order ($E_1<E_2<E_3\cdots$). 
For comparison, we calculate the level spacing ratio distribution, $P(r)$, for the random matrix in 
GOE, GUE and GSE and for the Hermitian AM, U(1) models (FIG. \ref{Pr_Hermitian}). 
A random matrix theory \cite{Atas13} tells that $P(r)$ in the metal phase is given by
\begin{align}
P(r)=\frac{1}{C_{\beta}}\frac{(r+r^2)^{\beta}}{(1+r+r^2)^{1+\frac{3}{2}\beta}}\Theta(1-r). \label{pr_metal}
\end{align}
Here $C_{\beta}$ is a constant, $\beta=1,2,4$ for GOE, GUE, and GSE, respectively, and $\Theta(x)$ is the 
Heaviside step function.

In insulator, $P(r)$ is given by \cite{Atas13}
\begin{align}
P(r)=\frac{2}{(1+r)^2}\Theta(1-r). \label{pr_insulator}
\end{align}
for all the three WD classes. 
FIG. \ref{Pr_Hermitian} shows that $P(r)$ in metal phase of the Hermitian AM and U(1) models are consistent 
with $P(r)$ of GOE and GUE, respectively. It also shows that $P(r)$ in the insulator phase has the same distribution as 
in Eq.~(\ref{pr_insulator}) for  both models.

\begin{figure}[b]
	\centering
	\includegraphics[width=0.7\linewidth]{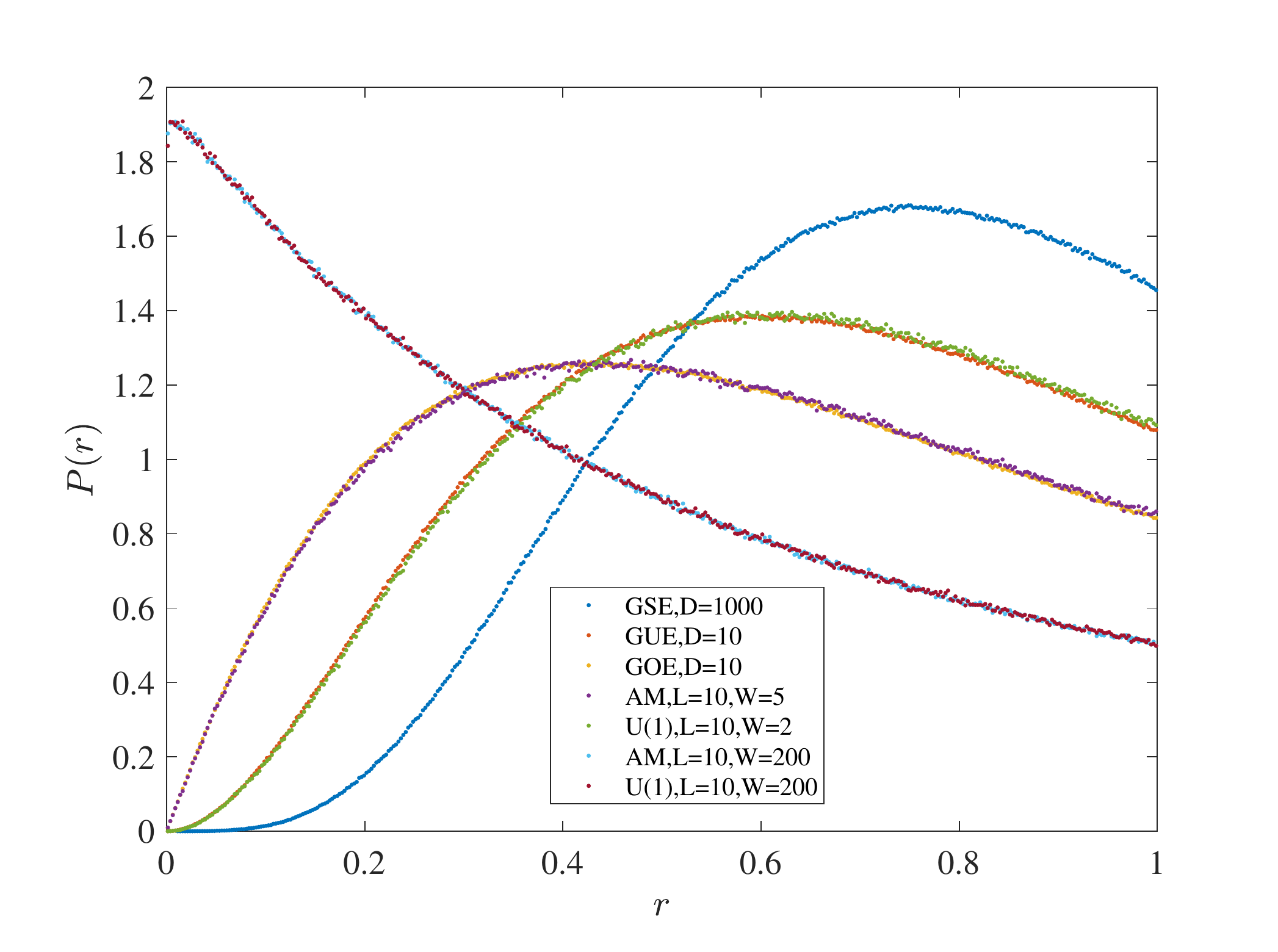}
	\caption{Level spacing ratio distribution $P(r)$ for GOE, GUE, GSE, the Hermitian Anderson model (AM) and U(1) model. 
We set the matrix dimension $D=10$ with $6\times 10^6$ realizations for GOE, GUE, and $D=1000$ with $64000$ realizations for GSE. 
For the Hermitian Anderson model and U(1) model, $P(r)$ is obtained from $10\%$ eigenvalues around $E=0$ calculated 
with the system size $L=10$ and $10^5$ samples.}
	\label{Pr_Hermitian}
\end{figure}

\subsection{Level spacing ratio distribution for non-Hermitian system}
For the non-Hermitian case, we can consider not only the level spacing ratio $r$ but also the angle of $z_i$,
\begin{align}
\theta_i\equiv {\arg} (z_i),
\end{align}
where $z_i$ is defined in Eq. (\ref{z_i}). To see distributions of $r$ and $\theta$ in metal phase, 
we calculate $10\%$ eigenvalues around $E=0$ in the complex 
Euler plane for NH AM and U(1) model at $W=3$. We take the statistics over 6400 disorder realizations, 
to obtain the level spacing ratio distributions, $P(r)$ and $P(\theta)$. 
$P(r)$ and $P(\theta)$ in metal phase of the NH U(1) model are consistent with those of the GinUE. 
However, $P(r)$ and $P(\theta)$ in metal phase of the NH AM behave quite differently from the GinUE. 
This supports the conclusion of Ref. \cite{Hamazaki20} that the metal phase of the 
NH AM and the metal phase in the NH U(1) model belong to two different universality classes. 
Noted that FIG. \ref{Pr_NH_O1_U1_Metal} also shows a small deviation between $P(\theta)$ of the NH U(1) model 
and $P(\theta)$ of the GinUE. We speculate that $\theta$ is more sensitive to the finite-system-size effect 
than $r$, because of boundary effects \cite{Sa20}.

FIG. \ref{Pr_Ptheta_NH_O1_U1_Metal_insulator} (a)-(d) show behaviors of $P(r)$ and $P(\theta)$ from metal phase 
to insulator phase in the NH AM and NH U(1) model.
$P(r)$ becomes linear in $r$ in the insulator phase for both models. This observation is 
consistent with that in Ref. \cite{Sa20}. On the other hand, $P(\theta)$ shows a small peak at $\theta=0$ in the 
insulator phase for both models (FIG. \ref{Pr_Ptheta_NH_O1_U1_Metal_insulator} (e)). This observation is different from Ref. \cite{Sa20}. 
We speculate that the small peak in $P(\theta)$ come from the boundary effect, as pointed out in Ref \cite{Sa20}. 
Namely, those $E_i$ around the boundary of the $10\%$ circular energy window have higher 
chance to give smaller $\theta_i$, because such $E_i$ is apt to find its nearest ($E_{\rm NN}$) and next nearest 
neighbor ($E_{\rm NNN}$) in the same direction (an inner direction of the circular window; toward $E=0$). For 
calculations with the smaller system size, these eigenvalues near the circular boundary have considerable effect, 
causing a small peak at $\theta=0$ in $P(\theta)$. To uphold this speculation, we also calculate all the  
eigenvalues of the NH U(1) model with $L=20$ at $W=100$, and take the statistics over 640 samples.  
$P(\theta)$ thus obtained is flat in $\theta$ as expected (FIG. \ref{Pr_Ptheta_NH_O1_U1_Metal_insulator}(f)).

FIG. \ref{Pr_Ptheta_NH_O1_U1_c} shows critical $P(r)$ and $P(\theta)$ for the two NH models. 
$P(\theta)$ shows some amount of size dependences as $|\theta|$ approaches $\pi$, where 
$P(\theta)$ for the large $\theta$ has a tendency to be larger for larger system size. In other 
words, $P(\theta)$ for the small $\theta$ tends to be smaller for the larger system. We speculate 
that this size dependence also partially comes from the boundary effect mentioned above. 
Note also that $P(r)$ and $P(\theta)$ are almost identical in the two models, except for 
small deviations observed in $P(r)$ at smaller $r$ (FIG. \ref{Pr_Ptheta_NH_O1_U1_c} (f)) and 
$P(\theta)$ at larger $\theta$ (FIG. \ref{Pr_Ptheta_NH_O1_U1_c} (e)). We conclude that 
it is hard to distinguish the two different universality classes in NH AM and NH U(1) models 
in terms of critical distributions of $P(r)$ and $P(\theta)$. 

\begin{figure}[h]
	\centering
		\subfigure[ $P(r)$]{
		\begin{minipage}[t]{0.5\linewidth}
			\centering
			\includegraphics[width=1\linewidth]{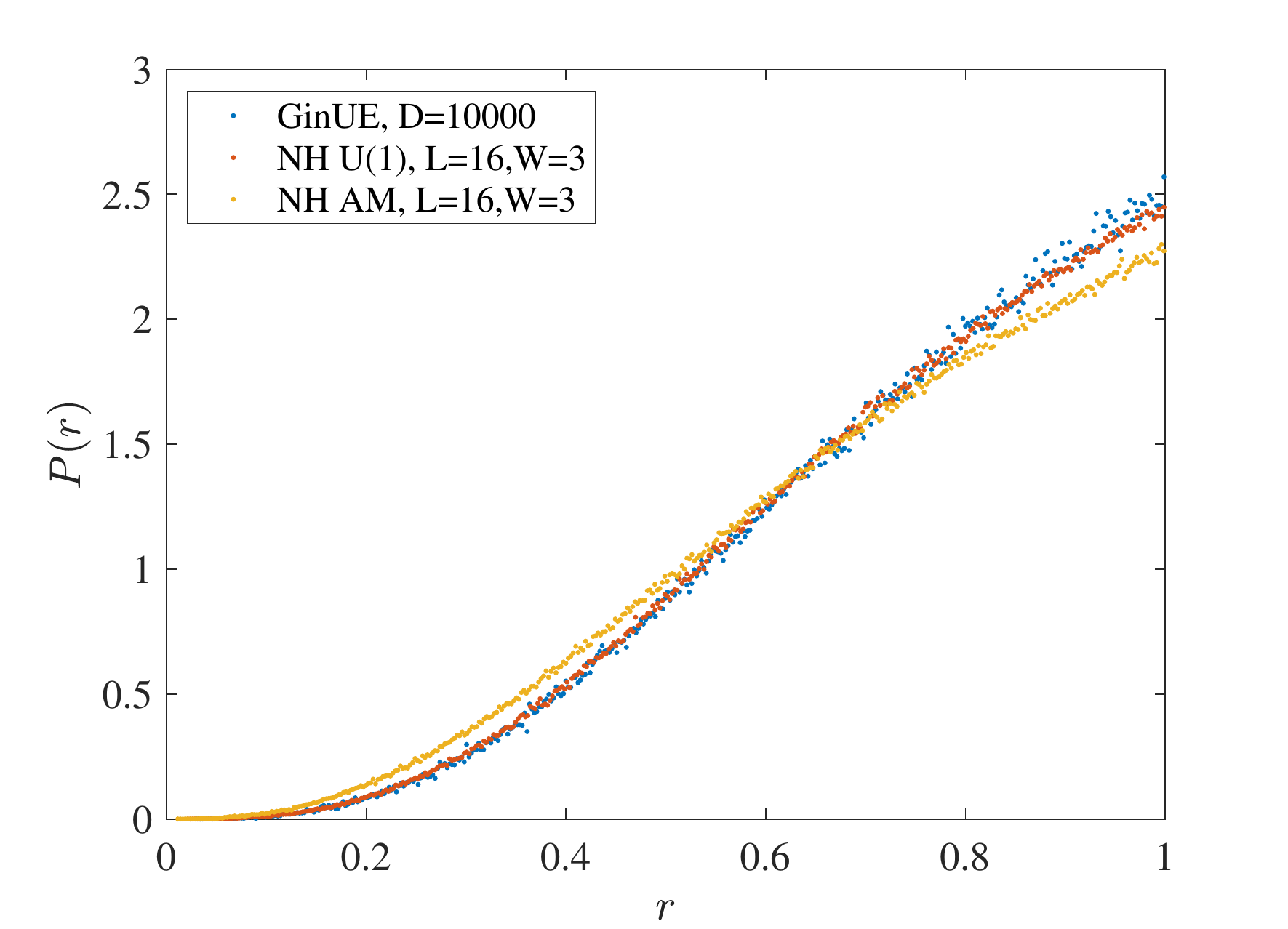}
		\end{minipage}%
	}%
	\subfigure[ $P(\theta)$]{
		\begin{minipage}[t]{0.5\linewidth}
			\centering
			\includegraphics[width=1\linewidth]{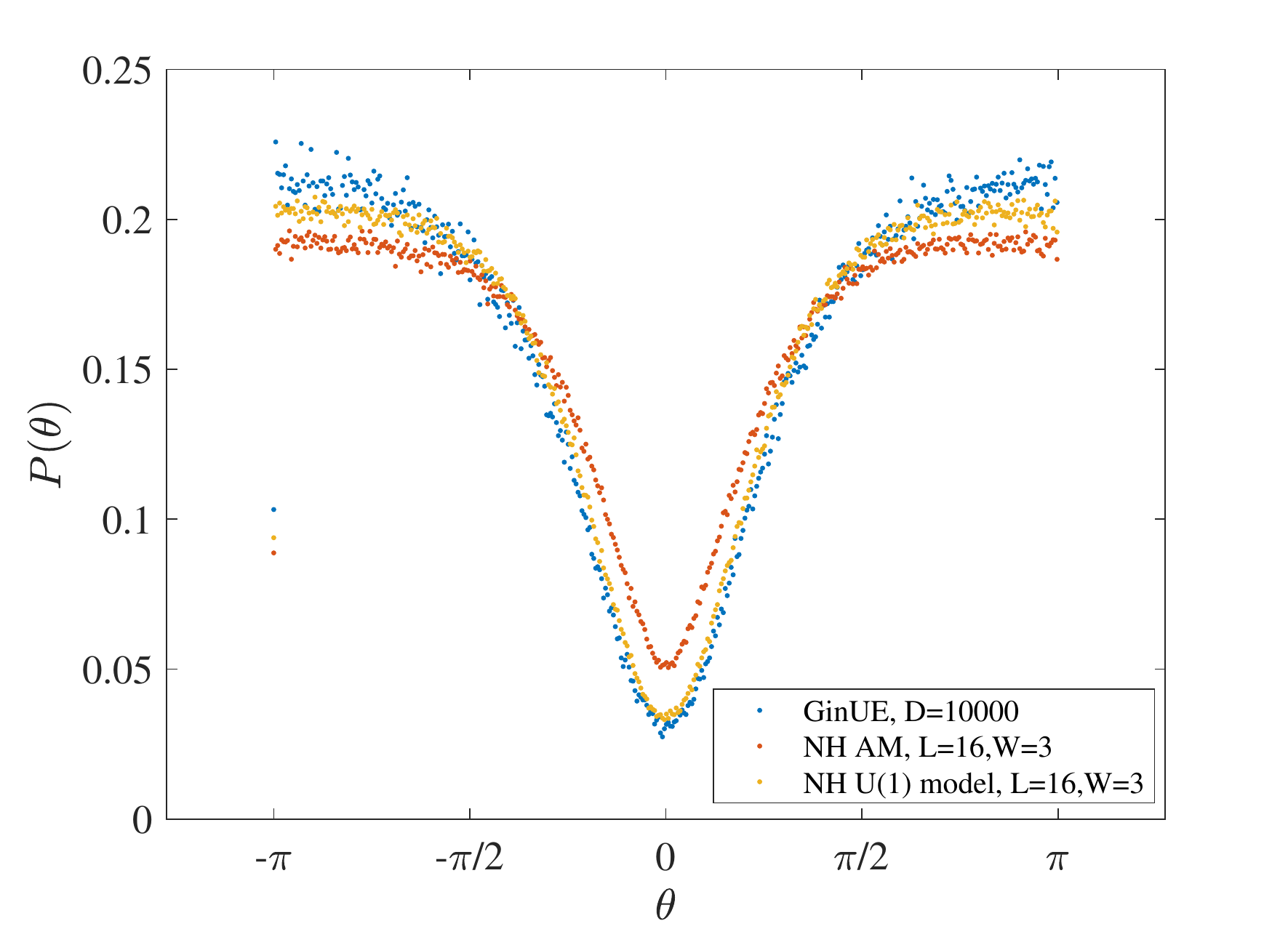}
		\end{minipage}%
	}%
	\caption{$P(r)$ and $P(\theta)$ for the Ginibre unitary ensemble, non-Hermitian (NH) Anderson model ($W=3$), 
        NH U(1) model ($W=3$). We take 64 samples for the GinUE, $10\%$ eigenvalues around $E=0$ over 
        6400 samples for the NH Anderson model and U(1) model. It is guaranteed that the $10\%$ eigenvalues 
        around $E=0$ in these two NH models are in the metal phase at $W=3$.}
	\label{Pr_NH_O1_U1_Metal}
\end{figure}

\begin{figure}
	\centering
	\subfigure[$P(r)$ for NH AM in metal phase ($W=1$), critical point ($W=6.3$) and insulator phase ($W=7.4, 100$)]{
		\begin{minipage}[t]{0.5\linewidth}
			\centering
			\includegraphics[width=1\linewidth]{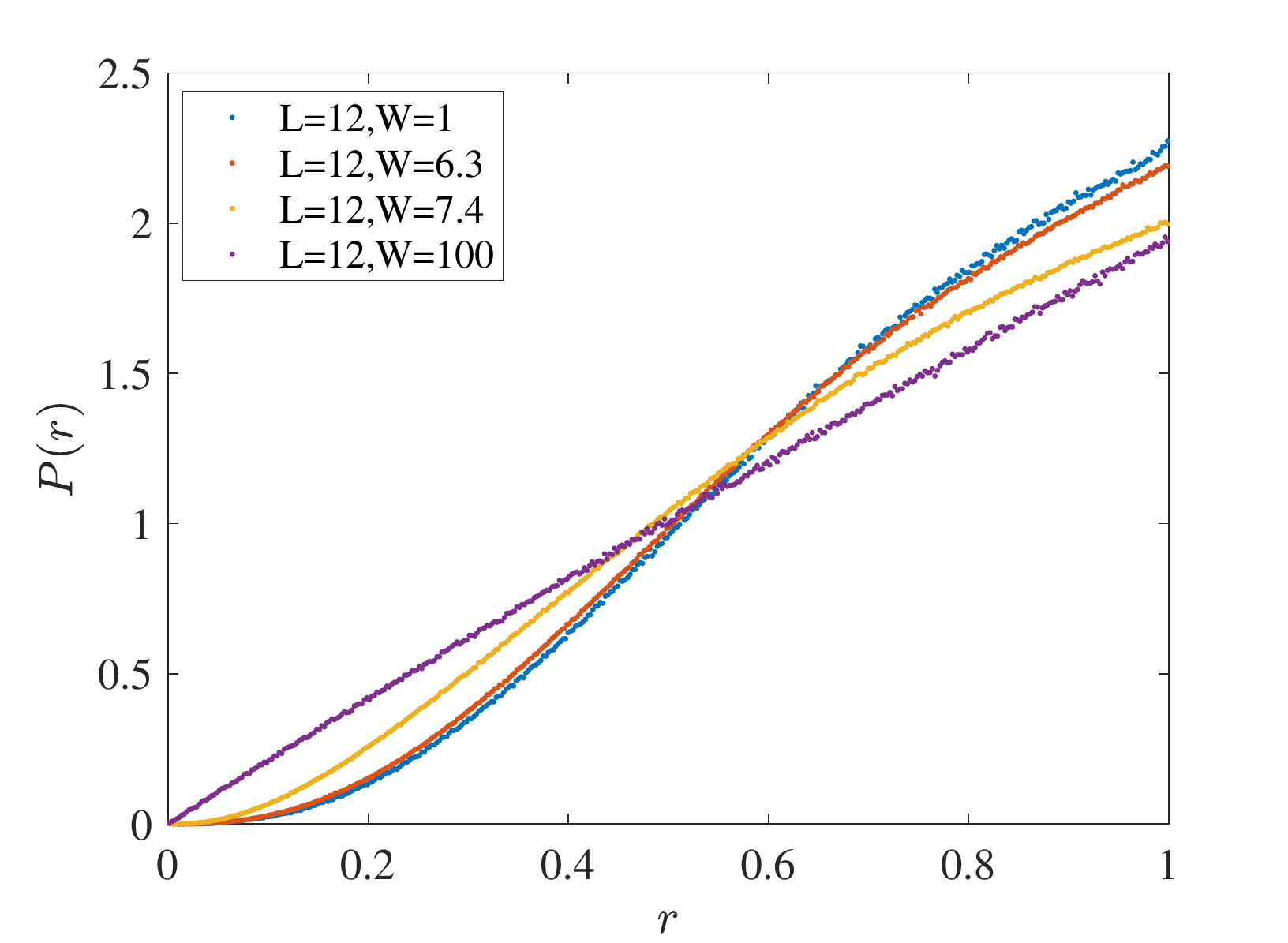}
		\end{minipage}%
	}%
	\subfigure[$P(r)$ for NH U(1) model in metal phase ($W=1, 6.24$) and insulator phase ($W=8.72, 100$)]{
		\begin{minipage}[t]{0.5\linewidth}
			\centering
			\includegraphics[width=1\linewidth]{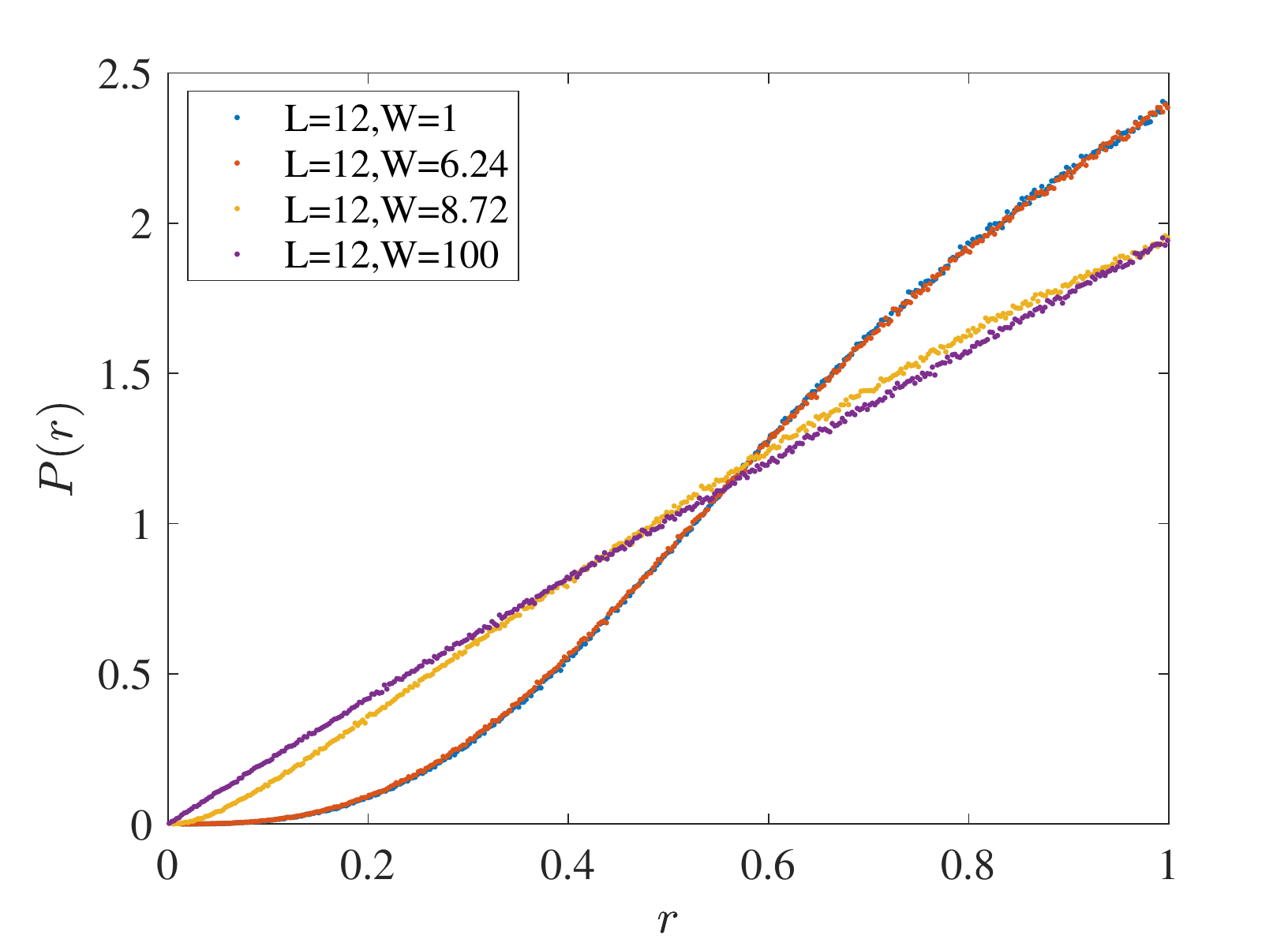}
		\end{minipage}%
	}%

\subfigure[$P(\theta)$  for NH AM]{
	\begin{minipage}[t]{0.5\linewidth}
		\centering
		\includegraphics[width=1\linewidth]{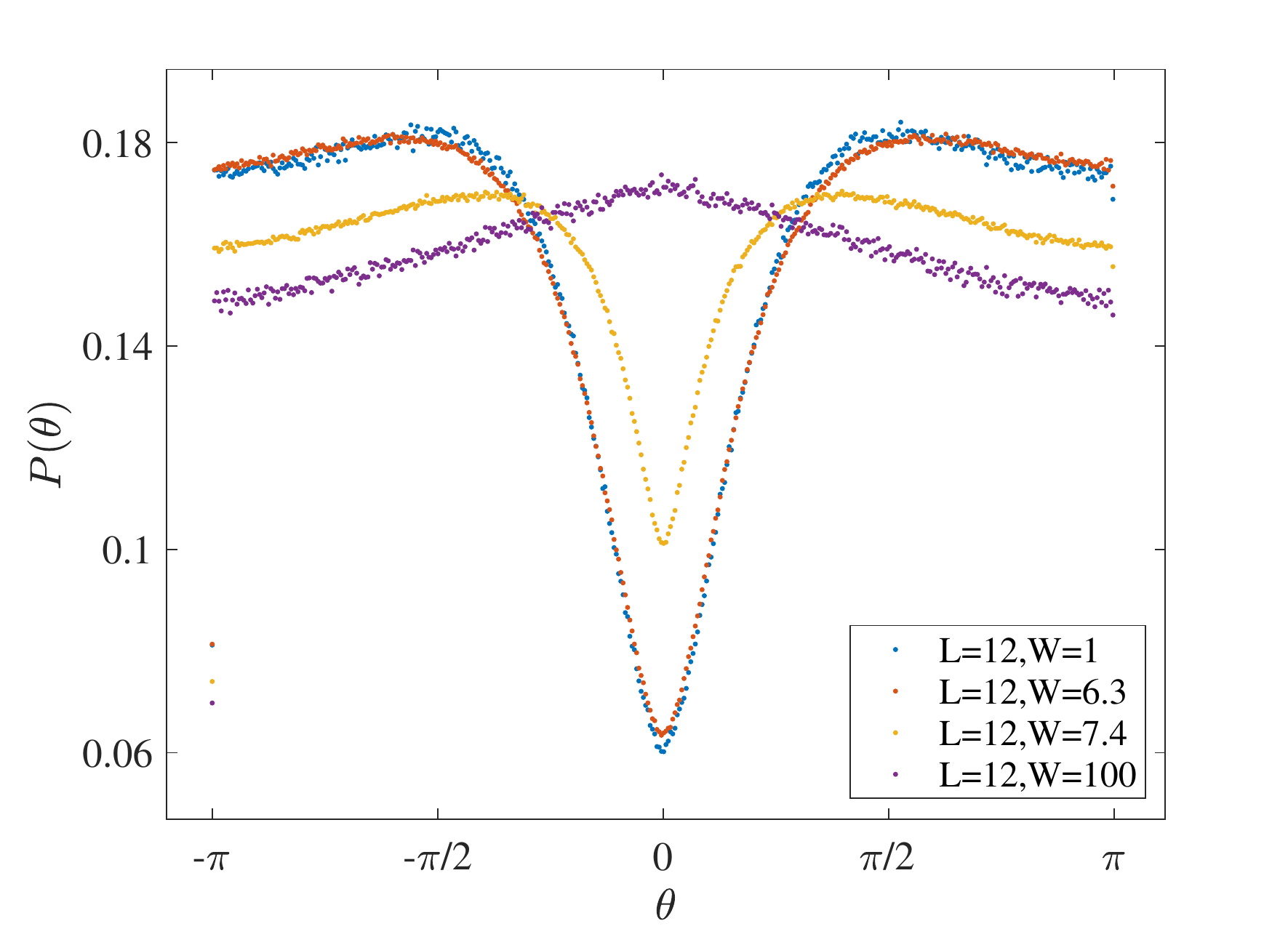}
	\end{minipage}%
}%
\subfigure[$P(\theta)$  for NH U(1) model]{
	\begin{minipage}[t]{0.5\linewidth}
		\centering
		\includegraphics[width=1\linewidth]{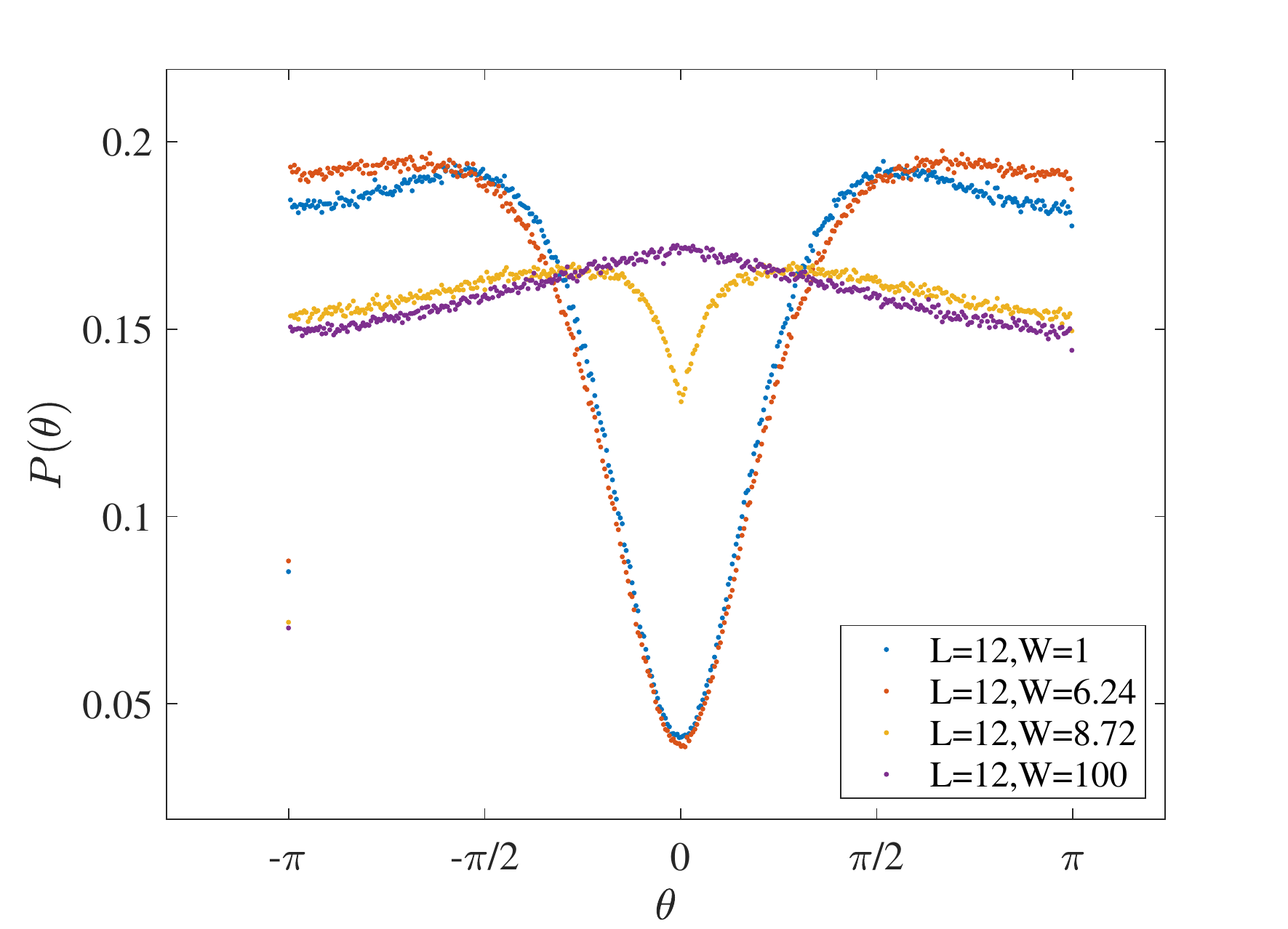}
	\end{minipage}%
}%

	\subfigure[$P(\theta)$  for NH AM and U(1) model]{
	\begin{minipage}[t]{0.5\linewidth}
		\centering
		\includegraphics[width=1\linewidth]{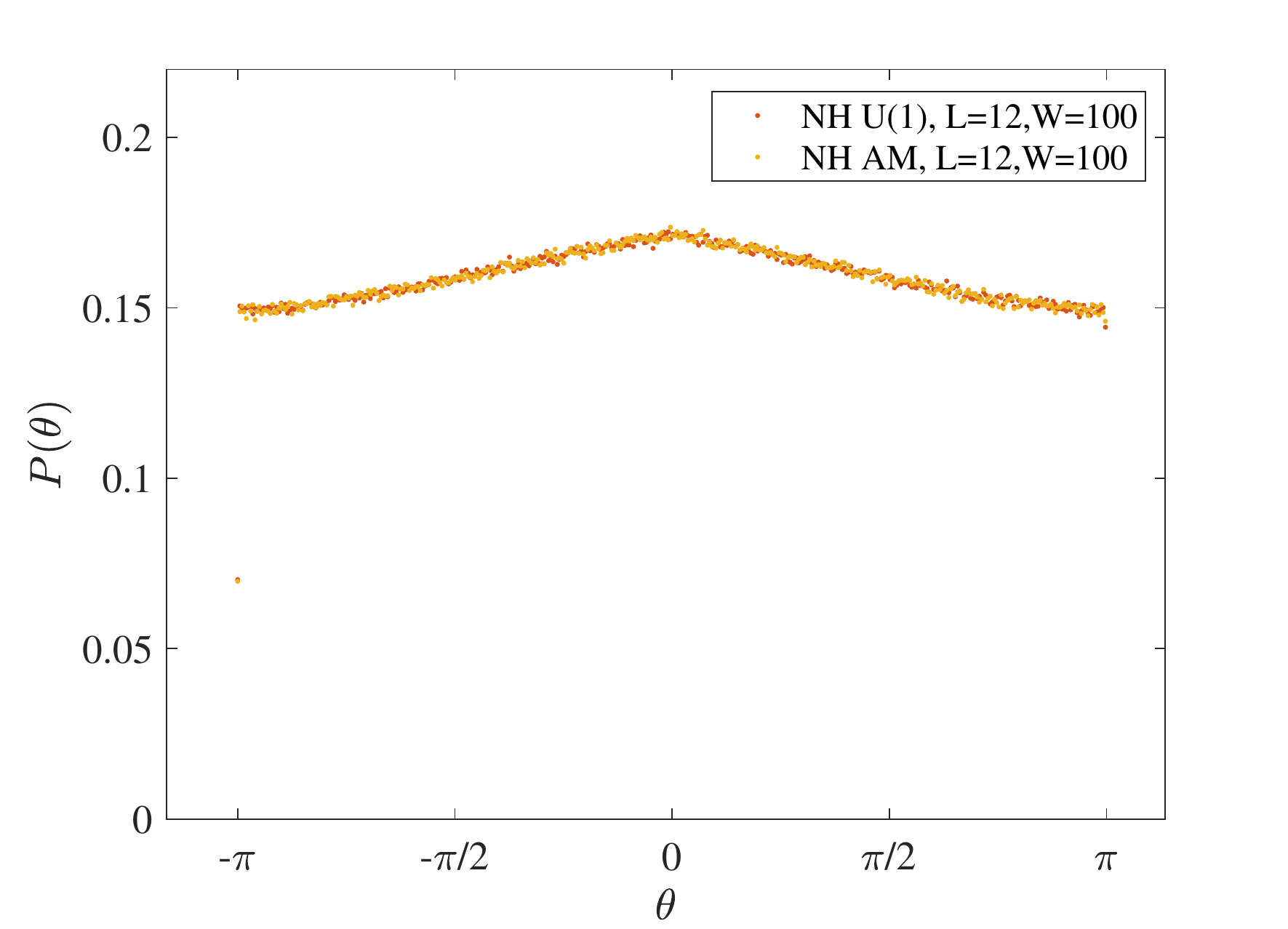}
	\end{minipage}%
}%
	\subfigure[$P(\theta)$ for NH U(1) model with $L=20$, $W=100$, $100\%$ eigenvalues.]{
	\begin{minipage}[t]{0.5\linewidth}
		\centering
		\includegraphics[width=1\linewidth]{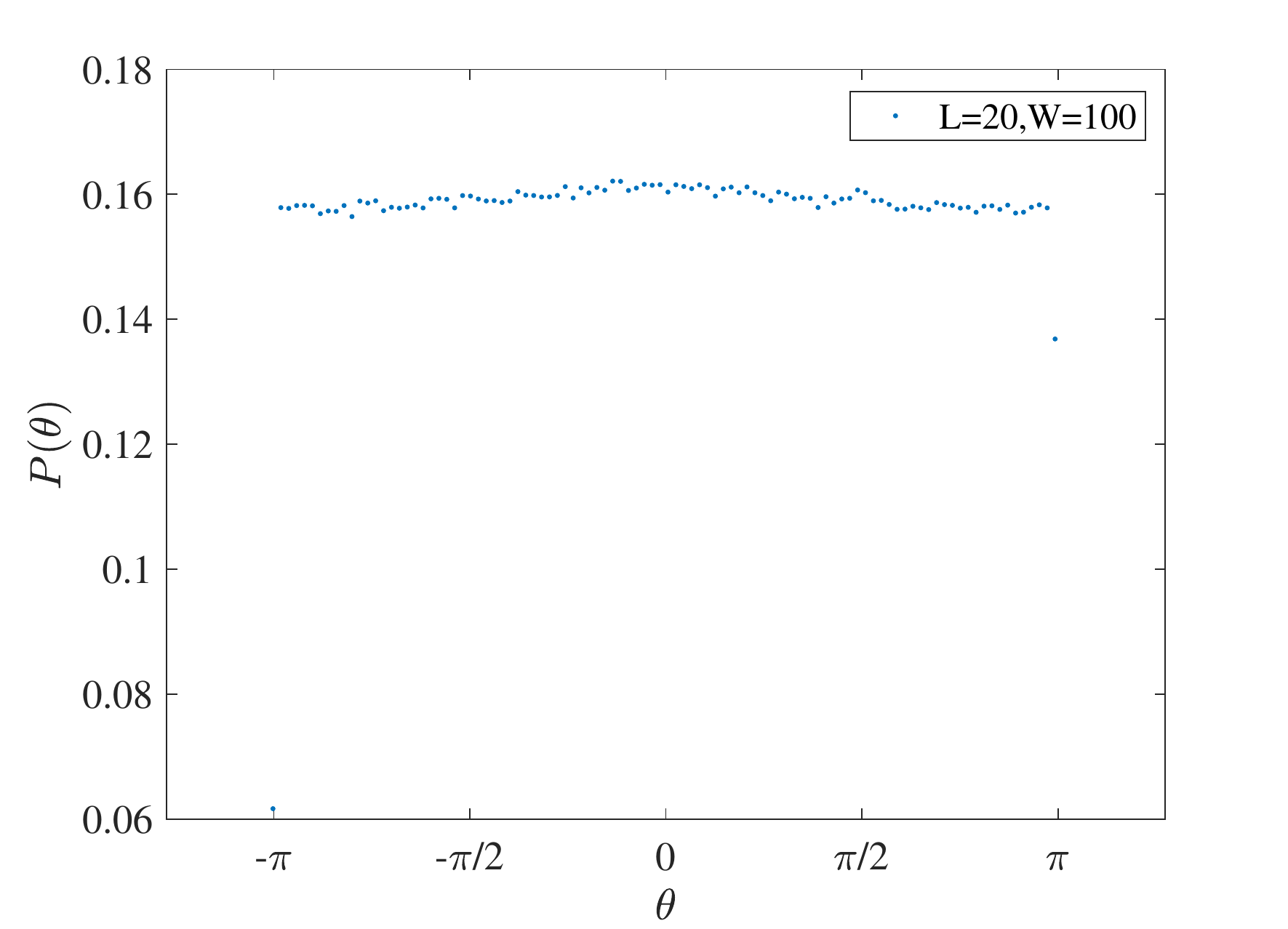}
	\end{minipage}%
}%
	\caption{$P(r)$ and $P(\theta)$  from metal to insulator phase (a)-(d), and (e) comparison of $P(\theta)$ between non-Hermitian (NH) 
    Anderson model (AM) and U(1) model. We take $10\%$ eigenvalues around $E=0$ over $6\times 10^4$ samples (NH U(1) model), 
    $3\times 10^5$ samples (NH AM at $W=6.3, 7.4$), and $6\times 10^4$ samples (NH AM at $W=1,100$). (f) 
   $P(\theta)$ in insulator phase of the NH U(1) model with $100\%$ eigenvalues.}
	\label{Pr_Ptheta_NH_O1_U1_Metal_insulator}
\end{figure}

\begin{figure}
	\centering
	\subfigure[$P(r)$ of the NH AM at the critical point]{
		\begin{minipage}[t]{0.5\linewidth}
			\centering
			\includegraphics[width=1\linewidth]{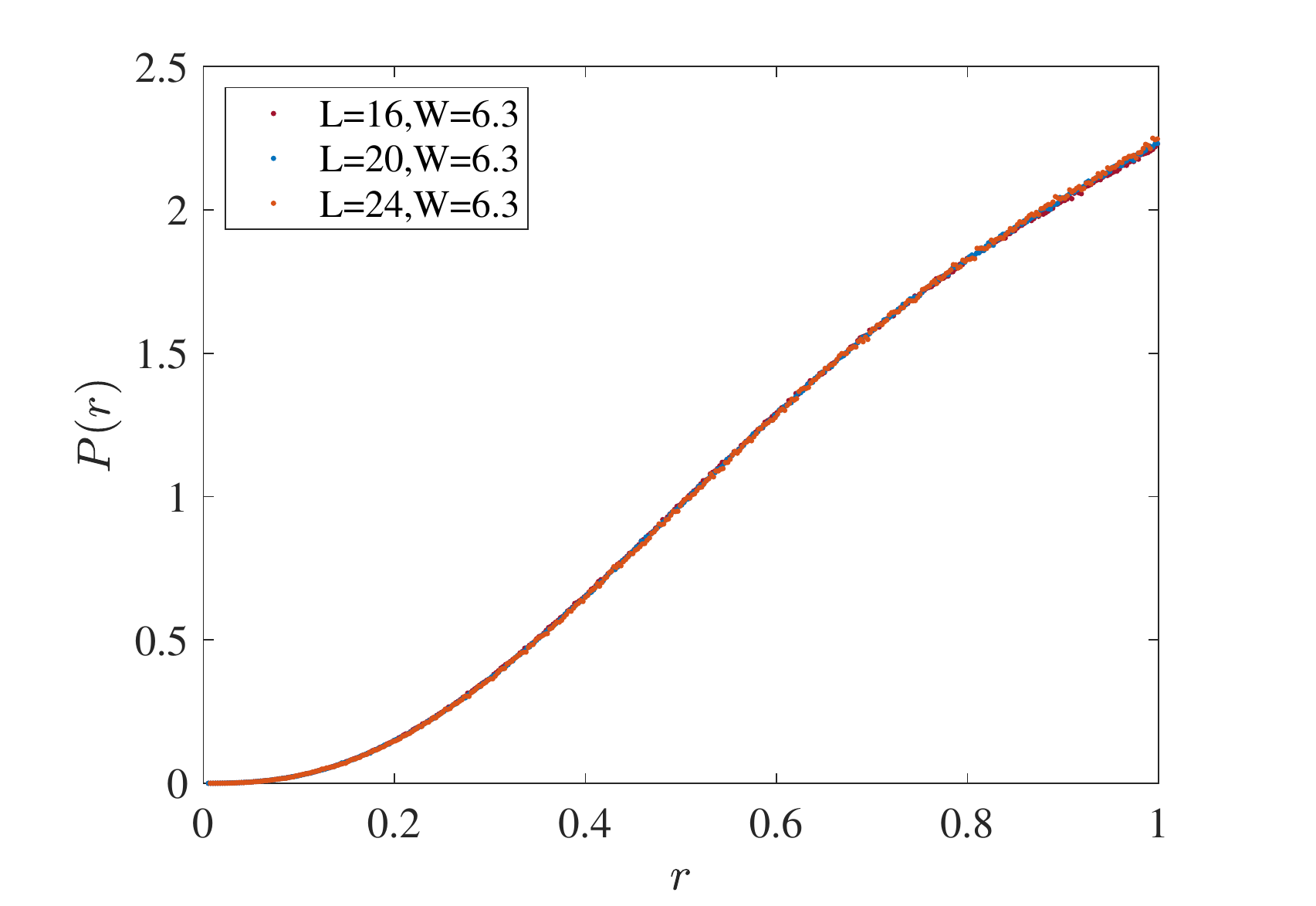}
		\end{minipage}%
	}%
	\subfigure[$P(r)$ of the NH U(1) model at the critical point]{
		\begin{minipage}[t]{0.5\linewidth}
			\centering
			\includegraphics[width=1\linewidth]{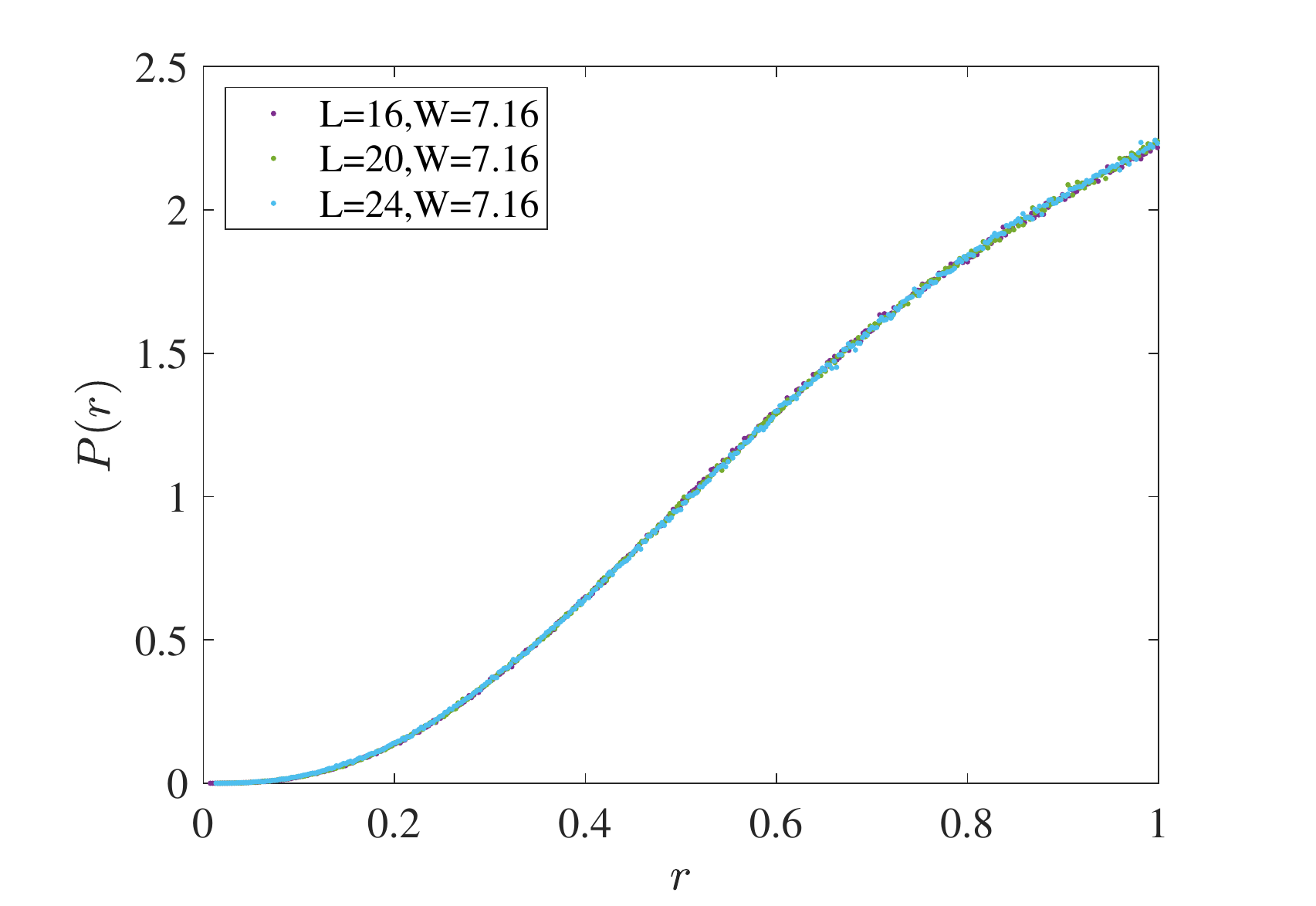}
		\end{minipage}%
	}%

	\subfigure[$P(\theta)$ of the NH AM at the critical point]{
	\begin{minipage}[t]{0.5\linewidth}
		\centering
		\includegraphics[width=1\linewidth]{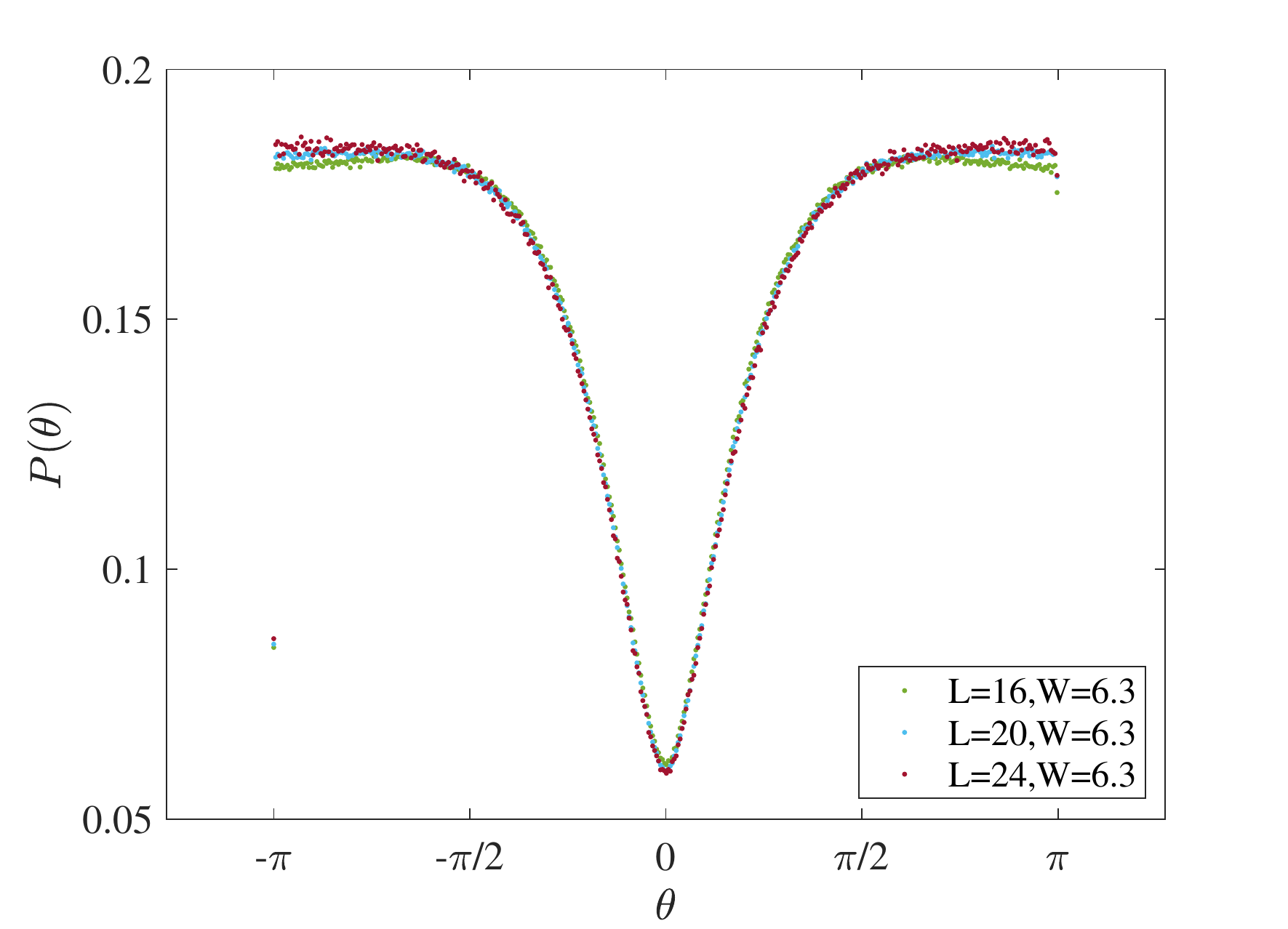}
	\end{minipage}%
}%
\subfigure[$P(\theta)$ of the NH U(1) model at the critical point]{
	\begin{minipage}[t]{0.5\linewidth}
		\centering
		\includegraphics[width=1\linewidth]{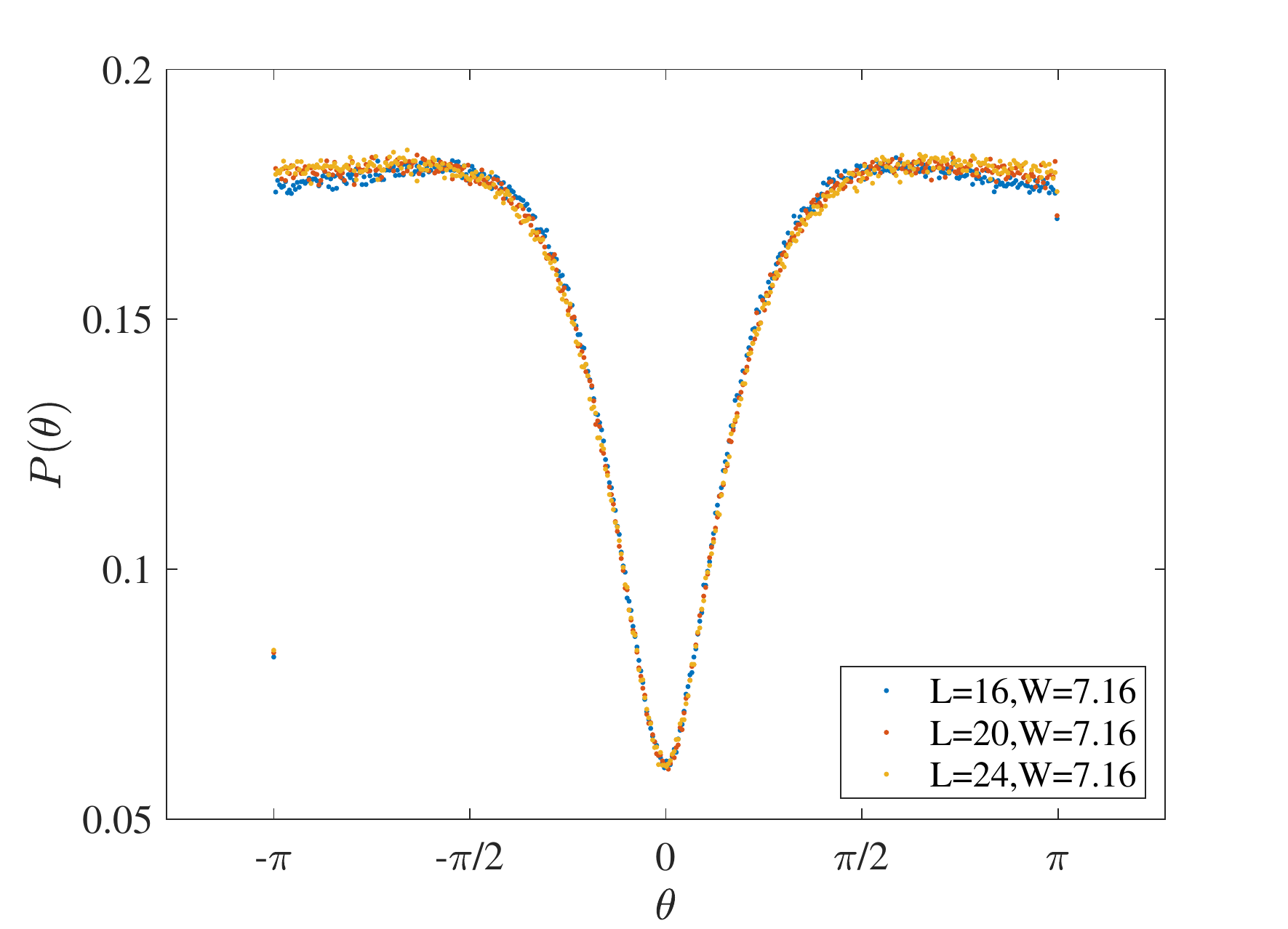}
	\end{minipage}%
}%

	\subfigure[critical $P(\theta)$ of the NH AM and U(1) model]{
	\begin{minipage}[t]{0.5\linewidth}
		\centering
		\includegraphics[width=1\linewidth]{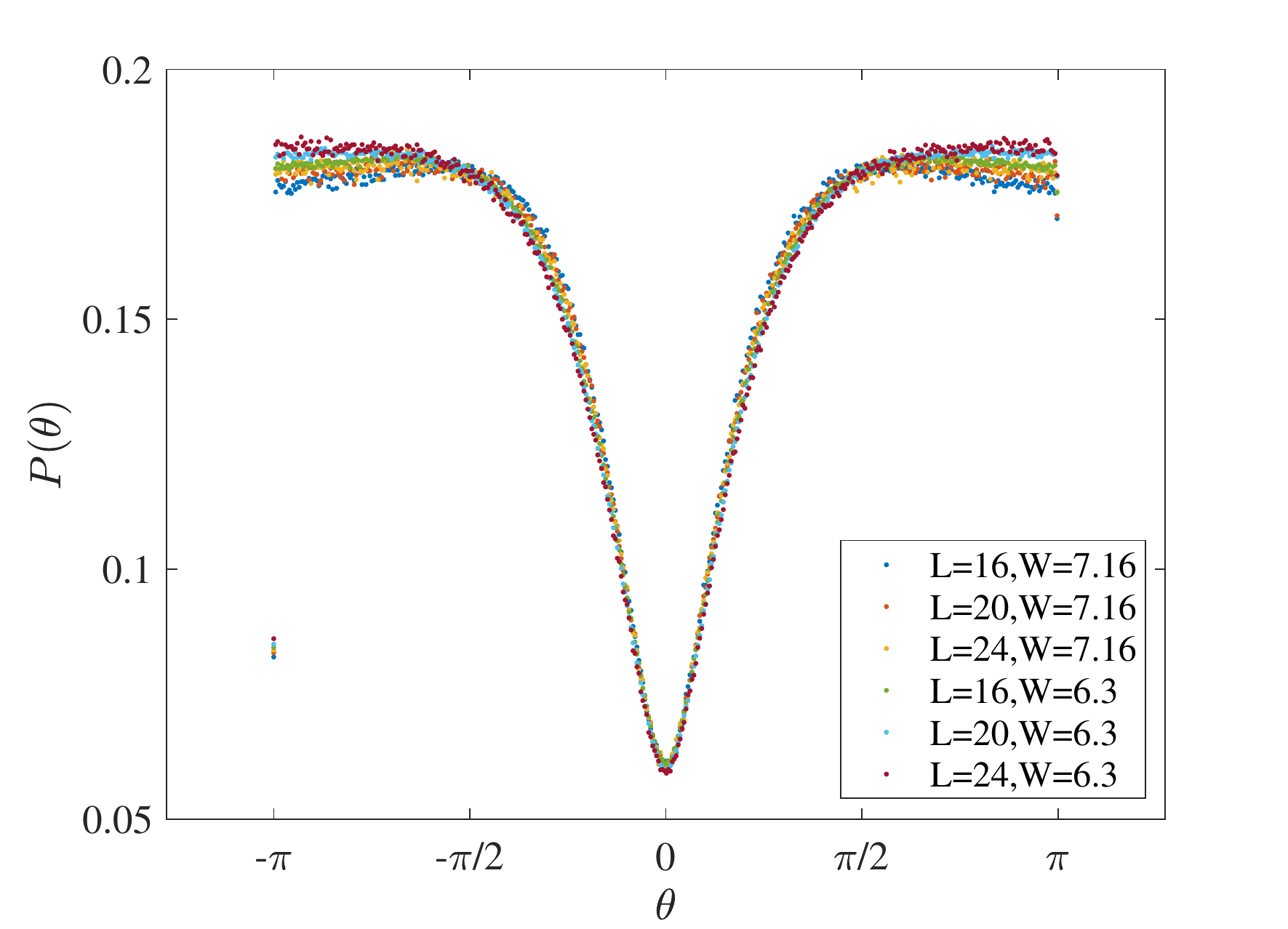}
	\end{minipage}%
}%
\subfigure[critical $P(r)$ of the NH AM and U(1) model]{
	\begin{minipage}[t]{0.5\linewidth}
		\centering
		\includegraphics[width=1\linewidth]{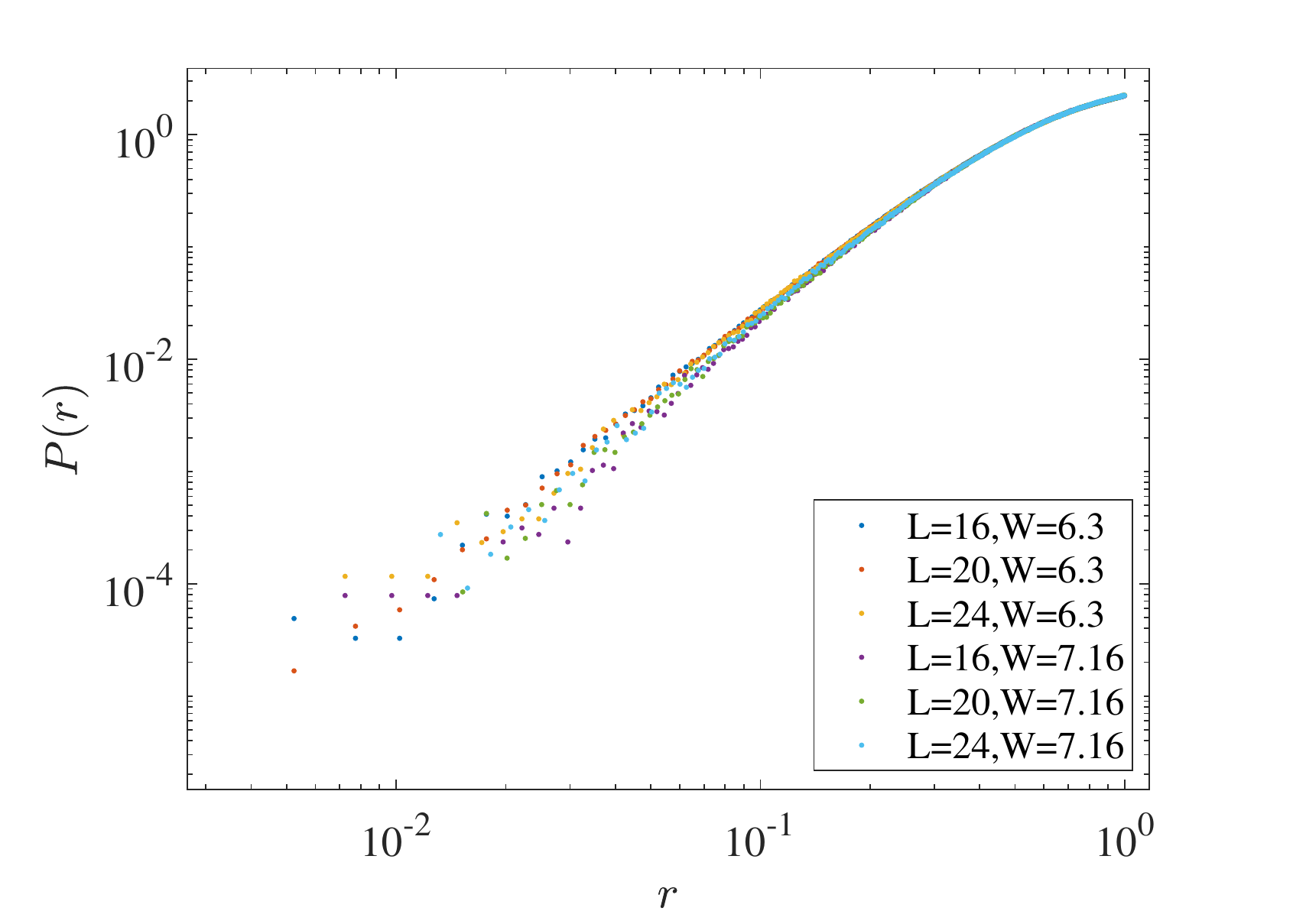}
	\end{minipage}%
}%
	\caption{ $P(r)$ and $P(\theta)$  at the critical point. The statistics for the non-Hermitian (NH) Anderson model (AM) are taken over 
    $M=1.2\times 10^5$, $6\times10^4$, $10^4$ samples for the system size $L=16$, $20$, $24$, respectively. The statistics for 
     NH U(1) model are taken over $M=2.5\times 10^4$, $1.2\times10^4$, $6400$ samples for the system size 
   $L=16$, $20$, $24$, respectively.}
	\label{Pr_Ptheta_NH_O1_U1_c}
\end{figure}
\end{widetext}

\end{document}